\documentclass[%
reprint,
superscriptaddress,
amsmath,amssymb,
aps,
prb,
]{revtex4-1}
\usepackage[T1]{fontenc}
\usepackage{xcolor}
\usepackage{float}
\usepackage{isotope}
\usepackage{graphicx}
\usepackage{dcolumn}
\usepackage{bm}

\begin{document}
	
	\title{Interface-related magnetic and vibrational
		properties in Fe/MgO heterostructures from nuclear resonant spectroscopy and first-principles calculations}
	
	\author{Benedikt Eggert}
	\thanks{corresponding author: Benedikt.Eggert@uni-due.de}
	\affiliation{Faculty of Physics and Center for Nanointegration Duisburg-Essen (CENIDE), University of Duisburg-Essen, Lotharstr. 1, D-47057, Duisburg, Germany}
	
	\author{Markus E. Gruner}
	\affiliation{Faculty of Physics and Center for Nanointegration Duisburg-Essen (CENIDE), University of Duisburg-Essen, Lotharstr. 1, D-47057, Duisburg, Germany}
	
	\author{Katharina Ollefs}
	\affiliation{Faculty of Physics and Center for Nanointegration Duisburg-Essen (CENIDE), University of Duisburg-Essen, Lotharstr. 1, D-47057, Duisburg, Germany}
	
	\author{Ellen Schuster}
	\affiliation{Faculty of Physics and Center for Nanointegration Duisburg-Essen (CENIDE), University of Duisburg-Essen, Lotharstr. 1, D-47057, Duisburg, Germany}
	
	\author{Nico Rothenbach}
	\affiliation{Faculty of Physics and Center for Nanointegration Duisburg-Essen (CENIDE), University of Duisburg-Essen, Lotharstr. 1, D-47057, Duisburg, Germany}
	
	\author{Michael Y. Hu}
	\affiliation{Advanced Photon Source, Argonne National Laboratory, Argonne, IL 60439, USA}
	
	\author{Jiyong Zhao}
	\affiliation{Advanced Photon Source, Argonne National Laboratory, Argonne, IL 60439, USA}
	
	\author{Thomas S. Toellner}
	\affiliation{Advanced Photon Source, Argonne National Laboratory, Argonne, IL 60439, USA}
	
	\author{Wolfgang Sturhahn}
	\affiliation{Advanced Photon Source, Argonne National Laboratory, Argonne, IL 60439, USA}
	\affiliation{Division of Geophysical and Planetary Sciences, California Institute of Technology, Pasadena, CA 91125, USA}
	
	\author{Rossitza Pentcheva}
	\affiliation{Faculty of Physics and Center for Nanointegration Duisburg-Essen (CENIDE), University of Duisburg-Essen, Lotharstr. 1, D-47057, Duisburg, Germany}
	
	\author{Beatriz Roldan Cuenya}
	\affiliation{Department of Physics, University of Central Florida, Orlando, Florida 32816, USA}
	\affiliation{Department of Interface Science, Fritz-Haber-Institute of the Max Planck Society, 14195 Berlin, Germany}
	
	\author{Esen E. Alp}
	\affiliation{Advanced Photon Source, Argonne National Laboratory, Argonne, IL 60439, USA}
	
	\author{Heiko Wende}
	\affiliation{Faculty of Physics and Center for Nanointegration Duisburg-Essen (CENIDE), University of Duisburg-Essen, Lotharstr. 1, D-47057, Duisburg, Germany}
	
	\author{Werner Keune}
	\affiliation{Faculty of Physics and Center for Nanointegration Duisburg-Essen (CENIDE), University of Duisburg-Essen, Lotharstr. 1, D-47057, Duisburg, Germany}

	\date{\today}
	
	\begin{abstract}
		We combine \isotope[57]Fe Mössbauer spectroscopy and \isotope[57]Fe nuclear resonant inelastic x-ray scattering (NRIXS) in nanoscale polycrystalline [bcc-\isotope[57]Fe/MgO] multilayers with various Fe layer thicknesses and layer-resolved density-functional-theory (DFT) based first-principles calculations of a (001)-oriented [Fe(8\,ML)/MgO(8\,ML)](001) heterostructure to unravel the interface-related atomic vibrational properties of a multilayer system. In theory and experiment, we observe consistently enhanced hyperfine magnetic fields compared to bulk which is associated with the Fe/MgO interface layers. NRIXS and DFT both reveal a strong reduction of the longitudinal acoustic (LA) phonon peak in combination with an enhancement of the low-energy vibrational density of states (VDOS) suggesting that the presence of interfaces and the associated increase in the layer-resolved magnetic moments results in drastic changes in the Fe-partial VDOS. From the experimental and calculated VDOS, vibrational thermodynamic properties have been determined as a function of Fe thickness and were found to be in excellent agreement. 
	\end{abstract}
	
	\keywords{Mössbauer spectroscopy, nuclear resonant inelastic x-ray scattering, phonon density of states, vibrational thermodynamics, density functional theory, interfaces, Fe/MgO multilayers}
	\maketitle
	
	\section{Introduction}
	
	Nanoscale Fe/MgO(001) multilayers have evolved as a paradigmatic example of a metal-insulator heterostructure. 
	The system has raised attention in the past as it combines a high abundance of its constituents with well understood epitaxial growth properties \cite{cn:Valeri07} and potential application scenarios in leading technology areas, such as data storage in terms of giant magnetoresistive elements in hard disk read heads \cite{cn:Parkin04,cn:Yuasa04} or for spintronics applications as a spin diode or rectifier to be used in magnetic logic elements \cite{cn:Iovan08}, spin light-emitting diodes (spin LEDs) \cite{Schuster2010}, or for devices with high output voltages \cite{cn:Tiusan06}. Such applications are essentially based on the large tunneling magnetoresistance (TMR) effect in Fe/MgO multilayer systems. A substantial fraction of previous research is therefore devoted to the electronic transport characteristics of Fe/MgO/Fe(001) trilayer tunnel junctions \cite{cn:Butler01,cn:Mathon01,cn:Tiusan04,cn:Waldron06,cn:Belashchenko05,cn:Heiliger08,cn:Peralta08,cn:Rungger09,cn:Feng09,cn:Abedi10,cn:Raza11}. One major outcome in the importance of these studies is the relevance of resonant interface states for the spin-dependent transport \cite{cn:Butler01}. Here, TMR is controlled by the electronic interface states in the minority spin channel  \cite{cn:Tiusan04,cn:Belashchenko05}. Complementary work was concerned with the electronic structure, magnetic moments of Fe/MgO(001) stackings and also magnetocrystalline anisotropy \cite{cn:Colonna09,cn:Choi11,cn:Bose2016,cn:Gu17,Koo2013,Okabayashi2014,KozioRachwa2014,cn:Balogh13,Suwardy2018}. The early density functional theory work of Li and Freeman investigated the properties of single and double layers of Fe on a MgO(001) surface \cite{cn:Li91}. The authors find enhanced surface moments of ~3\,$\mu_{\rm B}$. For the reverse setup, a single layer of MgO on Fe(001), DFT predicts a somewhat smaller but still enhanced moment in the interface Fe layer of 2.64\,$\mu_{\rm B}$ \cite{cn:Yu06}, while Ozeki {\em et al.} calculated 2.75\,$\mu_{\rm B}$ for a Fe$_5$/(MgO)$_5$ heterostructure \cite{cn:Ozeki07}. An enhanced moment of the Fe interface layer was also confirmed in experiment \cite{cn:Sicot03,cn:Miyokawa05,cn:Jal15}, although no interfacial Fe moment enhancement was inferred from other experiments \cite{Okabayashi2014}. As pointed out by Feng {\em et al.} \cite{cn:Feng09}, electronic structure, magnetization of the interface layer and consequently also the electronic transport properties are strongly affected by the interface relaxation and sensitive to the choice of the exchange-correlation functional in the calculations. 
	
	Most of the work described above was performed on epitaxial Fe/MgO(001) heterostructures. Besides, the investigation of magnetic properties of polycrystalline [Fe/MgO] multilayers can be found in the literature. For example, in the pioneering \isotope[57]Fe Mössbauer-effect study by Hine et al. \cite{Hine1979}, interface effects in ultrahigh-vacuum (UHV) deposited nanoscale polycrystalline [\isotope[56]Fe/MgO] multilayers were enhanced by inserting isotopically enriched Mössbauer active \isotope[57]Fe probe layers at the \isotope[56]Fe/MgO interfaces, and the interface hyperfine magnetic field in the magnetic ground state was observed to be enhanced relative to that of bulk bcc Fe. Later, Koyano et al. \cite{Koyano1988} reported enhanced hyperfine magnetic fields, enhanced magnetic moments and perpendicular Fe spin texture in nanoscale polycrystalline [Fe/MgO] multilayered films. More recently, a vortex-like magnetic domain structure in polycrystalline [Fe/MgO] multilayers was inferred from \isotope[57]Fe conversion electron Mössbauer spectroscopy (CEMS) and magnetooptical Kerr effect (MOKE) studies by Koziol-Rachwal et al. \cite{KozioRachwa2014}. All of these investigations show that the structural and magnetic properties of the interfaces in Fe/MgO heterostructures are of paramount importance for controlling the properties of this system. 
	
	Although a vast amount of literature reports on the Fe/MgO system is motivated by its fundamental and technological relevance, there is no experimental investigation of the lattice dynamics and vibrational thermodynamics of the Fe/MgO system, to the best of our knowledge. The phonon dispersion relations and the vibrational (phonon) density of states (VDOS, g(E)) of bulk MgO are well known experimentally from inelastic neutron diffraction \cite{Sangster1970} and inelastic x-ray scattering \cite{Ghose2006} and theoretically from DFT-based computations \cite{Parlinski2000}, while the dispersion and phonon spectra of bulk bcc Fe have been reported in Refs. \onlinecite{Hellwege1981,cn:Neuhaus97,cn:Koermann08,cn:Leonov12,cn:Neuhaus14}.
	The fundamental question of how the VDOS in nanoscale metal/metal or metal/insulator multilayers is modified (as compared to bulk materials) is rather unexplored. The methods of inelastic neutron scattering or inelastic x-ray scattering remain a challenge because of insufficient sensitivity. On the other hand, the isotope-selective method of \isotope[57]Fe nuclear resonant inelastic x-ray scattering (NRIXS) has been shown to have monolayer (ML) sensitivity with respect to \isotope[57]Fe ultrathin films in order to determine the partial Fe specific VDOS, \cite{Slzak2007}. Applying \isotope[57]Fe NRIXS to various nanoscale [Fe/metal] multilayers, the observed impact of the Fe film thickness and type of metal on the shape of the partial-Fe g(E) has been interpreted by confinement of high energy Fe phonons in the Fe films \cite{RoldanCuenya2008}. Moreover, the influence of interfaces on the VDOS of nanoscale metallic multilayers in terms of phonon confinement and phonon localization could be determined by \isotope[57]Fe probe-layer NRIXS and atomic-layer resolved DFT-based calculations \cite{Keune2018}. First-principles calculations for monolayer-scale Fe(001)/Au(001) superlattices predicted drastic variations of g(E) with individual Fe and Au thicknesses \cite{Sternik2006}. As it has been outlined in Ref. \onlinecite{Keune2018,cn:Rothenbach19}, the modifications of g(E) in nanoscale multilayers are anticipated to influence the phonon transport and the vibrational thermodynamic properties in these systems. Therefore, \isotope[57]Fe NRIXS measurements and DFT-based computations of the phononic properties of [Fe/MgO] multilayers are highly desirable, in particular in view of the fact that Fe/MgO is of interest as a prototype system for the study of time-resolved phenomena on the femtosecond scale \cite{cn:Rothenbach19,cn:GrunerRTTDDFT}. 
	
	In the present work, we focus on the investigation of the Fe-specific atomic vibrational dynamics and thermodynamics in polycrystalline nanoscale [Fe/MgO] multilayers by \isotope[57]Fe NRIXS, accompanied by element- and layer-resolved state-of-the-art
	DFT-based computations of the layer-resolved phonon dispersions.
	In both experiment and theory, we observe distinct differences between the phononic properties of the multilayers and the bulk material, which change consistently with layer thickness.
	The multilayers were characterized with respect to their magnetic properties and hyperfine magnetic properties by magnetometry and \isotope[57]Fe CEMS, respectively, while the DFT-based calculations provided Fe layer resolved magnetic moments and electronic properties. The predictions of the DFT calculations are found to be in
	good agreement with our experimental results.


	\section{Experimental Details}
	
\begin{figure}
		\includegraphics[width=.85\linewidth]{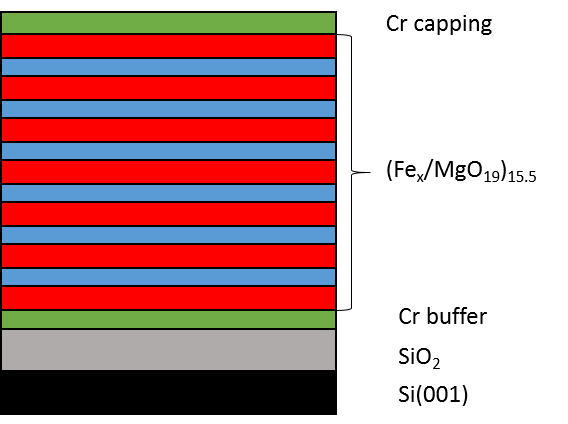}
		\caption{Schematic representation of the hetereostructure. For three studied multilayers we have varied the number of Fe monolayers ($t_{Fe}$=55, 28 and 10\,ML), while keeping the MgO thickness (effectively 19\,ML) constant. A fourth multilayer had the composition [Fe(7\,ML)MgO(5\,ML)]$_{15.5}$.}
		\label{fig:layer-scheme}
		\vspace{-15pt}
	\end{figure}

	Polycrystalline $\left[\mathrm{\isotope[57]Fe/MgO}\right]_{15.5}$ multilayers, with Fe thicknesses $t_{Fe}$= 1.5\,nm, 4\,nm and 8\,nm and with a MgO-thickness $t_{MgO}$ of 4\,nm were grown at room temperature in ultrahigh vacuum (UHV) by thermal evaporation on naturally oxidized Si(001) substrates covered with a 4-nm thick Cr buffer layer. All samples were capped with 5\,nm Cr for protection. A fourth sample had thinner layers of $t_{Fe}$=1\,nm and $t_{MgO}$ = 1\,nm. Every sample consists of 15.5 bilayers. The \isotope[57]Fe isotopic enrichment is 95\%. For additional information concerning the sample preparation, we refer the reader to our Supplemental Material \cite{Supplement}. The crystalline nature of the individual layers was confirmed with x-ray diffraction (XRD) \cite{Supplement}. For better comparability between the discussed theoretical and experimental results the thickness of the individual Fe and MgO layers can be expressed in units of atomic Fe(001) and MgO(001) geometrical monolayers, which corresponds to 1.45\,\AA{} for Fe and 2.1\,\AA{} for MgO, respectively. This conversion means that for this work a multilayer structure with a \isotope[57]Fe layer thickness of 10\,ML, 28\,ML and 55\,ML with 19\,ML for MgO have been used. For the fourth sample this conversion resulted in 7\,ML \isotope[57]Fe and $\sim$ 5\,ML MgO
	For a schematic model of the heterostructure, we refer to Fig \ref{fig:layer-scheme}. 
	
	\isotope[57]Fe Mössbauer spectroscopy in zero external field at perpendi\-cular incidence of the $\gamma$-rays onto the film surface was performed by detection of conversion electrons (CEMS). For the detection of the electrons, the sample was installed in a proportional gas counter, i.e. housing with a continuous He gas flow mixed with 4\% CH$_4$ to avoid ionization processes. In addition, low temperature CEMS at T = 80\,K was feasible by using a channeltron detector. For the measurement, a constant acceleration Mössbauer driving unit was used with a $^{57}$Co source embedded in a Rh matrix, while the velocity of the spectrometer was calibrated with a $\alpha$-Fe foil reference sample at room temperature. The experimental spectra were evaluated by a least-squares fitting routine using the $Pi$ program package \cite{PiLink}, and the discussed isomer shifts $\delta_{iso}$ are given relative to bulk bcc-Fe at room-temperature. Additional field-dependent magnetization measurements were performed using a Quantum Design PPMS DynaCool with a VSM option applying an external magnetic field up to 9\,T achieving a temperature range between 1.8\,K up to 400\,K. Nuclear resonant inelastic x-ray scattering experiments were performed at the beamline 3-ID of the Advanced Photon Source, Argonne National Laboratory. The energy of the x-ray beam was tuned around the \isotope[57]Fe nuclear resonance energy of 14.412\,keV with an energetic bandwith of 1.3\,meV with the use of a high-resolution monochromator. The x-ray beam was focussed to 0.2 $\times$ 1.0 mm$^2$ (V$\times$H) after passing through a Kirckpatrick-Baez mirror, while the measurements were performed in grazing incidence relative to the film plane. An avalanche photodiode (APD) was used to detect the delayed incoherent inelastic signal by measuring the Fe $K_{\alpha}$ fluorescence radiation. Furthermore, the instrumental resolution function was determined by measuring the nuclear forward scattering intensity for each sample. For a detailed introduction into NRIXS and the corresponding data evaluation, we refer the reader to Refs. \cite{Seto1995,Sturhahn1995,Chumakov1995,Rhlsberger2005,Chen2007LatticeDynamics,Sturhahn2004}. In the present case, the Fe-partial VDOS were obtained from the NRIXS spectra employing the PHOENIX software (version 3.0.1\cite{PHOENIX}) by W. Sturhahn \cite{Sturhahn2000,Sturhahn2004}.
	
	\section{Computational Details}
	We compare the experimental results with first-principles calculations of a Fe$_8$/(MgO)$_8$(001) heterostructure in the framework of density functional theory (DFT). For structural relaxation of the 24 atom primitive cell with respect to the atomic positions and cell parameters and the subsequent calculation of the phonon dispersion, we employed the VASP code \cite{cn:VASP1,cn:VASP2} using PAW potentials with the electron configurations 3$p^6$3$d^7$4$s^1$ for Fe, 2$p^6$3$s^2$ for Mg and 2$s^2$2$p^4$ for O and an energy-cutoff of 580\,eV. For the exchange-correlation functional, we used the generalized gradient approximation (GGA) in the formulation of Perdew, Burke and Ernzerhof (PBE) \cite{cn:Perdew96}. Results were considered converged, when the energy fell below $10^{-8}$\,eV between two consecutive electronic and $10^{-6}$\,eV between two geometric optimization steps. Brillouin zone integration was carried out on 90 $k$-points in the irreducible zone (IBZ) in combination with a Gaussian-type Fermi-surface smearing with a width of $\sigma$$\,=\,$$0.1\,$eV. For the electronic DOS, we applied the tetrahedron method with Bl\"ochl-corrections \cite{cn:Bloechl94} and 408 $k$-points for the Brillouin-zone integration. The phonon dispersion and vibrational density of states were determined with the so-called direct approach based on the restoring forces obtained from central differences between $2\times 24$ displacements of inequivalent atoms in a 768 atom supercell, constructed as a $4\times 4\times 2$ replication of the 24 atom primitive cell. Here, a k-mesh of $5\times 5\times 1$ points in the full Brillouin zone (FBZ) in combination with Gaussian smearing of width $\sigma$$\,=\,$$0.05\,$eV was employed which guarantees the required accuracy. Finally, the dynamical matrix and the vibrational density of states were obtained after employing the PHON code by Dario Alf\`e \cite{cn:Alfe09PHON}. The methodological details are the same as reported in Ref.\ \onlinecite{cn:Rothenbach19}. 
	
	In addition, we calculated the M\"ossbauer hyperfine parameters based on the previously optimized geometry using the Elk full-potential augmented plane wave code, version 6.2.8 \cite{cn:Elk628}, with the PBE exchange-correlation functional and spin-orbit interaction included. 
	Core polarization was taken into account, and the semi-core Fe-3s states were moved to the core and thus treated with the full Dirac equation. In Elk version 6.2.8, dipole fields are included self-consistently and spin and current densities of the entire crystal are used to calculate spin and orbital dipole contributions. A cutoff parameter $R\,K_{\rm max}$$\,=\,$8 for the plane waves and a maximum angular momentum $l_{\rm max}$$\,=\,$9 for the APW functions was used in combination with Gaussian-type smearing with a width of $\sigma=0.0272\,$eV, 114 $k$-points in the IBZ (magnetization assumed collinear and perpendicular to the layering), and muffin tin radii of 1.058\,\AA{} for Fe, 1.141\,\AA{} for Mg and 0.934\,\AA{} for O. Convergence criterion for electronic selfconsistency was a root mean square change of $10^{-7}$ in the Kohn-Sham potential. To ensure convergence, we used a very large number of 2304 empty states, which is almost half the total number of 4850 valence states, adding in addition 2096 conduction state orbitals.
	

	\section{Results \& Discussion}

	\subsection{Magnetic characterization}
	
	Mössbauer spectroscopy \cite{G_tlich_2011} probes slight deviations in the energy levels of the \isotope[57]Fe nuclei due to the presence of hyperfine interactions, and reveals the valence state of the iron atom via the isomer shift $\delta_{iso}$ and the orientation of the Fe-spin relative to the incident $\gamma$-ray. The hyperfine magnetic field, $B_{hf}$, indicates magnetic ordering on a local scale. In some bulk Fe alloys, the hyperfine field is found to be proportional to the atomic Fe magnetic moment \cite{Vincze_Kaptas_Kemeny_Kiss_Balogh_1994}. For epitaxial Fe(001)/MgO(001), a 24\% increase of $\mu_{Fe}$ at the interfaces was experimentally observed \cite{cn:Jal15}, while hyperfine field values above 36\,T (i.e. larger than the bulk value) were reported in Ref. \cite{cn:Balogh13} Thus, an increase of $B_{hf}$ appears to be correlated with an enhancement of $\mu_{Fe}$ at the Fe/MgO interface. The intensity ratio of sextet lines number 2 (5) and 3 (4), in the following referred to as $A_{2,3}$-ratio, describes the average angle $\theta$ between the Fe-spin and the incident $\gamma$-ray, by the formula.
	\begin{equation}
	A_{2,3}=\dfrac{I_2}{I_3} = \dfrac{4\sin^2(\theta)}{1+\cos^2(\theta)}.
	\label{eq:canting-angle}
	\end{equation} 
	Therefore, the $A_{2,3}$-ratio can vary between 0 ($\Theta=0^{\circ}$ with spin orientation out-of-plane) and 4 ($\Theta=90^{\circ}$ with spin orientation in-plane), if the incident $\gamma$-ray is impinging the sample in a perpendicular geometry.
	
\begin{figure}[H]
		\includegraphics[width=.92\linewidth]{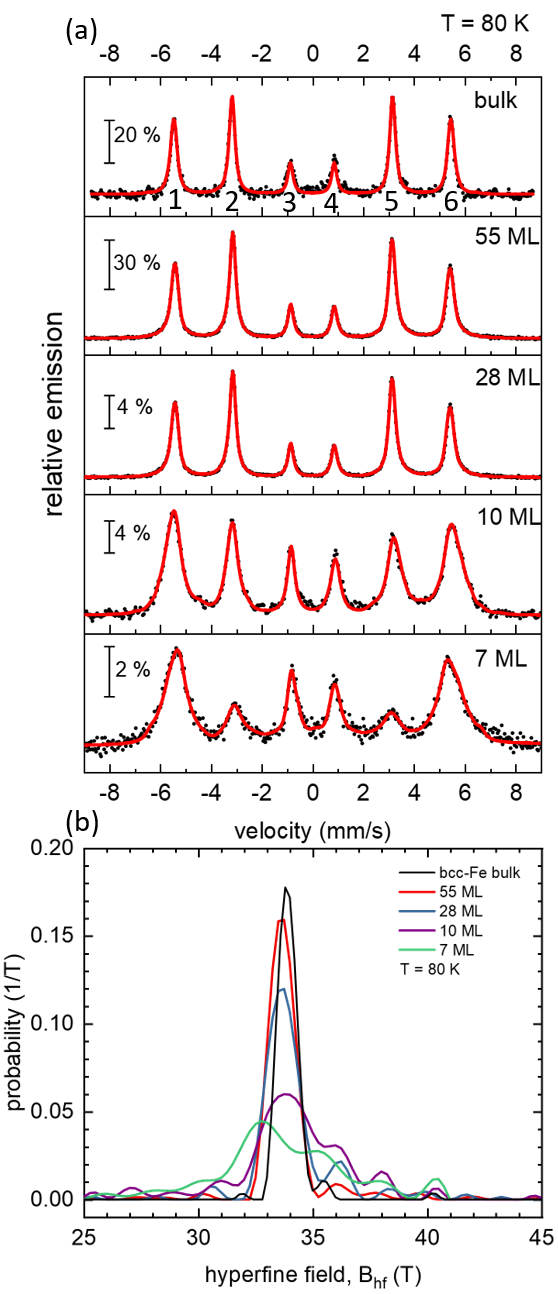}
		\vspace{-10pt}
		\caption{(a) Result of zero-field CEMS measurements at $T = 80\,\mathrm{K}$ for bulk bcc-Fe (top) and Fe/MgO multilayers with varying Fe thicknesses $t_{Fe}$ of 55, 28 and 10 monolayers (ML) and a MgO thickness of 19\,ML and for a multilayer with $t_{Fe}$ = 7\,ML and $t_{MgO}\approx$ 5\,ML. Black dots: experimental data; red lines: least-squares fitted curves using the corresponding hyperfine-field distributions $p\left(B_{hf}\right)$ obtained from the fittings and shown in (b). The sextet lines are numbered in (a) for bcc-Fe as an example. In (a), a clear reorientation of the Fe spins from preferred in-plane to preferred out-of-plane direction with decreasing Fe thickness is evident from the reduction of the relative intensity of lines \#2 and \#5. In (b), a broadening of $p\left(B_{hf}\right)$ with decreasing Fe thickness is observed, which extends up to large values $B_{hf}$ of about 40\,T.}
		\label{fig:CEMS}
	\end{figure}

	Due to observed thermal relaxation phenomena at room temperature (discussed in the Supplemental Material, Ref. \cite{Supplement}), zero-field CEMS measurements were performed at T=80\,K. These measurements reveal magnetic ordering for all investigated multilayer structures and for a bulk bcc-Fe foil (Fig\ref{fig:CEMS}(a)). Motivated by the increasing apparent linewidths with decreasing Fe thickness, each spectrum is analyzed in terms of a distribution of hyperfine fields $p\left(B_{hf}\right)$. The distributions are shown in Fig. \ref{fig:CEMS}(b) and a full summary of the obtained fitting parameters is given in Table \ref{tab:CEMS-parameter}.
	
	The bulk bcc-Fe foil shows a sextet structure with an average hyperfine splitting $\langle B_{hf}\rangle$ of 33.9\,T, while the standard deviation of the hyperfine field distribution yields a value of only 0.7\,T, combined with an intrinsic narrow linewidth $\Gamma$ of 0.236\,mm/s, indicating a single magnetic site, as expected. From the line intensity ratio $A_{2,3}$ we infer a preferential in-plane alignment of the Fe spins (canting angle of 74.7$^{\circ}$), while a complete in-plane spin orientation is suppressed, due to the bulk character of the sample and the corresponding magnetic multidomain structure. In addition, a slightly elevated isomer shift $\delta_{iso}$ of 0.11\,mm/s is present (relative to bulk bcc-Fe at 300\,K). This increased isomer shift, relative to bulk bcc-Fe at room temperature, is due to the low measurement temperature of 80\,K and the corresponding change of the second order Doppler (SOD) shift \cite{Greenwood_1971}. Taking into account a Debye temperature $\Theta_D$ of 420\,K for bulk bcc Fe, a reasonable agreement with the known room temperature values can be obtained.
	
	For the multilayer structures with an Fe thickness $t_{Fe}$ of 55 and 28ML similar sextet spectra appear, with the average field $\langle B_{hf}\rangle$ unchanged within error margins, (Table \ref{tab:CEMS-parameter}), while a broadening of the $p\left(B_{hf}\right)$ distribution towards lower and higher fields occurs with decreasing Fe thickness, resulting in an increased standard deviation $\sigma_{B_{hf}}$ of 1.5 and 1.8\,T, for intrinsic line widths of 0.258 and 0.262\,mm/s, respectively. This broadening is attributed to various local environments of \isotope[57]Fe atoms in the interfacial Fe/MgO region due to the inevitable interface roughness and due to different crystallite orientation in the polycrystalline samples.
	Noticeably, a peak at $B_{hf} =36$\,T occurs in the hyperfine field distribution, which can be ascribed to the Fe/MgO interface \cite{cn:Balogh13}. Also, films with these two thicknesses have a preferred in-plane spin orientation due to shape anisotropy, while a tendency of a slightly increased isomer shift $\delta_{iso}$ is present for decreasing Fe thicknesses relative to the bulk. This indicates very small electronic charge transfer between interfacial Fe and MgO atoms.
	
	For the multilayer sample with an Fe thickness of 10\,\,ML the distributions broadens even more, reaching a standard deviation $\sigma_{B_{hf}}$ of 2.7\,T and an intrinsic linewidth of 0.343\,mm/s, combined with a change of the line intensity ratio $A_{2,3}$ to 1.95 ($\Theta=54.1\,^{\mathrm{\circ}}$). Such a ratio is equivalent to a random spin orientation. The thinnest multilayer sample with an Fe thickness of 7\,\,ML reveals an even stronger broadening of the hyperfine field distribution $p\left(B_{hf}\right)$ ($\sigma_{B_{hf}}$ = 2.9\,T) and an increased linewidth $\Gamma$ of 0.401\,mm/s. For this sample the obtained spectrum reveals a preferential out-of-plane spin orientation ($A_{2,3} = 0.74$, $\Theta=34\,^{\mathrm{\circ}}$). Even though this is the lowest Fe-thickness investigated in this work, the metallic character of the Fe layers is proven for this sample (and for the other samples) indicating, that no iron oxide formation occurred, for example at the Fe/MgO-interface. In fact, no evidence of FeO (wustite) phase formation in our samples is provided by our CEMS spectra at 300 K (see Supplemental Material \cite{Supplement}), because FeO at 300 K is characterized by a paramagnetic asymmetric feature with a main peak at an isomer shift near + 1 mm/s \cite{McCammon1985}, which is not observed here. In Table \ref{tab:CEMS-parameter}, a detailed summary of the Mössbauer fitting parameters at T = 80\,K is shown.
	Summarizing, we have characterized the hyperfine magnetic and electronic properties with varying \isotope[57]Fe layer thicknesses, and we can demonstrate that the Fe layers in all of these samples exhibit metallic behaviour, similar to bcc-Fe, but with occasional trends.
	First, we discussed the broadening of the hyperfine field distribution $p(B_{hf})$ with decreasing Fe thickness $t_{Fe}$ and the occurrence of a 36 T high-field contribution – a signature contribution of the Fe/MgO interface \cite{cn:Balogh13,Hine1979,Koyano1988}. This increased hyperfine splitting (relative to the bulk) is also determined in our ab-initio DFT-calculations (section B). The changes of the hyperfine-field distribution width $\sigma_{B_{hf}}$ with $t_{Fe}$ can be qualitatively explained by interface roughness and crystallites at the Fe/MgO interfaces with different orientations. The stronger contribution of this Fe/MgO interface with decreasing Fe-thickness is also evident from the rising intrinsic linewidth $\Gamma$. Furthermore, the measurements reveal a clear spin reorientation in zero external fields from preferential in-plane to preferential out-of-plane orientation with decreasing Fe layer thickness $t_{Fe}$ \cite{Koyano1988,KozioRachwa2014}. This reorientation process can be explained by competing surface and volume anisotropy contributions, whereby the former dominates for smaller Fe layer thicknesses $t_{Fe}$. In addition, the observed tendency of a very small increase of the isomer shift $\delta_{iso}$ reveals a slightly increased charge transfer from Fe to MgO with decreasing Fe thickness, while a similar conclusion can be drawn 
	\noindent
	from the first principles calculations (section C on the layer-resolved hyperfine properties from DFT).
	
	\begin{table}[t!]
		\caption{Mössbauer parameters for the investigated samples obtained from least-squares fitting of the spectra seen in Figure \ref{fig:CEMS} based on the corresponding distribution of hyperfine fields, $p\left(B_{hf}\right)$ for the different samples with varying Fe-layer thickness $t_{Fe}$. $\langle\delta_{iso}\rangle$ is the average isomer shift relative to bulk bcc-Fe at room temperature, $\Gamma$ describes the intrinsic linewidth (FWHM) of the used sextets in the distribution, $\langle B_{hf}\rangle$ refers to the average magnetic hyperfine field between 25\,T and 45\,T, $\sigma_{B_{hf}}$ is the standard deviation of the obtained distribution, $A_{2,3}$ is obtained by the intensity ratio of lines 2 and 3 (or 5 and 4), $\langle\Theta\rangle$ describes the average angle between Fe spin direction and the incident $\gamma$-ray direction determined from the $A_{2,3}$-ratio shown in equation (\ref{eq:canting-angle}).}
		\begin{tabular}{l|c|c|c|c|c|c}
			$t_{Fe}$ & $\langle\delta_{iso}\rangle$ & $\Gamma$ & $\langle B_{hf}\rangle$ & $\sigma_{B_{hf}}$ & $A_{2,3}$ & $\langle\Theta\rangle$ \\ 
			& $\left(\mathrm{mm/s}\right)$ & $\left(\mathrm{mm/s}\right)$ & (T) & (T) & & $\left(^{\circ}\right)$ \\ \hline\hline
			bulk  &  0.11(1) &  0.236(1)  &  33.9(2) & 0.7 &  3.46(5)  &  74.7 
			\\
			55\,ML &  0.12(3) &  0.258(2)   &  33.8(5)  & 1.5 &  3.65(1) &  77.7 \\
			28\,ML &  0.12(1) &  0.262(5)   &  34.1(6)  & 1.8 &  3.72(6) &  79.1 \\
			10\,ML &  0.13(6) &  0.343(4)   &  34.1(9)  & 2.7 &  1.95(2) &  54.1 \\
			7\,ML  &  0.14(4) &  0.401(9)   &  33.7(3)  & 2.9 &  0.74(9) &  34.0 \\
		\end{tabular}
		\label{tab:CEMS-parameter}
		\vspace{-10pt}
	\end{table}
	
	\noindent

	\begin{figure}[ht]
		\vspace{-19pt}
		\includegraphics[width=\linewidth]{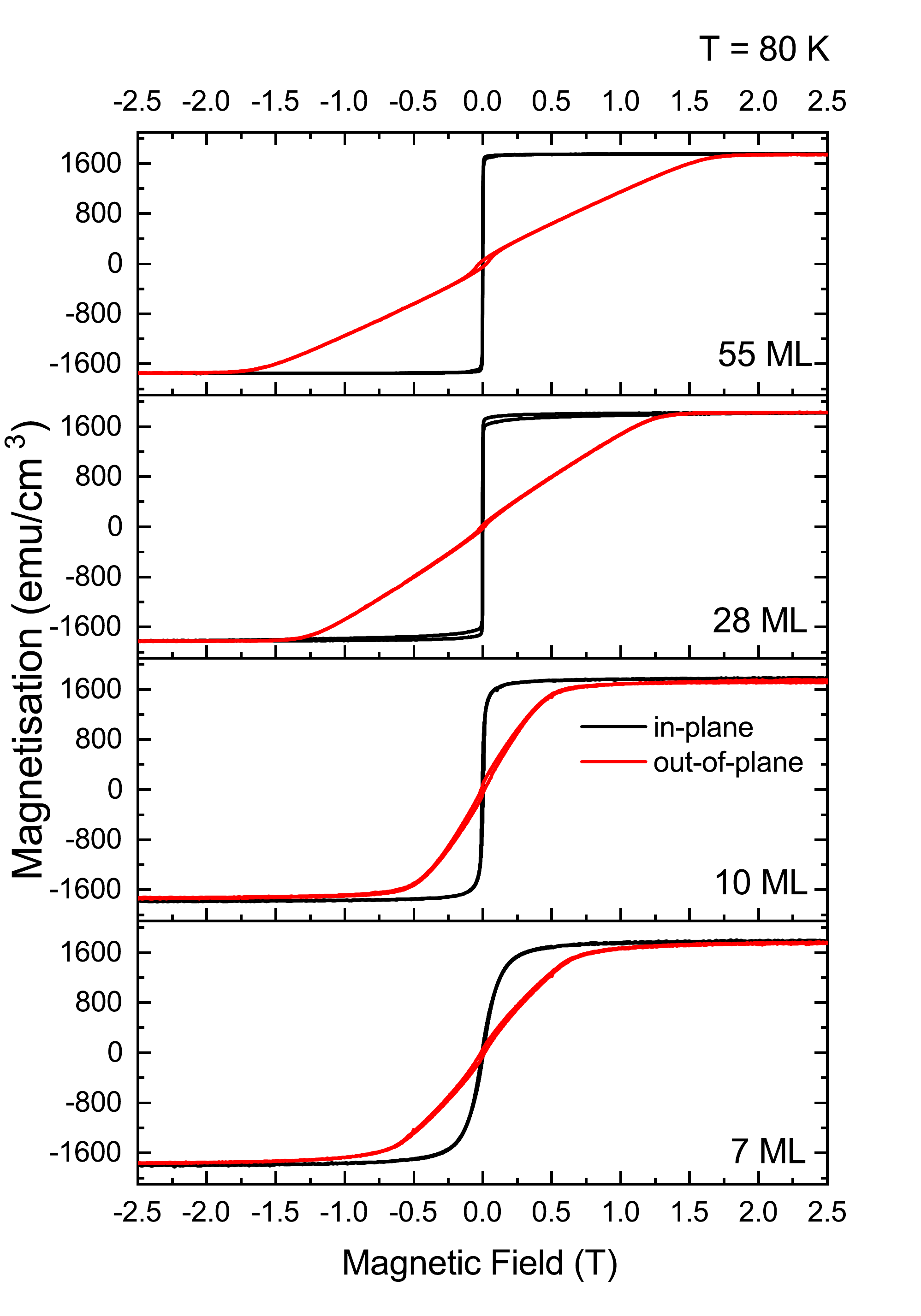}
		\vspace{-15pt}
		\caption{Field dependent magnetization curves measured for different Fe thicknesses for in-plane and out-of-plane geometry. Measurements have been performed at T = 80\,K. The MgO layer thickness was 19\,ML for the 55, 28 and 10\,ML Fe samples and $\sim$ 5\,ML for the 7\,ML Fe sample.}
		\label{fig:magnetometry-comp}
	\end{figure}
	\noindent
	Complementary magnetometry measurements performed at T=80\,K, with parallel (in-plane) and perpendicular (out-of-plane) applied field relative to the film surface, reveal a ferromagnetic ordering for all investigated multilayer samples, see Figure \ref{fig:magnetometry-comp}, with a saturation magnetisation similar to that of bcc-Fe  \cite{Danan_1968} ($M_s^{bulk}=1747\mathrm{\,emu/cm^3}$ at 0\,K and $M_s^{bulk}=1716\mathrm{\,emu/cm^3}$ at RT). In addition, a decrease of the anisotropy field from $\sim$1.6\,T for the 55\,ML Fe layer sample to $\sim$0.6\,T for the 10\,ML Fe sample is revealed, combined with a change of the hysteresis shape from the in-plane measurements. These results are in agreement with the findings of the previously discussed Mössbauer results and reveal the bcc-Fe like magnetic behaviour and the spin reorientation process resulting in the preferred out-of-plane spin orientation for the samples with thin \isotope[57]Fe layers that is also evident in different studies \cite{Koyano1988,Kozio-Rachwa2013}.
	
	\begin{table}[t!]
		\caption{Magnetic properties obtained from hysteresis curves performed at T=80\,K for varying Fe thickness $t_{Fe}$, where $M_s$ describes the saturation magnetisation, $H_k$ the anisotropy field and $H_c$ the coercivity field.}
		\centering
		\begin{tabular}{l|c|c|c|c}
			$t_{Fe}$  & $M_s$  & $\mu_0H_k$ & \multicolumn{2}{c}{$\mu_0H_c$}\\ \hline
			Fe ML    &  (emu/cm$^3$) & (T) & \multicolumn{2}{|c}{(mT)}\\
			&  &     & in-plane   &  out-of-plane \\ \hline\hline
			55    &  1751        & 1.607   & $<$2     &   10.5    \\
			28    &  1803        & 1.264   & $<$2     &   5.5     \\
			10    &  1740        & 0.51   & $<$2     &   3      \\
			7     &  1718        & 0.63   & 3     &   6     \\
		\end{tabular}
		\label{tab:Magnetometry}
	\end{table}
	
	\subsection{Hyperfine field parameters from DFT}
	
	\begin{table*}
		\caption{Layer-resolved and average M\"ossbauer parameters
			and electric field gradients from DFT-calculations of a Fe$_8$/(MgO)$_8$(001) heterostructure: Spin and orbital moments $m_\mathrm{spin}$ and $m_\mathrm{orb}$ (within the muffin-tin sphere), contact moment $m_\mathrm{contact}$, average contact charge density at the Fe sites $\rho_\mathrm{contact}$, isomer shift $\delta_{iso}$ relative to bcc-Fe, magnetic hyperfine field $B_\mathrm{hf}$ including spin and orbital dipole field and the in-plane ($V_{xx}=V_{yy}$) and cross-plane ($V_{zz}$) elements of the (trace-free) electric field gradient. For comparison, the corresponding values calculated for bulk Fe with a lattice constant $a=2.866\,$\AA{}
			with similar numerical settings are included as well.}
		\begin{tabular}{l|c|c|c|c|c|c|c|c}
			Fe-Layer  & $m_\mathrm{spin}$ & $m_\mathrm{orb}$  & $m_\mathrm{contact}$  & $\rho_\mathrm{contact}$  & $\delta_{iso}$ & $B_\mathrm{hf}$ & $V_{xx}$ & $V_{zz}$  \\ 
			& $\left(\mu_\mathrm{B}/\mathrm{Fe}\right)$ & $\left(\mu_\mathrm{B}/\mathrm{Fe}\right)$ & $\left(\mu_\mathrm{B}/\mathrm{Fe}\right)$ &
			$\left(a_\mathrm{Bohr}^{-3}\right)$ & (mm/s) & (T) & (V/\AA{}$^2$) & (V/\AA{}$^2$) \\ \hline\hline
			4 (center)  & 2.415 & 0.042 & 0.674 & 15422.610 & 0.046 & 31.34 & -3.664 & 7.328 \\
			3       & 2.457 & 0.046 & 0.685 & 15422.576 & 0.056 & 30.68 & -6.820 & 13.641 \\
			2       & 2.480 & 0.038 & 0.679 & 15422.707 & 0.019 & 32.65 & 8.555 & -17.109 \\
			1 (IF)    & 2.626 & 0.069 & 0.822 & 15422.277 & 0.146 & 39.75 & 1.075 & -2.150 \\ \hline
			8\,ML average & 2.495 & 0.049 & 0.715 & 15422.543 & 0.066 & 33.61 & -0.214 & 0.427 \\ \hline
			bcc (bulk)  & 2.292 & 0.042 & 0.646 & 15422.763 & $-$  & 30.27 & -0.057 & 0.115 \\
		\end{tabular}
		\label{tab:DFT-EFG}
	\end{table*}

	Hyperfine parameters calculated from first-principles are typically very sensitive to the technical details and the choice of the exchange-correlation potential. Nevertheless, the layer-resolved trends obtained from DFTin the calculated magnetic moments and M\"ossbauer parameters shown in Tab.\ \ref{tab:DFT-EFG} strongly confirm the interpretation of the experimental data.From the interface to the center of the Fe-slab, we observe a decrease in orbital and spin moments, $m_\mathrm{spin}$ and $m_\mathrm{orb}$, towards the bulk value. A corresponding trend appears also in the variation of the contact charge density at the Fe core and the contact magnetic moment $m_\mathrm{contact}$. However, bulk properties are not yet reached in the central layer. This is partially due to the tetragonal distortion induced by the epitaxial relation to the MgO slab. In thick slabs beyond 10\,ML, misfit dislocations are introduced to relieve the epitaxial stress, bringing the average values closer to the bcc-Fe around room temperature. This also applies to the isomer shift, hyperfine field and electric field gradients. The isomer shift is calculated from the difference between the contact charge density in the layer and bcc-Fe (source), $\delta_{iso}=\alpha\,\left(\rho_\mathrm{contact}^\mathrm{layer}-\rho_\mathrm{contact}^\mathrm{bcc}\right)$, using an approximate generic constant $\alpha=-0.3\,\mathrm{mm}\mathrm{s}^{-1}a_\mathrm{Bohr}^3$, which can be determined from the fit of calculations to experimental results of various Fe-based compounds \cite{cn:Duff66,cn:Blaha92,Neese2002,cn:Wdowik07,cn:Spiel09,Kurian2010Isomer}. Here, our results indicate that the large values are dominated by the interface layer, which exhibits a $\delta_{iso}$ about three times as large as the central layer. For decreasing temperatures, we also expect the bcc-Fe parts to be strained according to the different thermal expansion of Fe, bulk MgO and substrate. We expect this to result in an elevated isomer shift in the entire Fe slab explaining the experimental trend for $T=80\,$K, in addition to the SOD shift.
	
	The calculated hyperfine fields $B_\mathrm{hf}$ are underestimated by a few percent. 
	We observe a significant increase of $B_\mathrm{hf}$ only at the interface and a slight increase in the second layer, while the following layers are already quite close to the bulk value. However, the two layers close to the interface are characterized by a smaller number of magnetic neighbours in the first two coordination shells, where we expect the largest exchange interaction. This destabilizes the collinear ferromagnetic arrangement and enhances magnetic disorder at finite temperatures, which reduces the average $\langle B_{hf}\rangle$ and overcompensates the increase predicted from theory for perfect magnetic order. The reduced coordination (and thus symmetry) at the interface also enhances the impact of spin-orbit interaction as another source for non-collinearity at low temperatures, which is reflected in the considerable increase of the orbital moments in the interface layer. Still, the hyperfine field distribution at $T=80$\,K shown in Fig.\ \ref{fig:CEMS}(d) for the 7\,nm and 10\,nm slabs shows a characteristic broadening. This includes a considerable amount of sites with increased $B_\mathrm{hf}$ between 35\,T and 40\,T, which coincides with the value predicted from theory for the interface layer.
	
	The electric field gradient (EFG) is a measure of the asymmetry of the charge density around the nuclei which can be routinely calculated with DFT \cite{cn:Dufek95,cn:Petrilli98,cn:Martinez01,cn:Ebert05}. For bcc-Fe, the (small) discrepancy between $V_{xx}$ and $V_{zz}$ is solely a consequence of the spin-orbit interaction in combination with the alignment of the spin moments in $z$-direction, which breaks the cubic symmetry. As expected, the EFG in the heterostructure is significantly larger -- by one to two orders of magnitude. However, we also see a significant variation of the values from layer to layer, which is a signature of the large influence of the electronic structure. The largest absolute values are obtained in the second and third layer, with a change in sign in between, which indicates a qualitative change in the shape of the asymmetry of the charge cloud.     The consequence is that the effect largely cancels out if we average over the entire heterostructure, which makes it difficult to reproduce this behaviour experimentally, for instance through measurements of the quadrupole splitting.
	
	From the experimental spectrum, the magnitude of the EFG in a magnetically ordered state can be determined by the quadrupole splitting $\epsilon$ \cite{Chen2007LatticeDynamics}, which follows the relation
	\begin{equation}
	\epsilon = \dfrac{1}{8} eQV_{zz} \cdot\left(3\cos^2\left(\alpha\right)-1\right), 
	\label{eq:2epsilon}
	\end{equation}
	where $Q=1.6\cdot10^{-29}\mathrm{\,m}$ is the quadrupole moment of the first excited state \cite{cn:Dufek95} and $\alpha$ is the angle between $B_{hf}$ and the main component $V_{zz}$ of the EFG. Furthermore, equation (\ref{eq:2epsilon}) is simplified under the assumption of a vanishing asymmetry parameter $\eta =\left|\left(V_{xx}-V_{yy}\right)/V_{zz}\right|$ \cite{Kaufmann1979EFG}, due to the geometry of the sample. For a bulk sample one would therefore expect a quadrupole splitting between 0.004 and -0.002\,mm/s depending on the angle $\alpha$, which is below the experimental resolution.

	\subsection{Nuclear Resonant Inelastic x-ray Scattering}
	
	The method of \isotope[57]Fe nuclear resonant inelastic x-ray scattering (NRIXS) is sensitive to the \isotope[57]Fe resonant isotope and measures the phonon excitation probability, as described in Refs. \cite{Seto1995,Sturhahn1995,Chumakov1995,Rhlsberger2005,Chen2007LatticeDynamics,Chumakov1999}. NRIXS provides the Fe-partial (Fe specific) phonon (vibrational) density of states (VDOS) in the harmonic approximation \cite{Singwi1960} rather directly with a minimum of modeling \cite{Sturhahn2004,Sturhahn2000}. The incoherent cross section [probability W(E)] of nuclear resonance absorption for a particular phonon energy E and phonon momentum vector $\vec{q}$ is known to be proportional to $\lvert \vec{s} \cdot \vec{e}_j(q)\rvert^2$ , where $\vec{e}_j(q)$ is the polarization vector of vibrations for the Fe atom in the phonon mode $j$ and $\vec{s} = \vec{k}_0/\lvert \vec{k}_0\rvert$ is the unit vector in the incident photon momentum direction \cite{Kohn1998}, which is nearly in the film plane (off by $\sim$ 4$^{\circ}$ in our case). Thus, W(E) scales with $\cos^2\left(\Theta\right)$, where $\Theta$ is the angle between the phonon polarization vector $\vec{e}_j(q)$ and the x-ray direction $\vec{s}$. For a polycrystalline sample (as in our case), the spatial average of $\cos^2\left(\Theta\right)$ is constant, which means that the inherent anisotropy of our multilayer sample averages out. 
	
\begin{figure}[ht]
		\centering
		\includegraphics[width=1.0\linewidth]{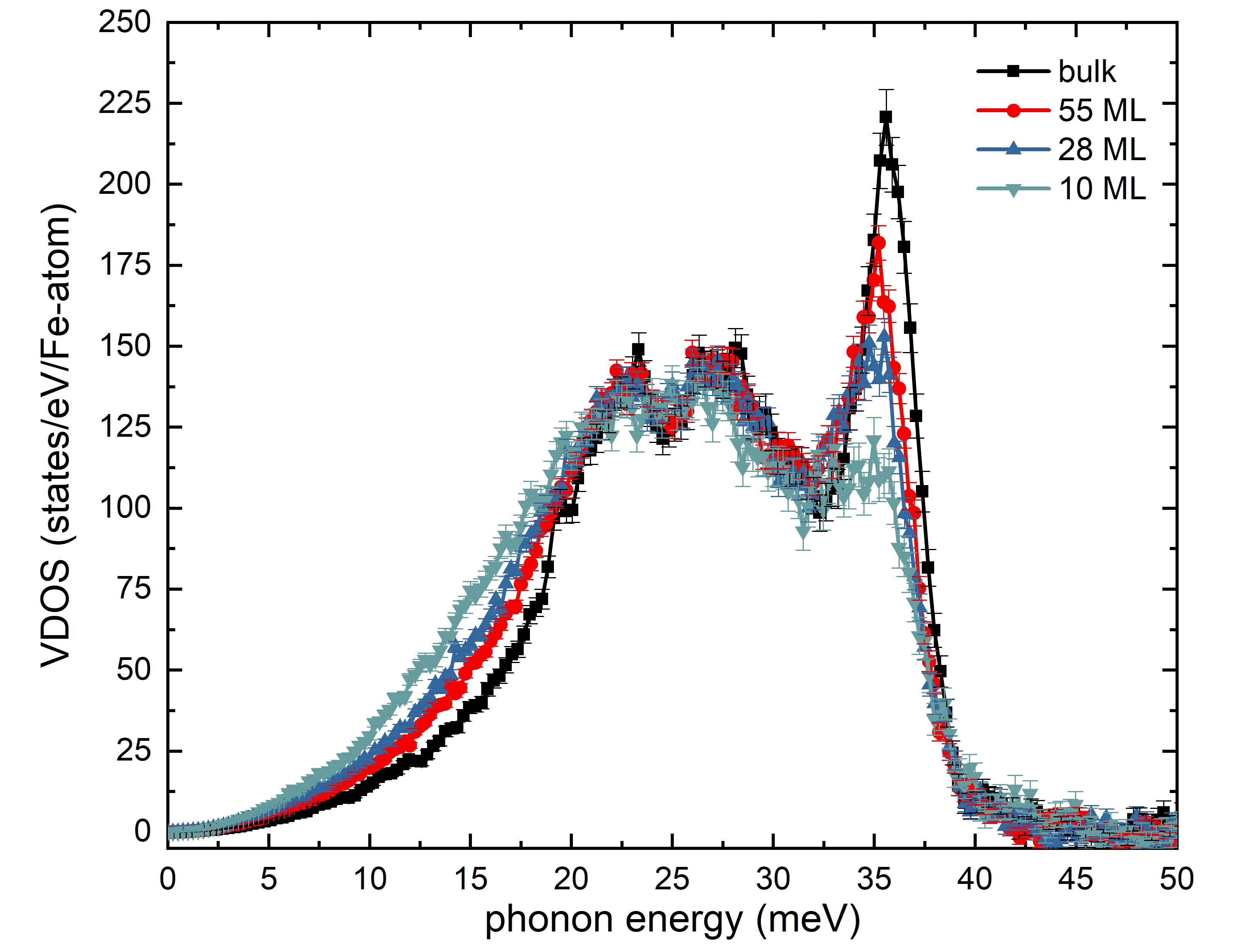}
		\caption{Fe-partial VDOS obtained from NRIXS for a bcc-Fe foil and Fe/MgO multilayer systems with varying Fe thicknesses $t_{Fe}$ of 55, 28 and 10\,ML and constant MgO layer thickness $t_{MgO}$ of 19\,ML. One can clearly observe a softening of the Fe partial VDOS and a reduction of the longitudinal acoustic phonon mode at 35\,meV. The VDOS of the bcc-Fe foil has been taken from Ref. \cite{Toellner1997}. All VDOS are area-normalized to 3 (3 vibrational degrees of freedom per Fe atom).}
		\label{fig:vdos}
	\end{figure}

	The Fe-partial VDOS for a polycrystalline bcc-Fe foil and the three Fe/MgO multilayers obtained at room temperature from NRIXS are exhibited in Fig. \ref{fig:vdos}. The NRIXS raw data and the phonon excitation probabilities (after subtraction of the zero-phonon (Mössbauer) peak and normalization) are found in the Supplemental Material \cite{Supplement}. The VDOS of the multilayers with the thickest Fe layers of 55 and 28 effective MLs show similarities with g(E) of the bulk bcc Fe foil: The latter is characterized by the dominant and sharp longitudinal-acoustic (LA) peak at 35.5\,meV and by the two resolved transverse-acoustic (TA) phonon peaks at 23.2 and 26.3\,meV, in agreement with literature reports \cite{Hellwege1981}.
	However, as compared to the LA peak height of about 220\,eV$^{-1}$(Fe-atom)$^{-1}$ for bulk bcc Fe, the heights of the corresponding peaks of the 55\,ML and 28\,ML Fe multilayer samples are systematically reduced to 181 and 151\,eV$^{-1}$(Fe-atom)$^{-1}$, respectively. Simultaneously, the position of the former LA peak remains relatively constant in the 55\,ML system (35.2\,meV), or is slightly shifted to lower energy (34.8\,meV) for the 28\,ML system. 
	
	The biggest change (relative to the bulk) occurs for the 10\,ML Fe sample.
	Here, the height of the former LA peak is drastically reduced to 113.3\,eV$^{-1}$(Fe-atom)$^{-1}$, and the peak position is shifted to 34.2\,meV. As to the two TA peaks, there is a tendency of a slight reduction in the peak height and peak position with decreasing Fe thickness, but the two TA peaks become less resolved and even dominant for the 10\,ML Fe sample. Interestingly, the VDOS below about 20\,meV is remarkably and systematically enhanced with decreasing Fe film thickness; this effect is most pronounced for the thinnest Fe layer (10\,ML) in the multilayer. For example, at 15\,meV, $g(E)$ increases from $\sim$39\,eV$^{-1}$(Fe-atom)$^{-1}$ for bulk Fe to $\sim$ 52, 57, and 74\,eV$^{-1}$(Fe-atom)$^{-1}$ for 55, 28 and 10\,ML Fe, respectively. All of these observations imply that the phonon features of the Fe layers in the multilayer experience a distinct overall redshift as compared to those of the Fe bulk sample, meaning that increasing lattice softening occurs with decreasing Fe layer thickness. However, the phonon cutoff energy of all samples remains constant at $\sim$ 40\,meV, including bulk bcc Fe. The typical features in the phonon spectra of the Fe/MgO multilayer are summarized in Table \ref{tab:NRIXS-Features}. 
	
	We like to mention that a qualitatively similar reduction and energy shift of the LA phonon peak combined with an enhancement of the low-energy VDOS part have been observed previously by layer selective \isotope[57]Fe NRIXS combined with layer- and element-selective DFT-based calculations for Fe/Ag multilayers \cite{Keune2018} and were interpreted in terms of high-energy phonon confinement in the Fe layers and phonon localization at the Fe/Ag interfaces, occuring as the result of the large energy mismatch between Ag and Fe LA phonons. For the energetically high Fe LA phonons (at 35.5\,meV) there is no counterpart in the $g(E)$ of the phononically soft Ag metal, because the upper phonon cutoff energy of Ag is located at only $\sim$ 21\,meV, much lower than that of bcc Fe ($\sim$40\,meV). In the DFT-based calculations for the Fe/Ag interface, a strong suppression of the Fe LA phonon peak is observed, accompanied by the appearance of a dominant Fe-partial $g(E)$ peak in the low-energy range at $\sim$ 15\,meV. As will be described for Fe/MgO multilayers in the next section (section D on the layer-resolved properties from DFT calculations), the contribution to $g(E)$ of O- and Mg-vibrations is rather small below $\sim$40\,meV, because the strong optical modes develop at and extend to higher energies. Thus, the concept of energy mismatch between the Fe-specific $g(E)$ in the Fe layers and the partial $g(E)$ curves of the O and Mg sublattices in the MgO layer appears to hold also for the nanoscale Fe/MgO multilayer system. 
	
	\begin{table*}[ht]
		\centering
		\caption{Properties of the Fe-partial g(E) for different Fe/MgO multilayers obtained from NRIXS. Given are the g(E) peak height and energetic position of the longitudinal-acoustic (LA) peak and the two transverse-acoustic (TA) peaks for different Fe layer thicknesses.}
		\begin{tabular}{l|c|c|c|c|c|c}
			& \multicolumn{2}{c|}{LA peak} & \multicolumn{2}{c|}{TA1 peak} & \multicolumn{2}{c}{TA2 peak} \\ \hline
			sample & energy & height & energy & height & energy & height \\
			(ML Fe) & (meV) & (eV$^{-1}$(Fe-atom)$^{-1}$) & (meV) & (eV$^{-1}$(Fe-atom)$^{-1}$) & (meV) & (eV$^{-1}$(Fe-atom)$^{-1}$) \\ \hline \hline
			bulk & 35.5 & 220.0 $\pm$ 8.6 & 23.2 & 149.0 $\pm$ 5.1 & 26.3 & 147.1 $\pm$ 5.8 \\
			55  & 35.2 & 181.9 $\pm$ 5.3 & 22.2 & 142.5 $\pm$ 3.2 & 27.5 & 141.6 $\pm$ 3.3 \\
			28  & 34.8 & 151.0 $\pm$ 5.5 & 22.5 & 139.4 $\pm$ 3.7 & 26.7 & 140.7 $\pm$ 4.2 \\
			10  & 34.2 & 113.3 $\pm$ 6.7 & 22.5 & 134.5 $\pm$ 4.9 & 26.4 & 139.7 $\pm$ 5.8 \\
		\end{tabular}%
		\label{tab:NRIXS-Features}
	\end{table*}
	
	Our $g(E)$ results obtained from NRIXS for the different Fe/MgO multilayers allow the calculation of Fe-specific vibrational thermodynamic quantities. These quantities are given in Table \ref{tab:VDOS-Parameter-Phoenixa}\,\&\,\ref{tab:VDOS-Parameter-Phoenixb}. The equations for the calculation of these thermodynamic quantities from $g(E)$ are given in the literature \cite{Grimvall1999,Fultz_2010,Stankov2010,Hu2013,cn:Gruner15PRL,cn:Landers18}. 
	One can notice in Table \ref{tab:VDOS-Parameter-Phoenixa}\,\&\,\ref{tab:VDOS-Parameter-Phoenixb} that all thermodynamic quantities are dependent on the thickness of the individual Fe layer in the multilayer. For example, the Lamb-Mössbauer factor $f_{LM}$, the average atomic force constant $\Phi_k$ and the Debye temperature $\Theta_D$ are reduced with decreasing Fe film thickness, while the vibrational entropy $S_{vib}$ is enhanced. At low phonon energies, the average Debye velocity of sound, $\langle v_D\rangle$, can be obtained from the $E$-dependence of the reduced VDOS $g(E)/E^2$, using the relation \cite{Achterhold2002,Hu2003}
	\begin{equation}
	\lim\limits_{E \to 0} \left(\dfrac{g\left(E\right)}{E^2}\right)= \dfrac{3m_{Fe}}{2\pi^2\rho \langle v_D\rangle^3\hbar^3},
	\label{eq:Schallgeschwindigkeit}
	\end{equation}
	
	where $m_{Fe}$ is the mass of the \isotope[57]Fe atom, $\hbar$ is the reduced Planck constant, and $\rho$ is the mass density of the Fe film, which we assume to be equal to the bulk Fe density. $g(E)$ is quadratic in $E$ at low phonon energies $E$. The factor of 3 appears because $g(E)$ is area-normalized to 3 (three vibrational degrees of freedom). The ratio $g(E)/E^2$ can be described at low energies by a constant called the Debye level, which can be obtained from the experimental VDOS in the limit $E \rightarrow 0$ (see our Supplemental Material, Ref. \cite{Supplement}). In this work, the average Debye sound velocity $\langle v_D\rangle$ is determined by an empirical power law combined with an Bayesian information approach. This procedure is described in detail in the work of Morrison et al. \cite{Morrison_2019}. As a crosscheck we have also determined $\langle v_D \rangle$ according to Eq. (\ref{eq:Schallgeschwindigkeit}). The $\langle v_D\rangle$ values obtained from both procedures are given in Table \ref{tab:VDOS-Parameter-Phoenixb}. One can notice that the velocity of sound decreases with decreasing Fe film thickness in the Fe/MgO multilayer and both procedures yield for these systems a similar result within a deviation of up to 1.7\,\%. 
	
	\begin{table*}[ht]
		\centering
		\caption{Selected thermodynamic properties extracted from the Fe partial VDOS, which were obtained from the experimental data with PHOENIX 3.0.1 \cite{Sturhahn2000}.Here $f_{LM}$ is the Lamb-Mössbauer factor, $S_{vib}$ the vibrational entropy, $C_{vib}$ the specific heat, $\langle v^2\rangle$ is the mean square velocity, $T_k$ the average kinetic energy, $\varPhi_k$ the average force constant, $F_{vib}$ free vibrational energy, $U_{vib}$ vibrational internal energy and $\Theta_\mathrm{D}^\mathrm{S}$ the entropy Debye temperature.}
		\begin{tabular}{l|c|c|c|c|c|c|c}
			Fe thickness t$_{Fe}$& $f_{LM}$  & $S_{vib}$ & $C_{vib}$ & $\varPhi_k$ & $F_{vib}$ & $U_{vib}$ & $\Theta_\mathrm{D}^\mathrm{S}$ \\
			& & $\left(k_{B}/\mathrm{atom}\right)$ &$\left(k_{B}/\mathrm{atom}\right)$ & $\left(\mathrm{N/m}\right)$ & $\left(\mathrm{meV}/\mathrm{atom}\right)$ & $\left(\mathrm{meV}/\mathrm{atom}\right)$& (K)\\ \hline \hline
			bcc (bulk)  & 0.7984(8)  & 3.07(1)   & 2.71(1)   & 178(3) & 5.04(25)     & 85.3(4) & 429(4) \\
			55\,ML          & 0.7708(5)  & 3.215(7)  & 2.736(7)  & 164(1) & 0.65(13)     & 84.7(2) & 408(2) \\
			28\,ML          & 0.7575(6)  & 3.277(8)  & 2.743(7)  & 159(1) & -1.27(12)    & 84.5(3) & 402(2)  \\
			10\,ML          & 0.730(1)   & 3.38(1)   & 2.75(1)   & 155(2) & -3.95(20)    & 84.3(3) & 398(3) \\
			
		\end{tabular}%
		\label{tab:VDOS-Parameter-Phoenixa}
	\end{table*}
	\begin{table*}[ht!]
		\centering
		\caption{Average Debye sound velocity $\langle v_D\rangle$ determined from the conventional method, denoted as $\langle v_D\rangle_{con}$, discussed in Eq. (\ref{eq:Schallgeschwindigkeit}) and originating from Refs. \cite{Achterhold2002,Hu2003}, and the procedure based on the empirical power law, denoted as $\langle v_D\rangle_{emp}$, presented in detail in Ref. \cite{Morrison_2019} and schematically described in Ref. \cite{Supplement}. The comparison of both methods yields a relatively good agreement.}
		\begin{tabular}{c|c|c|c}
			Fe thickness t$_{Fe}$ & $\langle v_D\rangle_{con}$  & $\langle v_D\rangle_{emp}$ & Deviation\\
			& (m/s) &   (m/s)   & (\%) \\ \hline\hline
			bcc (bulk)  &  3550(10) & 3568(8) & -0.50(36) \\
			55\,ML      &   3170(6) &   3164(12) & 0.19(43)\\
			28\,ML      &   2940(5) &   2991(8) &   -1.71(31)\\
			10\,ML      &   2800(10)&   2760(6) &   1.45(43)\\    
		\end{tabular}
		\label{tab:VDOS-Parameter-Phoenixb}    
	\end{table*}
	
	\subsection{Layer-resolved vibrational properties from DFT}
	\begin{figure*}[ht]
		\includegraphics[width=\linewidth]{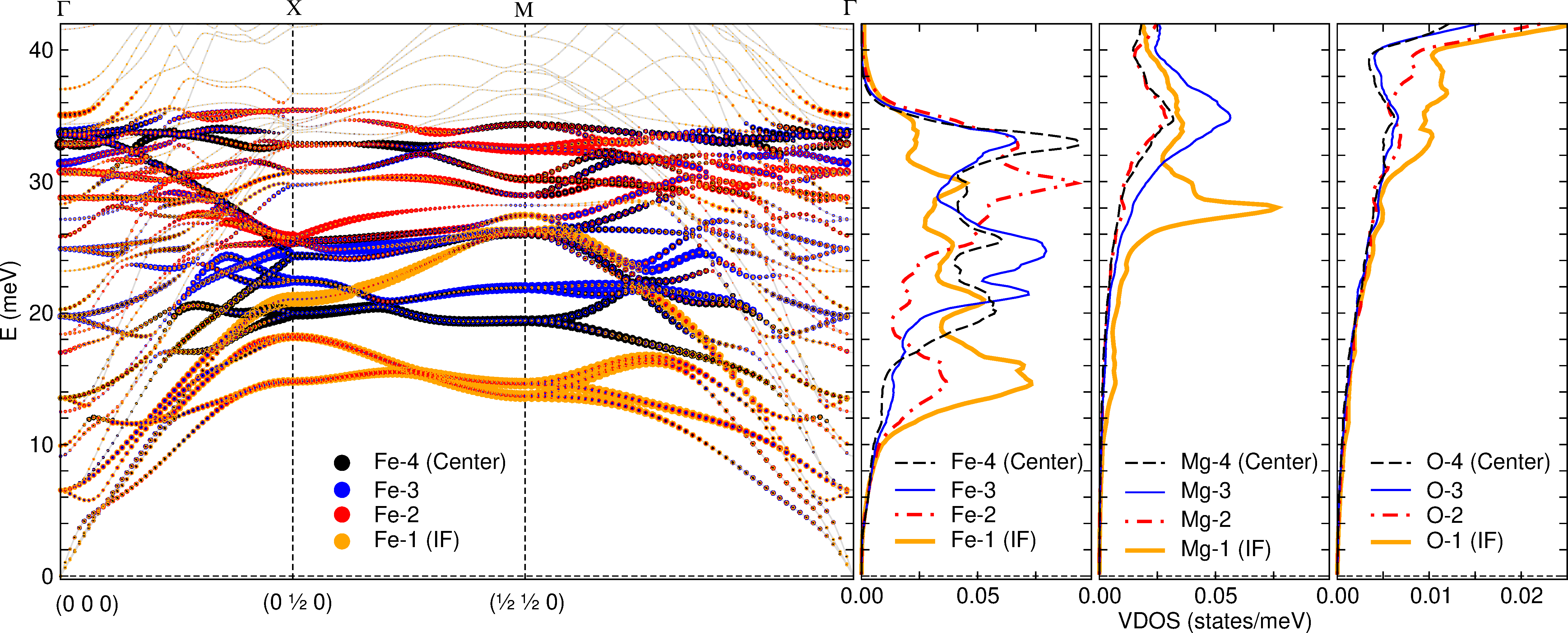}
		\caption{Layer-resolved Fe-projected phonon dispersion and VDOS obtained from our DFT
			calculations.
			The left panel shows the low-energy part of the phonon dispersion for
			reciprocal vectors lying in-plane.
			The size of the symbols indicates the contribution of the
			symmetry-inequivalent Fe-atoms (marked by different colours)
			to the eigenvectors corresponding to the respective $q$-vector. Contributions from
			Mg and O atoms are not shown.
			The rightmost three panels show the corresponding site-resolved VDOS (normalized to one)
			for all inequivalent Fe-, Mg- and O-atoms in the respective
			energy window of the Fe-phonons.}
		\label{fig:disp}
	\end{figure*}
	\begin{table*}[ht!]
		\caption{Thermodynamic properties for $T=300\,$K extracted from the Fe-partial DFT-VDOS. The notation of the thermodynamic properties is identical to Table \ref{tab:VDOS-Parameter-Phoenixa}\,\&\,\ref{tab:VDOS-Parameter-Phoenixb}.}
		\centering
		\begin{tabular}{l|c|c|c|c|c|c|c|c|c}
			Fe-Layer  & $f_{LM}$ & $S_{vib}$  & $C_{vib}$  &  $\left<v^2\right>$  & $\varPhi_k$  & $F_{vib}$  & $U_{vib}$ & $\Theta_\mathrm{D}^\mathrm{S}$ & $\langle v_\mathrm{D} \rangle_{con}$\\ 
			& & $\left(k_\mathrm{B}/\mathrm{atom}\right)$ & $\left(k_\mathrm{B}/\mathrm{atom}\right)$ & $\left(\mathrm{m}^2/\mathrm{s}^2\right)$  & $\left(\mathrm{N/m}\right)$  & $\left(\mathrm{meV}/\mathrm{atom}\right)$  & $\left(\mathrm{meV}/\mathrm{atom}\right)$ & (K) & (m/s)\\ \hline\hline
			4 (center)  & 0.763 & 3.30 & 2.76 & 47465 & 151 & ~~-1.16 & 84.12 & 395 & 3411\\
			3       & 0.758 & 3.34 & 2.76 & 47368 & 147 & ~~-2.28 & 83.95 & 390 & 3418\\
			2       & 0.748 & 3.33 & 2.75 & 47560 & 156 & ~~-1.80 & 84.29 & 391 & 3458\\
			1 (IF)    & 0.694 & 3.79 & 2.81 & 46689 & 120 &  -15.27 & 82.75 & 332 & 3472\\\hline
			8\,ML average & 0.740 & 3.43 & 2.77 & 47270 & 144 & ~~-5.13 & 83.78 & 376 & 3440\\
			
		\end{tabular}
		\label{tab:DFT-THERMO}
	\end{table*}
	In analogy to the magnetic subsystem discussed above, the layer-resolved contributions to the vibrational and thermodynamic properties are obtained from first-principles calculations of the harmonic lattice dynamics in the Fe$_8$/(MgO)$_8$(001). In this way, we obtain not only the layer resolved VDOS, but also the phonon dispersion, where we can identify the contribution and hybridization of particular element-resolved modes in reciprocal space. This is summarized in Fig.\ \ref{fig:disp}. We concentrate our discussion on the energy range below 40\,meV covering the Fe-modes, which are subject to the NRIXS measurements. In contrast, the contribution of the O- and Mg-vibrations to this energy range is rather small, as their optical modes start at comparatively high energies and extend up to 85\,meV (the full site-resolved VDOS is published elsewhere \cite{cn:Rothenbach19}). The shape from the layer-resolved VDOS of Fe differs substantially between the layers. For the two layers closest to the interface, we observe a shift of the spectral density to energies below 17\,meV. This is associated with a set of peaks, which are very dominant in the IF layer and originate from van-Hove singularities at $X$ (14.8 and 18.2\,meV) and $M$ (13.7 and 14.6\,meV). The character plot of the dispersion (left panel of Fig.\ \ref{fig:disp}) reveals that these modes hybridize with the motion of the 2nd layer of Fe-atoms. We also see a signature of this peak structure in the Mg-VDOS -- but not in the O-VDOS, which is plotted in Fig.\ \ref{fig:disp} with a four times enlarged scale. In turn, the contribution of the IF layer to the highest Fe modes, which rang up to 38\,meV is strongly suppressed. This energy window also contains optical modes involving the motion of atoms at both sides of the interface. We expect the hybridization of Fe and Mg modes to play an important role for the energy transfer (thermalization) between the Fe and MgO the subsystem, when the Fe slab is selectively heated up. This would, for instance, occur as a consequence of an optical excitation with an energy below the band gap of bulk MgO \cite{cn:Rothenbach19}. In contrast, the sharp peak in the interface Mg-VDOS at 28.2\,meV can be traced back to a rather extended, particularly flat mode around $M$, which does not exhibit significant Fe-character at the zone boundary. 
	
	From the layer resolved VDOS, we derive various thermodynamic quantities with atomic resolution using the usual textbook relations for $T=300\,$K, listed in Table\ \ref{tab:DFT-THERMO}. Comparing the experimental values of the 10\,ML system with the 8\,ML average over the layer-resolved calculations, we find a very close agreement. Although the shape of the VDOS differs significantly between all layers, we see a distinct change in the resulting thermodynamic properties only directly at the interface. Here, the overall red shift of the density of states results in softening of the system, as indicated by the 20\% decrease in the average force constant. As the Debye-temperature is nearly reached, this is accompanied by a slight increase in the specific heat, only. This explains, why the specific heat derived from the experimental VDOS is nearly independent of the slab thickness.
	
\begin{figure}[H]
		\includegraphics[width=\linewidth]{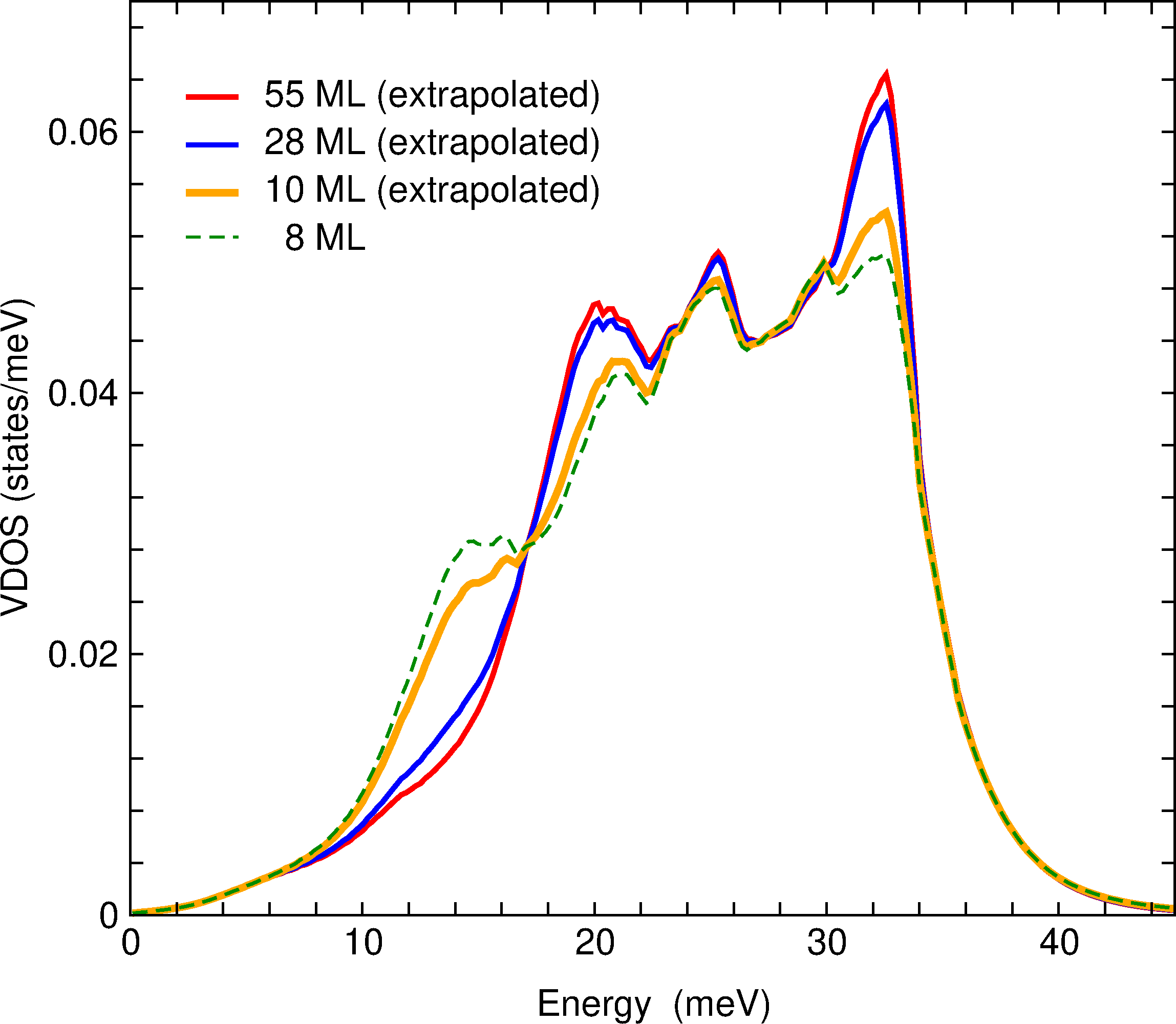}
		\caption{Extrapolation of the Fe-projected VDOS obtained from our DFT calculations (Fig.\ \protect\ref{fig:disp}) to the layer-thickness of the experimental sample corresponding to Fig.\ \protect\ref{fig:vdos}.
			To improve the comparison of the characteristic features with experiment,
			the Gaussian broadening has been increased to 4\,meV.
			The calculations reproduce the significant loss of spectral weight below 18\,meV for
			increasing slab thickness, which is accompanied by a moderate increase at the two peaks
			around 25\,meV and the evolution of a sharp peak above 30\,meV.}
		\label{fig:DFTvdos}
		\vspace{-8pt}
	\end{figure}

	
	In turn, we can also compare by extrapolating the calculated, layer-resolved results to the experimental slab thickness. This is done in Fig.\ \ref{fig:DFTvdos} by replicating the VDOS of the central layer in the average, which disregards the effect of changes in the equilibrium lattice parameters and strain relaxation mechanisms, like misfit dislocations, on the VDOS. Already with this simplified approach, we can reproduce the thickness-dependent changes of the main features of the VDOS in the multilayer structures. With decreasing thickness of the Fe subsystem, a shoulder appears between 10 and 16\,meV which arises mainly from the two layers closest to the interface. Additional weight in this energy range is also seen in the experimental spectra of the thinner Fe-slabs. In the experiment, this feature is apparently broader, which we ascribe to the rougher and less well-defined interfaces in the multilayer systems. At the same time, the distinct peak around 34\,meV, which is characteristic for bcc and bct coordinated Fe, diminishes, which again corresponds well to the experimental observation.
	
	The calculated Debye velocity of sound $\langle v_D\rangle$ associated with each layer is essentially constant for all layers (see Tab.\ \ref{tab:DFT-THERMO}). Unlike in the experimental multilayer systems, it does not decrease together with the layer-resolved Debye temperature $\Theta_\mathrm{D}^\mathrm{S}$ (cf.\ Tab.\ref{tab:VDOS-Parameter-Phoenixa}). This is not a contradiction, since the entropy Debye temperature $\Theta_\mathrm{D}^\mathrm{S}$ results from the logarithmic moment of $g(E)$, i.\,e., the integration of the entire VDOS, while $\langle v_D\rangle$ is derived from 
	\noindent
	its low energy limit. Indeed, we see from Fig.\ \ref{fig:disp} that the first optical modes close to $\Gamma$, which appear according to backfolding to the reduced Brillouin zone, start above 5\,meV. This means that the effective dispersion of the in-plane acoustic modes (and thus the sound velocity) is the same for all layers in the low energy limit. In turn, we ascribe the different pattern observed in experiment to the polycrystalline nature of the sample, which effectively smears out the low energy modes in the VDOS. Possible origins are a distribution in the interface-induced strain in the Fe-grains in combination with an increasing hardening from grain boundaries and the appearance of misfit-dislocations for larger thicknesses.

	\section{Conclusion}
	In our combined approach, we employ \isotope[57]Fe Mössbauer spectroscopy, \isotope[57]Fe nuclear resonant inelastic x-ray scattering (NRIXS) and layer-resolved density-functional-theory (DFT) based first-principles calculations to reveal the interface-related atomic vibrational properties of the prototype metal-insulator Fe/MgO multilayer system. Mössbauer spectroscopy at 80\,K indicates a Fe-layer thickness dependent spin reorientation from preferred in-plane to preferred out-of-plane orientation due to the influence of perpendicular magnetic anisotropy at the interfaces. Interestingly, relative to the bulk, enhanced hyperfine magnetic fields $B_{hf}$ (e.\,g., $B_{hf}$ = 36\,T and larger) are observed at the Fe/MgO interfaces, as is evident for the thinnest Fe layers. These experimental findings are supported by the layer-selective DFT-based calculations for a [Fe$_8$/MgO$_8$](001) heterostructure. From the obtained electronic and magnetic structure, an enhanced magnetic Fe moment and reduced contact charge density relative to bulk values is expected in all layers due to the tetragonal lattice distortion of the heterostructure. 
	
	NRIXS reveals the impact of interfaces by drastic changes in the Fe-partial vibrational (phonon) density of states (VDOS, g(E)) in nanoscale polycrystalline [bcc-\isotope[57]Fe/MgO] multilayers with decreasing Fe layer thickness: a reduction of the longitudinal acoustic (LA) phonon peak near $\sim$ 35\,meV and an enhancement of the low-energy part of g(E) below $\sim$ 20\,meV, leading to an overall red shift of g(E) and lattice softening. The DFT-based computations for (001)-oriented [Fe(8\,ML)/MgO(8\,ML)] multilayers (ML = monolayers) support the experimental findings and, moreover, predict distinct VDOS peaks below 20\,meV (i.e., at 14\,meV) for the Fe(001)/MgO(001) interface, which, however, are smeared in the experimental samples due to inevitable interface roughness and polycrystallinity.
	
	From the experimental and calculated VDOS, vibrational thermodynamic properties have been determined as a function of Fe thickness and are found to be in excellent agreement. For example, the Fe-specific thickness dependent average Debye velocity of sound has been obtained, which is an important property for heat transfer mediated through the lattice system, for example after optical excitation.
	
	The correlation between magnetic and lattice properties corroborates the conjecture from previous work on Fe/Ag multilayers \cite{Keune2018}, that the increasing softening of the phonons towards the interface originates from the enhancement of the Fe magnetic moment close to the interface and eventually modifies the separation of the energy ranges corresponding to the metal and insulator phonon modes. This suggests that controlling the magnetic state of the metallic interface layer can provide a mean to manipulate the transport of lattice excitations across the interface and modifies the thermalization of the entire multilayer system, for instance after an optical excitation which affects mainly one of the subsystems.

	\begin{acknowledgments}
		
		We acknowledge funding by the Deutsche Forschungsgemeinschaft (DFG, German Research Foundation) -– project number 278162697 -– SFB 1242 (subprojects A05 and C02) and WE2623/14-1. We thank U. von Hörsten for his outstanding technical assistance and help for sample preparation. Calculations were carried out on the MagnitUDE supercomputer system (DFG grants no.\ INST 20876/209-1 FUGG and INST 20876/243-1 FUGG). This research used resources of the Advanced Photon Source, a U.S. Department of Energy (DOE) Office of Science User Facility operated for the DOE Office of Science by Argonne National Laboratory under Contract No. DE-AC02-06CH11357
	\end{acknowledgments}
	

\begin{thebibliography}{94}%
		\makeatletter
		\providecommand \@ifxundefined [1]{%
			\@ifx{#1\undefined}
		}%
		\providecommand \@ifnum [1]{%
			\ifnum #1\expandafter \@firstoftwo
			\else \expandafter \@secondoftwo
			\fi
		}%
		\providecommand \@ifx [1]{%
			\ifx #1\expandafter \@firstoftwo
			\else \expandafter \@secondoftwo
			\fi
		}%
		\providecommand \natexlab [1]{#1}%
		\providecommand \enquote  [1]{``#1''}%
		\providecommand \bibnamefont  [1]{#1}%
		\providecommand \bibfnamefont [1]{#1}%
		\providecommand \citenamefont [1]{#1}%
		\providecommand \href@noop [0]{\@secondoftwo}%
		\providecommand \href [0]{\begingroup \@sanitize@url \@href}%
		\providecommand \@href[1]{\@@startlink{#1}\@@href}%
		\providecommand \@@href[1]{\endgroup#1\@@endlink}%
		\providecommand \@sanitize@url [0]{\catcode `\\12\catcode `\$12\catcode
			`\&12\catcode `\#12\catcode `\^12\catcode `\_12\catcode `\%12\relax}%
		\providecommand \@@startlink[1]{}%
		\providecommand \@@endlink[0]{}%
		\providecommand \url  [0]{\begingroup\@sanitize@url \@url }%
		\providecommand \@url [1]{\endgroup\@href {#1}{\urlprefix }}%
		\providecommand \urlprefix  [0]{URL }%
		\providecommand \Eprint [0]{\href }%
		\providecommand \doibase [0]{https://doi.org/}%
		\providecommand \selectlanguage [0]{\@gobble}%
		\providecommand \bibinfo  [0]{\@secondoftwo}%
		\providecommand \bibfield  [0]{\@secondoftwo}%
		\providecommand \translation [1]{[#1]}%
		\providecommand \BibitemOpen [0]{}%
		\providecommand \bibitemStop [0]{}%
		\providecommand \bibitemNoStop [0]{.\EOS\space}%
		\providecommand \EOS [0]{\spacefactor3000\relax}%
		\providecommand \BibitemShut  [1]{\csname bibitem#1\endcsname}%
		\let\auto@bib@innerbib\@empty
		\bibitem [{\citenamefont {Valeri}\ \emph {et~al.}(2007)\citenamefont {Valeri},
			\citenamefont {Benedetti},\ and\ \citenamefont {Luches}}]{cn:Valeri07}%
		\BibitemOpen
		\bibfield  {author} {\bibinfo {author} {\bibfnamefont {S.}~\bibnamefont
				{Valeri}}, \bibinfo {author} {\bibfnamefont {S.}~\bibnamefont {Benedetti}},\
			and\ \bibinfo {author} {\bibfnamefont {P.}~\bibnamefont {Luches}},\
		}\bibfield  {title} {\bibinfo {title} {Metals on oxides: Structure,
				morphology and interface chemistry},\ }\href
		{https://doi.org/10.1088/0953-8984/19/22/225002} {\bibfield  {journal}
			{\bibinfo  {journal} {J. Phys.: Condens. Matter}\ }\textbf {\bibinfo {volume}
				{19}},\ \bibinfo {pages} {225002} (\bibinfo {year} {2007})}\BibitemShut
		{NoStop}%
		\bibitem [{\citenamefont {Parkin}\ \emph {et~al.}(2004)\citenamefont {Parkin},
			\citenamefont {Kaiser}, \citenamefont {Panchula}, \citenamefont {Rice},
			\citenamefont {Hughes}, \citenamefont {Samant},\ and\ \citenamefont
			{Yang}}]{cn:Parkin04}%
		\BibitemOpen
		\bibfield  {author} {\bibinfo {author} {\bibfnamefont {S.~S.~P.}\
				\bibnamefont {Parkin}}, \bibinfo {author} {\bibfnamefont {C.}~\bibnamefont
				{Kaiser}}, \bibinfo {author} {\bibfnamefont {A.}~\bibnamefont {Panchula}},
			\bibinfo {author} {\bibfnamefont {P.~M.}\ \bibnamefont {Rice}}, \bibinfo
			{author} {\bibfnamefont {B.}~\bibnamefont {Hughes}}, \bibinfo {author}
			{\bibfnamefont {M.}~\bibnamefont {Samant}},\ and\ \bibinfo {author}
			{\bibfnamefont {S.-H.}\ \bibnamefont {Yang}},\ }\bibfield  {title} {\bibinfo
			{title} {Giant tunnelling magnetoresistance at room temperature with mgo(100)
				tunnel barriers},\ }\href {https://doi.org/10.1038/nmat1256} {\bibfield
			{journal} {\bibinfo  {journal} {Nat. Mater.}\ }\textbf {\bibinfo {volume}
				{3}},\ \bibinfo {pages} {862} (\bibinfo {year} {2004})}\BibitemShut {NoStop}%
		\bibitem [{\citenamefont {Yuasa}\ \emph {et~al.}(2004)\citenamefont {Yuasa},
			\citenamefont {Nagahama}, \citenamefont {Fukushima}, \citenamefont {Suzuki},\
			and\ \citenamefont {Ando}}]{cn:Yuasa04}%
		\BibitemOpen
		\bibfield  {author} {\bibinfo {author} {\bibfnamefont {S.}~\bibnamefont
				{Yuasa}}, \bibinfo {author} {\bibfnamefont {T.}~\bibnamefont {Nagahama}},
			\bibinfo {author} {\bibfnamefont {A.}~\bibnamefont {Fukushima}}, \bibinfo
			{author} {\bibfnamefont {Y.}~\bibnamefont {Suzuki}},\ and\ \bibinfo {author}
			{\bibfnamefont {K.}~\bibnamefont {Ando}},\ }\bibfield  {title} {\bibinfo
			{title} {Giant room-temperature magnetoresistance in single-crystal fe/mgo/fe
				magnetic tunnel junctions},\ }\href {https://doi.org/10.1038/nmat1257}
		{\bibfield  {journal} {\bibinfo  {journal} {Nat. Mater.}\ }\textbf {\bibinfo
				{volume} {3}},\ \bibinfo {pages} {868} (\bibinfo {year} {2004})}\BibitemShut
		{NoStop}%
		\bibitem [{\citenamefont {Iovan}\ \emph {et~al.}(2008)\citenamefont {Iovan},
			\citenamefont {Andersson}, \citenamefont {Naidyuk}, \citenamefont {Vedyaev},
			\citenamefont {Dieny},\ and\ \citenamefont {Korenivski}}]{cn:Iovan08}%
		\BibitemOpen
		\bibfield  {author} {\bibinfo {author} {\bibfnamefont {A.}~\bibnamefont
				{Iovan}}, \bibinfo {author} {\bibfnamefont {S.}~\bibnamefont {Andersson}},
			\bibinfo {author} {\bibfnamefont {Y.~G.}\ \bibnamefont {Naidyuk}}, \bibinfo
			{author} {\bibfnamefont {A.}~\bibnamefont {Vedyaev}}, \bibinfo {author}
			{\bibfnamefont {B.}~\bibnamefont {Dieny}},\ and\ \bibinfo {author}
			{\bibfnamefont {V.}~\bibnamefont {Korenivski}},\ }\bibfield  {title}
		{\bibinfo {title} {Spin diode based on fe/mgo double tunnel junction},\
		}\href {https://doi.org/10.1021/nl072676z} {\bibfield  {journal} {\bibinfo
				{journal} {Nano Letters}\ }\textbf {\bibinfo {volume} {8}},\ \bibinfo {pages}
			{805} (\bibinfo {year} {2008})}\BibitemShut {NoStop}%
		\bibitem [{\citenamefont {Schuster}\ \emph {et~al.}(2010)\citenamefont
			{Schuster}, \citenamefont {Brand}, \citenamefont {Stromberg}, \citenamefont
			{Ludwig}, \citenamefont {Reuter}, \citenamefont {Wieck}, \citenamefont
			{H\"{o}vel}, \citenamefont {Gerhardt}, \citenamefont {Hofmann}, \citenamefont
			{Wende},\ and\ \citenamefont {Keune}}]{Schuster2010}%
		\BibitemOpen
		\bibfield  {author} {\bibinfo {author} {\bibfnamefont {E.}~\bibnamefont
				{Schuster}}, \bibinfo {author} {\bibfnamefont {R.~A.}\ \bibnamefont {Brand}},
			\bibinfo {author} {\bibfnamefont {F.}~\bibnamefont {Stromberg}}, \bibinfo
			{author} {\bibfnamefont {A.}~\bibnamefont {Ludwig}}, \bibinfo {author}
			{\bibfnamefont {D.}~\bibnamefont {Reuter}}, \bibinfo {author} {\bibfnamefont
				{A.~D.}\ \bibnamefont {Wieck}}, \bibinfo {author} {\bibfnamefont
				{S.}~\bibnamefont {H\"{o}vel}}, \bibinfo {author} {\bibfnamefont {N.~C.}\
				\bibnamefont {Gerhardt}}, \bibinfo {author} {\bibfnamefont {M.~R.}\
				\bibnamefont {Hofmann}}, \bibinfo {author} {\bibfnamefont {H.}~\bibnamefont
				{Wende}},\ and\ \bibinfo {author} {\bibfnamefont {W.}~\bibnamefont {Keune}},\
		}\bibfield  {title} {\bibinfo {title} {Epitaxial growth and interfacial
				magnetism of spin aligner for remanent spin injection:
				{[Fe/Tb]n/Fe/}{MgO}/{GaAs}-light emitting diode as a prototype system},\
		}\href {https://doi.org/10.1063/1.3476265} {\bibfield  {journal} {\bibinfo
				{journal} {J. Appl. Phys.}\ }\textbf {\bibinfo {volume} {108}},\ \bibinfo
			{pages} {063902} (\bibinfo {year} {2010})}\BibitemShut {NoStop}%
		\bibitem [{\citenamefont {Tiusan}\ \emph {et~al.}(2006)\citenamefont {Tiusan},
			\citenamefont {Sicot}, \citenamefont {Hehn}, \citenamefont {Belouard},
			\citenamefont {Andrieu}, \citenamefont {Montaigne},\ and\ \citenamefont
			{Schuhl}}]{cn:Tiusan06}%
		\BibitemOpen
		\bibfield  {author} {\bibinfo {author} {\bibfnamefont {C.}~\bibnamefont
				{Tiusan}}, \bibinfo {author} {\bibfnamefont {M.}~\bibnamefont {Sicot}},
			\bibinfo {author} {\bibfnamefont {M.}~\bibnamefont {Hehn}}, \bibinfo {author}
			{\bibfnamefont {C.}~\bibnamefont {Belouard}}, \bibinfo {author}
			{\bibfnamefont {S.}~\bibnamefont {Andrieu}}, \bibinfo {author} {\bibfnamefont
				{F.}~\bibnamefont {Montaigne}},\ and\ \bibinfo {author} {\bibfnamefont
				{A.}~\bibnamefont {Schuhl}},\ }\bibfield  {title} {\bibinfo {title} {Fe/mgo
				interface engineering for high-output-voltage device applications},\ }\href
		{https://doi.org/10.1063/1.2172717} {\bibfield  {journal} {\bibinfo
				{journal} {Appl. Phys. Lett.}\ }\textbf {\bibinfo {volume} {88}},\ \bibinfo
			{pages} {062512} (\bibinfo {year} {2006})}\BibitemShut {NoStop}%
		\bibitem [{\citenamefont {Butler}\ \emph {et~al.}(2001)\citenamefont {Butler},
			\citenamefont {Zhang}, \citenamefont {Schulthess},\ and\ \citenamefont
			{MacLaren}}]{cn:Butler01}%
		\BibitemOpen
		\bibfield  {author} {\bibinfo {author} {\bibfnamefont {W.~H.}\ \bibnamefont
				{Butler}}, \bibinfo {author} {\bibfnamefont {X.-G.}\ \bibnamefont {Zhang}},
			\bibinfo {author} {\bibfnamefont {T.~C.}\ \bibnamefont {Schulthess}},\ and\
			\bibinfo {author} {\bibfnamefont {J.~M.}\ \bibnamefont {MacLaren}},\
		}\bibfield  {title} {\bibinfo {title} {Spin-dependent tunneling conductance
				of $\mathrm{Fe}|\mathrm{MgO}|\mathrm{Fe}$ sandwiches},\ }\href
		{https://doi.org/10.1103/PhysRevB.63.054416} {\bibfield  {journal} {\bibinfo
				{journal} {Phys. Rev. B}\ }\textbf {\bibinfo {volume} {63}},\ \bibinfo
			{pages} {054416} (\bibinfo {year} {2001})}\BibitemShut {NoStop}%
		\bibitem [{\citenamefont {Mathon}\ and\ \citenamefont
			{Umerski}(2001)}]{cn:Mathon01}%
		\BibitemOpen
		\bibfield  {author} {\bibinfo {author} {\bibfnamefont {J.}~\bibnamefont
				{Mathon}}\ and\ \bibinfo {author} {\bibfnamefont {A.}~\bibnamefont
				{Umerski}},\ }\bibfield  {title} {\bibinfo {title} {Theory of tunneling
				magnetoresistance of an epitaxial fe/mgo/fe(001) junction},\ }\href
		{https://doi.org/10.1103/PhysRevB.63.220403} {\bibfield  {journal} {\bibinfo
				{journal} {Phys. Rev. B}\ }\textbf {\bibinfo {volume} {63}},\ \bibinfo
			{pages} {220403} (\bibinfo {year} {2001})}\BibitemShut {NoStop}%
		\bibitem [{\citenamefont {Tiusan}\ \emph {et~al.}(2004)\citenamefont {Tiusan},
			\citenamefont {Faure-Vincent}, \citenamefont {Bellouard}, \citenamefont
			{Hehn}, \citenamefont {Jouguelet},\ and\ \citenamefont
			{Schuhl}}]{cn:Tiusan04}%
		\BibitemOpen
		\bibfield  {author} {\bibinfo {author} {\bibfnamefont {C.}~\bibnamefont
				{Tiusan}}, \bibinfo {author} {\bibfnamefont {J.}~\bibnamefont
				{Faure-Vincent}}, \bibinfo {author} {\bibfnamefont {C.}~\bibnamefont
				{Bellouard}}, \bibinfo {author} {\bibfnamefont {M.}~\bibnamefont {Hehn}},
			\bibinfo {author} {\bibfnamefont {E.}~\bibnamefont {Jouguelet}},\ and\
			\bibinfo {author} {\bibfnamefont {A.}~\bibnamefont {Schuhl}},\ }\bibfield
		{title} {\bibinfo {title} {Interfacial resonance state probed by
				spin-polarized tunneling in epitaxial fe/mgo/fe tunnel junctions},\ }\href
		{https://doi.org/10.1103/PhysRevLett.93.106602} {\bibfield  {journal}
			{\bibinfo  {journal} {Phys. Rev. Lett.}\ }\textbf {\bibinfo {volume} {93}},\
			\bibinfo {pages} {106602} (\bibinfo {year} {2004})}\BibitemShut {NoStop}%
		\bibitem [{\citenamefont {Waldron}\ \emph {et~al.}(2006)\citenamefont
			{Waldron}, \citenamefont {Timoshevskii}, \citenamefont {Hu}, \citenamefont
			{Xia},\ and\ \citenamefont {Guo}}]{cn:Waldron06}%
		\BibitemOpen
		\bibfield  {author} {\bibinfo {author} {\bibfnamefont {D.}~\bibnamefont
				{Waldron}}, \bibinfo {author} {\bibfnamefont {V.}~\bibnamefont
				{Timoshevskii}}, \bibinfo {author} {\bibfnamefont {Y.}~\bibnamefont {Hu}},
			\bibinfo {author} {\bibfnamefont {K.}~\bibnamefont {Xia}},\ and\ \bibinfo
			{author} {\bibfnamefont {H.}~\bibnamefont {Guo}},\ }\bibfield  {title}
		{\bibinfo {title} {First principles modeling of tunnel magnetoresistance of
				$\mathrm{Fe}/\mathrm{MgO}/\mathrm{Fe}$ trilayers},\ }\href
		{https://doi.org/10.1103/PhysRevLett.97.226802} {\bibfield  {journal}
			{\bibinfo  {journal} {Phys. Rev. Lett.}\ }\textbf {\bibinfo {volume} {97}},\
			\bibinfo {pages} {226802} (\bibinfo {year} {2006})}\BibitemShut {NoStop}%
		\bibitem [{\citenamefont {Belashchenko}\ \emph {et~al.}(2005)\citenamefont
			{Belashchenko}, \citenamefont {Velev},\ and\ \citenamefont
			{Tsymbal}}]{cn:Belashchenko05}%
		\BibitemOpen
		\bibfield  {author} {\bibinfo {author} {\bibfnamefont {K.~D.}\ \bibnamefont
				{Belashchenko}}, \bibinfo {author} {\bibfnamefont {J.}~\bibnamefont
				{Velev}},\ and\ \bibinfo {author} {\bibfnamefont {E.~Y.}\ \bibnamefont
				{Tsymbal}},\ }\bibfield  {title} {\bibinfo {title} {Effect of interface
				states on spin-dependent tunneling in
				$\mathrm{Fe}∕\mathrm{Mg}\mathrm{O}∕\mathrm{Fe}$ tunnel junctions},\
		}\href {https://doi.org/10.1103/PhysRevB.72.140404} {\bibfield  {journal}
			{\bibinfo  {journal} {Phys. Rev. B}\ }\textbf {\bibinfo {volume} {72}},\
			\bibinfo {pages} {140404} (\bibinfo {year} {2005})}\BibitemShut {NoStop}%
		\bibitem [{\citenamefont {Heiliger}\ \emph {et~al.}(2008)\citenamefont
			{Heiliger}, \citenamefont {Zahn}, \citenamefont {Yavorsky},\ and\
			\citenamefont {Mertig}}]{cn:Heiliger08}%
		\BibitemOpen
		\bibfield  {author} {\bibinfo {author} {\bibfnamefont {C.}~\bibnamefont
				{Heiliger}}, \bibinfo {author} {\bibfnamefont {P.}~\bibnamefont {Zahn}},
			\bibinfo {author} {\bibfnamefont {B.~Y.}\ \bibnamefont {Yavorsky}},\ and\
			\bibinfo {author} {\bibfnamefont {I.}~\bibnamefont {Mertig}},\ }\bibfield
		{title} {\bibinfo {title} {Thickness dependence of the tunneling current in
				the coherent limit of transport},\ }\href
		{https://doi.org/10.1103/PhysRevB.77.224407} {\bibfield  {journal} {\bibinfo
				{journal} {Phys. Rev. B}\ }\textbf {\bibinfo {volume} {77}},\ \bibinfo
			{pages} {224407} (\bibinfo {year} {2008})}\BibitemShut {NoStop}%
		\bibitem [{\citenamefont {Peralta-Ramos}\ \emph {et~al.}(2008)\citenamefont
			{Peralta-Ramos}, \citenamefont {Llois}, \citenamefont {Rungger},\ and\
			\citenamefont {Sanvito}}]{cn:Peralta08}%
		\BibitemOpen
		\bibfield  {author} {\bibinfo {author} {\bibfnamefont {J.}~\bibnamefont
				{Peralta-Ramos}}, \bibinfo {author} {\bibfnamefont {A.~M.}\ \bibnamefont
				{Llois}}, \bibinfo {author} {\bibfnamefont {I.}~\bibnamefont {Rungger}},\
			and\ \bibinfo {author} {\bibfnamefont {S.}~\bibnamefont {Sanvito}},\
		}\bibfield  {title} {\bibinfo {title} {I-v curves of fe/mgo (001) single- and
				double-barrier tunnel junctions},\ }\href
		{https://doi.org/10.1103/PhysRevB.78.024430} {\bibfield  {journal} {\bibinfo
				{journal} {Phys. Rev. B}\ }\textbf {\bibinfo {volume} {78}},\ \bibinfo
			{pages} {024430} (\bibinfo {year} {2008})}\BibitemShut {NoStop}%
		\bibitem [{\citenamefont {Rungger}\ \emph {et~al.}(2009)\citenamefont
			{Rungger}, \citenamefont {Mryasov},\ and\ \citenamefont
			{Sanvito}}]{cn:Rungger09}%
		\BibitemOpen
		\bibfield  {author} {\bibinfo {author} {\bibfnamefont {I.}~\bibnamefont
				{Rungger}}, \bibinfo {author} {\bibfnamefont {O.}~\bibnamefont {Mryasov}},\
			and\ \bibinfo {author} {\bibfnamefont {S.}~\bibnamefont {Sanvito}},\
		}\bibfield  {title} {\bibinfo {title} {Resonant electronic states and i-v
				curves of fe/mgo/fe(100) tunnel junctions},\ }\href
		{https://doi.org/10.1103/PhysRevB.79.094414} {\bibfield  {journal} {\bibinfo
				{journal} {Phys. Rev. B}\ }\textbf {\bibinfo {volume} {79}},\ \bibinfo
			{pages} {094414} (\bibinfo {year} {2009})}\BibitemShut {NoStop}%
		\bibitem [{\citenamefont {Feng}\ \emph {et~al.}(2009)\citenamefont {Feng},
			\citenamefont {Bengone}, \citenamefont {Alouani}, \citenamefont {Lebègue},
			\citenamefont {Rungger},\ and\ \citenamefont {Sanvito}}]{cn:Feng09}%
		\BibitemOpen
		\bibfield  {author} {\bibinfo {author} {\bibfnamefont {X.}~\bibnamefont
				{Feng}}, \bibinfo {author} {\bibfnamefont {O.}~\bibnamefont {Bengone}},
			\bibinfo {author} {\bibfnamefont {M.}~\bibnamefont {Alouani}}, \bibinfo
			{author} {\bibfnamefont {S.}~\bibnamefont {Lebègue}}, \bibinfo {author}
			{\bibfnamefont {I.}~\bibnamefont {Rungger}},\ and\ \bibinfo {author}
			{\bibfnamefont {S.}~\bibnamefont {Sanvito}},\ }\bibfield  {title} {\bibinfo
			{title} {Effects of structural relaxation on calculations of the interface
				and transport properties of fe/mgo(001) tunnel junctions},\ }\href
		{https://doi.org/10.1103/PhysRevB.79.174414} {\bibfield  {journal} {\bibinfo
				{journal} {Phys. Rev. B}\ }\textbf {\bibinfo {volume} {79}},\ \bibinfo
			{pages} {174414} (\bibinfo {year} {2009})}\BibitemShut {NoStop}%
		\bibitem [{\citenamefont {Abedi~Ravan}\ \emph {et~al.}(2010)\citenamefont
			{Abedi~Ravan}, \citenamefont {Shokri},\ and\ \citenamefont
			{Yazdani}}]{cn:Abedi10}%
		\BibitemOpen
		\bibfield  {author} {\bibinfo {author} {\bibfnamefont {B.}~\bibnamefont
				{Abedi~Ravan}}, \bibinfo {author} {\bibfnamefont {A.~A.}\ \bibnamefont
				{Shokri}},\ and\ \bibinfo {author} {\bibfnamefont {A.}~\bibnamefont
				{Yazdani}},\ }\bibfield  {title} {\bibinfo {title} {Spin-dependent tunneling
				characteristics in fe/mgo/fe trilayers: First-principles calculations},\
		}\href {https://doi.org/10.1016/j.ssc.2009.10.040} {\bibfield  {journal}
			{\bibinfo  {journal} {Solid State Commun.}\ }\textbf {\bibinfo {volume}
				{150}},\ \bibinfo {pages} {214} (\bibinfo {year} {2010})}\BibitemShut
		{NoStop}%
		\bibitem [{\citenamefont {Raza}\ \emph {et~al.}(2011)\citenamefont {Raza},
			\citenamefont {Cerdá},\ and\ \citenamefont {Raza}}]{cn:Raza11}%
		\BibitemOpen
		\bibfield  {author} {\bibinfo {author} {\bibfnamefont {T.~Z.}\ \bibnamefont
				{Raza}}, \bibinfo {author} {\bibfnamefont {J.~I.}\ \bibnamefont {Cerdá}},\
			and\ \bibinfo {author} {\bibfnamefont {H.}~\bibnamefont {Raza}},\ }\bibfield
		{title} {\bibinfo {title} {Three-dimensional extended h\"uckel
				theory-nonequilibrium green's function spin polarized transport model for
				fe/mgo/fe heterostructures},\ }\href {https://doi.org/10.1063/1.3525703}
		{\bibfield  {journal} {\bibinfo  {journal} {J. Appl. Phys.}\ }\textbf
			{\bibinfo {volume} {109}},\ \bibinfo {pages} {023705} (\bibinfo {year}
			{2011})}\BibitemShut {NoStop}%
		\bibitem [{\citenamefont {Colonna}\ \emph {et~al.}(2009)\citenamefont
			{Colonna}, \citenamefont {Cricenti}, \citenamefont {Luches}, \citenamefont
			{Valeri}, \citenamefont {Boscherini}, \citenamefont {Qi}, \citenamefont
			{Xu},\ and\ \citenamefont {Tolk}}]{cn:Colonna09}%
		\BibitemOpen
		\bibfield  {author} {\bibinfo {author} {\bibfnamefont {S.}~\bibnamefont
				{Colonna}}, \bibinfo {author} {\bibfnamefont {A.}~\bibnamefont {Cricenti}},
			\bibinfo {author} {\bibfnamefont {P.}~\bibnamefont {Luches}}, \bibinfo
			{author} {\bibfnamefont {S.}~\bibnamefont {Valeri}}, \bibinfo {author}
			{\bibfnamefont {F.}~\bibnamefont {Boscherini}}, \bibinfo {author}
			{\bibfnamefont {J.}~\bibnamefont {Qi}}, \bibinfo {author} {\bibfnamefont
				{Y.}~\bibnamefont {Xu}},\ and\ \bibinfo {author} {\bibfnamefont
				{N.}~\bibnamefont {Tolk}},\ }\bibfield  {title} {\bibinfo {title}
			{Fe/nio(100) and fe/mgo(100) interfaces studied by x-ray absorption
				spectroscopy and non-linear kerr effect},\ }\href
		{https://doi.org/10.1016/j.spmi.2009.01.013} {\bibfield  {journal} {\bibinfo
				{journal} {Superlattices Microstruct.}\ }\textbf {\bibinfo {volume} {46}},\
			\bibinfo {pages} {107} (\bibinfo {year} {2009})}\BibitemShut {NoStop}%
		\bibitem [{\citenamefont {Choi}\ \emph {et~al.}(2011)\citenamefont {Choi},
			\citenamefont {Lee}, \citenamefont {Cho}, \citenamefont {Yoo},\ and\
			\citenamefont {Chung}}]{cn:Choi11}%
		\BibitemOpen
		\bibfield  {author} {\bibinfo {author} {\bibfnamefont {H.}~\bibnamefont
				{Choi}}, \bibinfo {author} {\bibfnamefont {E.-K.}\ \bibnamefont {Lee}},
			\bibinfo {author} {\bibfnamefont {S.~B.}\ \bibnamefont {Cho}}, \bibinfo
			{author} {\bibfnamefont {D.~S.}\ \bibnamefont {Yoo}},\ and\ \bibinfo {author}
			{\bibfnamefont {Y.-C.}\ \bibnamefont {Chung}},\ }\bibfield  {title} {\bibinfo
			{title} {Interface-dependent spin-reorientation energy barrier in
				fe/{MgO}(001) thin film},\ }\href {https://doi.org/10.1109/led.2011.2160148}
		{\bibfield  {journal} {\bibinfo  {journal} {{IEEE} Electron Device Letters}\
			}\textbf {\bibinfo {volume} {32}},\ \bibinfo {pages} {1287} (\bibinfo {year}
			{2011})}\BibitemShut {NoStop}%
		\bibitem [{\citenamefont {Bose}\ \emph {et~al.}(2016)\citenamefont {Bose},
			\citenamefont {Cuadrado}, \citenamefont {Evans}, \citenamefont {Chepulskii},
			\citenamefont {Apalkov},\ and\ \citenamefont {Chantrell}}]{cn:Bose2016}%
		\BibitemOpen
		\bibfield  {author} {\bibinfo {author} {\bibfnamefont {T.}~\bibnamefont
				{Bose}}, \bibinfo {author} {\bibfnamefont {R.}~\bibnamefont {Cuadrado}},
			\bibinfo {author} {\bibfnamefont {R.~F.~L.}\ \bibnamefont {Evans}}, \bibinfo
			{author} {\bibfnamefont {R.~V.}\ \bibnamefont {Chepulskii}}, \bibinfo
			{author} {\bibfnamefont {D.}~\bibnamefont {Apalkov}},\ and\ \bibinfo {author}
			{\bibfnamefont {R.~W.}\ \bibnamefont {Chantrell}},\ }\bibfield  {title}
		{\bibinfo {title} {First-principles study of the fe | mgo(0 0 1) interface:
				Magnetic anisotropy},\ }\href
		{https://doi.org/10.1088/0953-8984/28/15/156003} {\bibfield  {journal}
			{\bibinfo  {journal} {J. Phys.: Condens. Matter}\ }\textbf {\bibinfo {volume}
				{28}},\ \bibinfo {pages} {156003} (\bibinfo {year} {2016})}\BibitemShut
		{NoStop}%
		\bibitem [{\citenamefont {Gu}\ \emph {et~al.}(2017)\citenamefont {Gu},
			\citenamefont {Takahashi},\ and\ \citenamefont {Maekawa}}]{cn:Gu17}%
		\BibitemOpen
		\bibfield  {author} {\bibinfo {author} {\bibfnamefont {B.}~\bibnamefont
				{Gu}}, \bibinfo {author} {\bibfnamefont {S.}~\bibnamefont {Takahashi}},\ and\
			\bibinfo {author} {\bibfnamefont {S.}~\bibnamefont {Maekawa}},\ }\bibfield
		{title} {\bibinfo {title} {Enhanced magneto-optical kerr effect at
				fe/insulator interfaces},\ }\href
		{https://doi.org/10.1103/PhysRevB.96.214423} {\bibfield  {journal} {\bibinfo
				{journal} {Phys. Rev. B}\ }\textbf {\bibinfo {volume} {96}},\ \bibinfo
			{pages} {214423} (\bibinfo {year} {2017})}\BibitemShut {NoStop}%
		\bibitem [{\citenamefont {Koo}\ \emph {et~al.}(2013)\citenamefont {Koo},
			\citenamefont {Mitani}, \citenamefont {Sasaki}, \citenamefont {Sukegawa},
			\citenamefont {Wen}, \citenamefont {Ohkubo}, \citenamefont {Niizeki},
			\citenamefont {Inomata},\ and\ \citenamefont {Hono}}]{Koo2013}%
		\BibitemOpen
		\bibfield  {author} {\bibinfo {author} {\bibfnamefont {J.~W.}\ \bibnamefont
				{Koo}}, \bibinfo {author} {\bibfnamefont {S.}~\bibnamefont {Mitani}},
			\bibinfo {author} {\bibfnamefont {T.~T.}\ \bibnamefont {Sasaki}}, \bibinfo
			{author} {\bibfnamefont {H.}~\bibnamefont {Sukegawa}}, \bibinfo {author}
			{\bibfnamefont {Z.~C.}\ \bibnamefont {Wen}}, \bibinfo {author} {\bibfnamefont
				{T.}~\bibnamefont {Ohkubo}}, \bibinfo {author} {\bibfnamefont
				{T.}~\bibnamefont {Niizeki}}, \bibinfo {author} {\bibfnamefont
				{K.}~\bibnamefont {Inomata}},\ and\ \bibinfo {author} {\bibfnamefont
				{K.}~\bibnamefont {Hono}},\ }\bibfield  {title} {\bibinfo {title} {Large
				perpendicular magnetic anisotropy at fe/{MgO} interface},\ }\href
		{https://doi.org/10.1063/1.4828658} {\bibfield  {journal} {\bibinfo
				{journal} {Appl. Phys. Lett.}\ }\textbf {\bibinfo {volume} {103}},\ \bibinfo
			{pages} {192401} (\bibinfo {year} {2013})}\BibitemShut {NoStop}%
		\bibitem [{\citenamefont {Okabayashi}\ \emph {et~al.}(2014)\citenamefont
			{Okabayashi}, \citenamefont {Koo}, \citenamefont {Sukegawa}, \citenamefont
			{Mitani}, \citenamefont {Takagi},\ and\ \citenamefont
			{Yokoyama}}]{Okabayashi2014}%
		\BibitemOpen
		\bibfield  {author} {\bibinfo {author} {\bibfnamefont {J.}~\bibnamefont
				{Okabayashi}}, \bibinfo {author} {\bibfnamefont {J.~W.}\ \bibnamefont {Koo}},
			\bibinfo {author} {\bibfnamefont {H.}~\bibnamefont {Sukegawa}}, \bibinfo
			{author} {\bibfnamefont {S.}~\bibnamefont {Mitani}}, \bibinfo {author}
			{\bibfnamefont {Y.}~\bibnamefont {Takagi}},\ and\ \bibinfo {author}
			{\bibfnamefont {T.}~\bibnamefont {Yokoyama}},\ }\bibfield  {title} {\bibinfo
			{title} {{Perpendicular magnetic anisotropy at the interface between
					ultrathin Fe film and MgO studied by angular-dependent x-ray magnetic
					circular dichroism}},\ }\href {https://doi.org/10.1063/1.4896290} {\bibfield
			{journal} {\bibinfo  {journal} {Appl. Phys. Lett.}\ }\textbf {\bibinfo
				{volume} {105}},\ \bibinfo {pages} {122408} (\bibinfo {year}
			{2014})}\BibitemShut {NoStop}%
		\bibitem [{\citenamefont {Kozio{\l}-Rachwa{\l}}\ \emph
			{et~al.}(2014)\citenamefont {Kozio{\l}-Rachwa{\l}}, \citenamefont
			{{\'{S}}l{\k{e}}zak}, \citenamefont {Matlak}, \citenamefont {Ku{\'{s}}wik},
			\citenamefont {Urbaniak}, \citenamefont {Stobiecki}, \citenamefont {Yao},
			\citenamefont {van Dijken},\ and\ \citenamefont {Korecki}}]{KozioRachwa2014}%
		\BibitemOpen
		\bibfield  {author} {\bibinfo {author} {\bibfnamefont {A.}~\bibnamefont
				{Kozio{\l}-Rachwa{\l}}}, \bibinfo {author} {\bibfnamefont {T.}~\bibnamefont
				{{\'{S}}l{\k{e}}zak}}, \bibinfo {author} {\bibfnamefont {K.}~\bibnamefont
				{Matlak}}, \bibinfo {author} {\bibfnamefont {P.}~\bibnamefont
				{Ku{\'{s}}wik}}, \bibinfo {author} {\bibfnamefont {M.}~\bibnamefont
				{Urbaniak}}, \bibinfo {author} {\bibfnamefont {F.}~\bibnamefont {Stobiecki}},
			\bibinfo {author} {\bibfnamefont {L.~D.}\ \bibnamefont {Yao}}, \bibinfo
			{author} {\bibfnamefont {S.}~\bibnamefont {van Dijken}},\ and\ \bibinfo
			{author} {\bibfnamefont {J.}~\bibnamefont {Korecki}},\ }\bibfield  {title}
		{\bibinfo {title} {Tunable magnetic properties of monoatomic metal-oxide
				fe/{MgO} multilayers},\ }\bibfield  {journal} {\bibinfo  {journal} {Phys.
				Rev. B}\ }\textbf {\bibinfo {volume} {90}},\ \href
		{https://doi.org/10.1103/physrevb.90.045428} {10.1103/physrevb.90.045428}
		(\bibinfo {year} {2014})\BibitemShut {NoStop}%
		\bibitem [{\citenamefont {Balogh}\ \emph {et~al.}(2013)\citenamefont {Balogh},
			\citenamefont {D\'ezsi}, \citenamefont {Fetzer}, \citenamefont {Korecki},
			\citenamefont {Kozioł-Rachwal}, \citenamefont {Mly\'nczak},\ and\
			\citenamefont {Nakanishi}}]{cn:Balogh13}%
		\BibitemOpen
		\bibfield  {author} {\bibinfo {author} {\bibfnamefont {J.}~\bibnamefont
				{Balogh}}, \bibinfo {author} {\bibfnamefont {I.}~\bibnamefont {D\'ezsi}},
			\bibinfo {author} {\bibfnamefont {C.}~\bibnamefont {Fetzer}}, \bibinfo
			{author} {\bibfnamefont {J.}~\bibnamefont {Korecki}}, \bibinfo {author}
			{\bibfnamefont {A.}~\bibnamefont {Kozioł-Rachwal}}, \bibinfo {author}
			{\bibfnamefont {E.}~\bibnamefont {Mly\'nczak}},\ and\ \bibinfo {author}
			{\bibfnamefont {A.}~\bibnamefont {Nakanishi}},\ }\bibfield  {title} {\bibinfo
			{title} {Magnetic properties of the fe-mgo interface studied by m\"ossbauer
				spectroscopy},\ }\href {https://doi.org/10.1103/PhysRevB.87.174415}
		{\bibfield  {journal} {\bibinfo  {journal} {Phys. Rev. B}\ }\textbf {\bibinfo
				{volume} {87}},\ \bibinfo {pages} {174415} (\bibinfo {year}
			{2013})}\BibitemShut {NoStop}%
		\bibitem [{\citenamefont {Suwardy}\ \emph {et~al.}(2018)\citenamefont
			{Suwardy}, \citenamefont {Nawaoka}, \citenamefont {Cho}, \citenamefont
			{Goto}, \citenamefont {Suzuki},\ and\ \citenamefont {Miwa}}]{Suwardy2018}%
		\BibitemOpen
		\bibfield  {author} {\bibinfo {author} {\bibfnamefont {J.}~\bibnamefont
				{Suwardy}}, \bibinfo {author} {\bibfnamefont {K.}~\bibnamefont {Nawaoka}},
			\bibinfo {author} {\bibfnamefont {J.}~\bibnamefont {Cho}}, \bibinfo {author}
			{\bibfnamefont {M.}~\bibnamefont {Goto}}, \bibinfo {author} {\bibfnamefont
				{Y.}~\bibnamefont {Suzuki}},\ and\ \bibinfo {author} {\bibfnamefont
				{S.}~\bibnamefont {Miwa}},\ }\bibfield  {title} {\bibinfo {title}
			{{Voltage-controlled magnetic anisotropy and voltage-induced
					Dzyaloshinskii-Moriya interaction change at the epitaxial Fe(001)/MgO(001)
					interface engineered by Co and Pd atomic-layer insertion}},\ }\href
		{https://doi.org/10.1103/PhysRevB.98.144432} {\bibfield  {journal} {\bibinfo
				{journal} {Phys. Rev. B}\ }\textbf {\bibinfo {volume} {98}},\ \bibinfo
			{pages} {144432} (\bibinfo {year} {2018})}\BibitemShut {NoStop}%
		\bibitem [{\citenamefont {Li}\ and\ \citenamefont {Freeman}(1991)}]{cn:Li91}%
		\BibitemOpen
		\bibfield  {author} {\bibinfo {author} {\bibfnamefont {C.}~\bibnamefont
				{Li}}\ and\ \bibinfo {author} {\bibfnamefont {A.~J.}\ \bibnamefont
				{Freeman}},\ }\bibfield  {title} {\bibinfo {title} {Giant monolayer
				magnetization of fe on mgo: A nearly ideal two-dimensional magnetic system},\
		}\href {https://doi.org/10.1103/PhysRevB.43.780} {\bibfield  {journal}
			{\bibinfo  {journal} {Phys. Rev. B}\ }\textbf {\bibinfo {volume} {43}},\
			\bibinfo {pages} {780} (\bibinfo {year} {1991})}\BibitemShut {NoStop}%
		\bibitem [{\citenamefont {Yu}\ and\ \citenamefont {Kim}(2006)}]{cn:Yu06}%
		\BibitemOpen
		\bibfield  {author} {\bibinfo {author} {\bibfnamefont {B.~D.}\ \bibnamefont
				{Yu}}\ and\ \bibinfo {author} {\bibfnamefont {J.-S.}\ \bibnamefont {Kim}},\
		}\bibfield  {title} {\bibinfo {title} {Ab initio study of ultrathin
				$\mathrm{MgO}$ films on $\mathrm{Fe}(001)$: Influence of interfacial
				structures},\ }\href {https://doi.org/10.1103/PhysRevB.73.125408} {\bibfield
			{journal} {\bibinfo  {journal} {Phys. Rev. B}\ }\textbf {\bibinfo {volume}
				{73}},\ \bibinfo {pages} {125408} (\bibinfo {year} {2006})}\BibitemShut
		{NoStop}%
		\bibitem [{\citenamefont {Ozeki}\ \emph {et~al.}(2007)\citenamefont {Ozeki},
			\citenamefont {Itoh},\ and\ \citenamefont {Inoue}}]{cn:Ozeki07}%
		\BibitemOpen
		\bibfield  {author} {\bibinfo {author} {\bibfnamefont {J.}~\bibnamefont
				{Ozeki}}, \bibinfo {author} {\bibfnamefont {H.}~\bibnamefont {Itoh}},\ and\
			\bibinfo {author} {\bibfnamefont {J.}~\bibnamefont {Inoue}},\ }\bibfield
		{title} {\bibinfo {title} {Interfacial electronic structure and oxygen
				vacancy states in fe/mgo/fe tunneling junctions},\ }\href
		{https://doi.org/10.1016/j.jmmm.2006.10.685} {\bibfield  {journal} {\bibinfo
				{journal} {J. Magn. Magn. Mater.}\ }\textbf {\bibinfo {volume} {310}},\
			\bibinfo {pages} {e644} (\bibinfo {year} {2007})}\BibitemShut {NoStop}%
		\bibitem [{\citenamefont {Sicot}\ \emph {et~al.}(2003)\citenamefont {Sicot},
			\citenamefont {Andrieu}, \citenamefont {Turban}, \citenamefont
			{Fagot-Revurat}, \citenamefont {Cercellier}, \citenamefont {Tagliaferri},
			\citenamefont {De~Nadai}, \citenamefont {Brookes}, \citenamefont {Bertran},\
			and\ \citenamefont {Fortuna}}]{cn:Sicot03}%
		\BibitemOpen
		\bibfield  {author} {\bibinfo {author} {\bibfnamefont {M.}~\bibnamefont
				{Sicot}}, \bibinfo {author} {\bibfnamefont {S.}~\bibnamefont {Andrieu}},
			\bibinfo {author} {\bibfnamefont {P.}~\bibnamefont {Turban}}, \bibinfo
			{author} {\bibfnamefont {Y.}~\bibnamefont {Fagot-Revurat}}, \bibinfo {author}
			{\bibfnamefont {H.}~\bibnamefont {Cercellier}}, \bibinfo {author}
			{\bibfnamefont {A.}~\bibnamefont {Tagliaferri}}, \bibinfo {author}
			{\bibfnamefont {C.}~\bibnamefont {De~Nadai}}, \bibinfo {author}
			{\bibfnamefont {N.~B.}\ \bibnamefont {Brookes}}, \bibinfo {author}
			{\bibfnamefont {F.}~\bibnamefont {Bertran}},\ and\ \bibinfo {author}
			{\bibfnamefont {F.}~\bibnamefont {Fortuna}},\ }\bibfield  {title} {\bibinfo
			{title} {Polarization of fe(001) covered by mgo analyzed by spin-resolved
				x-ray photoemission spectroscopy},\ }\href
		{https://doi.org/10.1103/PhysRevB.68.184406} {\bibfield  {journal} {\bibinfo
				{journal} {Phys. Rev. B}\ }\textbf {\bibinfo {volume} {68}},\ \bibinfo
			{pages} {184406} (\bibinfo {year} {2003})}\BibitemShut {NoStop}%
		\bibitem [{\citenamefont {Miyokawa}\ \emph {et~al.}(2005)\citenamefont
			{Miyokawa}, \citenamefont {Saito}, \citenamefont {Katayama}, \citenamefont
			{Saito}, \citenamefont {Kamino}, \citenamefont {Hanashima}, \citenamefont
			{Suzuki}, \citenamefont {Mamiya}, \citenamefont {Koide},\ and\ \citenamefont
			{Yuasa}}]{cn:Miyokawa05}%
		\BibitemOpen
		\bibfield  {author} {\bibinfo {author} {\bibfnamefont {K.}~\bibnamefont
				{Miyokawa}}, \bibinfo {author} {\bibfnamefont {S.}~\bibnamefont {Saito}},
			\bibinfo {author} {\bibfnamefont {T.}~\bibnamefont {Katayama}}, \bibinfo
			{author} {\bibfnamefont {T.}~\bibnamefont {Saito}}, \bibinfo {author}
			{\bibfnamefont {T.}~\bibnamefont {Kamino}}, \bibinfo {author} {\bibfnamefont
				{K.}~\bibnamefont {Hanashima}}, \bibinfo {author} {\bibfnamefont
				{Y.}~\bibnamefont {Suzuki}}, \bibinfo {author} {\bibfnamefont
				{K.}~\bibnamefont {Mamiya}}, \bibinfo {author} {\bibfnamefont
				{T.}~\bibnamefont {Koide}},\ and\ \bibinfo {author} {\bibfnamefont
				{S.}~\bibnamefont {Yuasa}},\ }\bibfield  {title} {\bibinfo {title} {X-ray
				absorption and x-ray magnetic circular dichroism studies of a monatomic
				fe(001) layer facing a single-crystalline mgo(001) tunnel barrier},\
		}\href@noop {} {\bibfield  {journal} {\bibinfo  {journal} {J. Appl. Phys.}\
			}\textbf {\bibinfo {volume} {44}},\ \bibinfo {pages} {L9} (\bibinfo {year}
			{2005})}\BibitemShut {NoStop}%
		\bibitem [{\citenamefont {Jal}\ \emph {et~al.}(2015)\citenamefont {Jal},
			\citenamefont {Kortright}, \citenamefont {Chase}, \citenamefont {Liu},
			\citenamefont {Gray}, \citenamefont {Shafer}, \citenamefont {Arenholz},
			\citenamefont {Xu}, \citenamefont {Jeong}, \citenamefont {Samant},
			\citenamefont {Parkin},\ and\ \citenamefont {D\"urr}}]{cn:Jal15}%
		\BibitemOpen
		\bibfield  {author} {\bibinfo {author} {\bibfnamefont {E.}~\bibnamefont
				{Jal}}, \bibinfo {author} {\bibfnamefont {J.~B.}\ \bibnamefont {Kortright}},
			\bibinfo {author} {\bibfnamefont {T.}~\bibnamefont {Chase}}, \bibinfo
			{author} {\bibfnamefont {T.}~\bibnamefont {Liu}}, \bibinfo {author}
			{\bibfnamefont {A.~X.}\ \bibnamefont {Gray}}, \bibinfo {author}
			{\bibfnamefont {P.}~\bibnamefont {Shafer}}, \bibinfo {author} {\bibfnamefont
				{E.}~\bibnamefont {Arenholz}}, \bibinfo {author} {\bibfnamefont
				{P.}~\bibnamefont {Xu}}, \bibinfo {author} {\bibfnamefont {J.}~\bibnamefont
				{Jeong}}, \bibinfo {author} {\bibfnamefont {M.~G.}\ \bibnamefont {Samant}},
			\bibinfo {author} {\bibfnamefont {S.~S.~P.}\ \bibnamefont {Parkin}},\ and\
			\bibinfo {author} {\bibfnamefont {H.~A.}\ \bibnamefont {D\"urr}},\ }\bibfield
		{title} {\bibinfo {title} {Interface fe magnetic moment enhancement in
				mgo/fe/mgo trilayers},\ }\href {https://doi.org/10.1063/1.4929990} {\bibfield
			{journal} {\bibinfo  {journal} {Appl. Phys. Lett.}\ }\textbf {\bibinfo
				{volume} {107}},\ \bibinfo {pages} {092404} (\bibinfo {year}
			{2015})}\BibitemShut {NoStop}%
		\bibitem [{\citenamefont {Hine}\ \emph {et~al.}(1979)\citenamefont {Hine},
			\citenamefont {Shinjo},\ and\ \citenamefont {Takada}}]{Hine1979}%
		\BibitemOpen
		\bibfield  {author} {\bibinfo {author} {\bibfnamefont {S.}~\bibnamefont
				{Hine}}, \bibinfo {author} {\bibfnamefont {T.}~\bibnamefont {Shinjo}},\ and\
			\bibinfo {author} {\bibfnamefont {T.}~\bibnamefont {Takada}},\ }\bibfield
		{title} {\bibinfo {title} {{The Surface Magnetism of Thick Fe Films Coated by
					MgO, MgF {\textless}sub{\textgreater}2{\textless}/sub{\textgreater} and
					Sb}},\ }\href {https://doi.org/10.1143/JPSJ.47.767} {\bibfield  {journal}
			{\bibinfo  {journal} {J. Phys. Soc. Jpn.}\ }\textbf {\bibinfo {volume}
				{47}},\ \bibinfo {pages} {767} (\bibinfo {year} {1979})}\BibitemShut
		{NoStop}%
		\bibitem [{\citenamefont {Koyano}\ \emph {et~al.}(1988)\citenamefont {Koyano},
			\citenamefont {Kuroiwa}, \citenamefont {Kita}, \citenamefont {Saegusa},
			\citenamefont {Ohshima},\ and\ \citenamefont {Tasaki}}]{Koyano1988}%
		\BibitemOpen
		\bibfield  {author} {\bibinfo {author} {\bibfnamefont {T.}~\bibnamefont
				{Koyano}}, \bibinfo {author} {\bibfnamefont {Y.}~\bibnamefont {Kuroiwa}},
			\bibinfo {author} {\bibfnamefont {E.}~\bibnamefont {Kita}}, \bibinfo {author}
			{\bibfnamefont {N.}~\bibnamefont {Saegusa}}, \bibinfo {author} {\bibfnamefont
				{K.}~\bibnamefont {Ohshima}},\ and\ \bibinfo {author} {\bibfnamefont
				{A.}~\bibnamefont {Tasaki}},\ }\bibfield  {title} {\bibinfo {title} {{The
					enhanced magnetic moment and structural study of Fe/MgO multilayered
					films}},\ }\href {https://doi.org/10.1063/1.342251} {\bibfield  {journal}
			{\bibinfo  {journal} {J. Appl. Phys.}\ }\textbf {\bibinfo {volume} {64}},\
			\bibinfo {pages} {5763} (\bibinfo {year} {1988})}\BibitemShut {NoStop}%
		\bibitem [{\citenamefont {Sangster}\ \emph {et~al.}(1970)\citenamefont
			{Sangster}, \citenamefont {Peckham},\ and\ \citenamefont
			{Saunderson}}]{Sangster1970}%
		\BibitemOpen
		\bibfield  {author} {\bibinfo {author} {\bibfnamefont {M.~J.~L.}\
				\bibnamefont {Sangster}}, \bibinfo {author} {\bibfnamefont {G.}~\bibnamefont
				{Peckham}},\ and\ \bibinfo {author} {\bibfnamefont {D.~H.}\ \bibnamefont
				{Saunderson}},\ }\bibfield  {title} {\bibinfo {title} {{Lattice dynamics of
					magnesium oxide}},\ }\href {https://doi.org/10.1088/0022-3719/3/5/017}
		{\bibfield  {journal} {\bibinfo  {journal} {J. Phys. C: Solid State Phys.}\
			}\textbf {\bibinfo {volume} {3}},\ \bibinfo {pages} {1026} (\bibinfo {year}
			{1970})}\BibitemShut {NoStop}%
		\bibitem [{\citenamefont {Ghose}\ \emph {et~al.}(2006)\citenamefont {Ghose},
			\citenamefont {Krisch}, \citenamefont {Oganov}, \citenamefont {Beraud},
			\citenamefont {Bosak}, \citenamefont {Gulve}, \citenamefont {Seelaboyina},
			\citenamefont {Yang},\ and\ \citenamefont {Saxena}}]{Ghose2006}%
		\BibitemOpen
		\bibfield  {author} {\bibinfo {author} {\bibfnamefont {S.}~\bibnamefont
				{Ghose}}, \bibinfo {author} {\bibfnamefont {M.}~\bibnamefont {Krisch}},
			\bibinfo {author} {\bibfnamefont {A.~R.}\ \bibnamefont {Oganov}}, \bibinfo
			{author} {\bibfnamefont {A.}~\bibnamefont {Beraud}}, \bibinfo {author}
			{\bibfnamefont {A.}~\bibnamefont {Bosak}}, \bibinfo {author} {\bibfnamefont
				{R.}~\bibnamefont {Gulve}}, \bibinfo {author} {\bibfnamefont
				{R.}~\bibnamefont {Seelaboyina}}, \bibinfo {author} {\bibfnamefont
				{H.}~\bibnamefont {Yang}},\ and\ \bibinfo {author} {\bibfnamefont {S.~K.}\
				\bibnamefont {Saxena}},\ }\bibfield  {title} {\bibinfo {title} {{Lattice
					Dynamics of MgO at High Pressure: Theory and Experiment}},\ }\href
		{https://doi.org/10.1103/PhysRevLett.96.035507} {\bibfield  {journal}
			{\bibinfo  {journal} {Phys. Rev. Lett.}\ }\textbf {\bibinfo {volume} {96}},\
			\bibinfo {pages} {035507} (\bibinfo {year} {2006})}\BibitemShut {NoStop}%
		\bibitem [{\citenamefont {Parlinski}\ \emph {et~al.}(2000)\citenamefont
			{Parlinski}, \citenamefont {{\L}a{\.{z}}ewski},\ and\ \citenamefont
			{Kawazoe}}]{Parlinski2000}%
		\BibitemOpen
		\bibfield  {author} {\bibinfo {author} {\bibfnamefont {K.}~\bibnamefont
				{Parlinski}}, \bibinfo {author} {\bibfnamefont {J.}~\bibnamefont
				{{\L}a{\.{z}}ewski}},\ and\ \bibinfo {author} {\bibfnamefont
				{Y.}~\bibnamefont {Kawazoe}},\ }\bibfield  {title} {\bibinfo {title} {{Ab
					initio studies of phonons in MgO by the direct method including LO mode}},\
		}\href {https://doi.org/10.1016/S0022-3697(99)00226-7} {\bibfield  {journal}
			{\bibinfo  {journal} {J. Phys. Chem. Solids}\ }\textbf {\bibinfo {volume}
				{61}},\ \bibinfo {pages} {87} (\bibinfo {year} {2000})}\BibitemShut {NoStop}%
		\bibitem [{\citenamefont {Schober}\ and\ \citenamefont
			{Dederichs}(1981)}]{Hellwege1981}%
		\BibitemOpen
		\bibfield  {author} {\bibinfo {author} {\bibfnamefont {H.~R.}\ \bibnamefont
				{Schober}}\ and\ \bibinfo {author} {\bibfnamefont {P.~H.}\ \bibnamefont
				{Dederichs}},\ }\href {https://doi.org/10.1007/b19988} {}edited by\ \bibinfo
		{editor} {\bibfnamefont {K.-H.}\ \bibnamefont {Hellwege}}\ and\ \bibinfo
		{editor} {\bibfnamefont {J.~L.}\ \bibnamefont {Olsen}}\ (\bibinfo
		{publisher} {Springer-Verlag},\ \bibinfo {year} {1981})\BibitemShut {NoStop}%
		\bibitem [{\citenamefont {Neuhaus}\ \emph {et~al.}(1997)\citenamefont
			{Neuhaus}, \citenamefont {Petry},\ and\ \citenamefont
			{Krimmel}}]{cn:Neuhaus97}%
		\BibitemOpen
		\bibfield  {author} {\bibinfo {author} {\bibfnamefont {J.}~\bibnamefont
				{Neuhaus}}, \bibinfo {author} {\bibfnamefont {W.}~\bibnamefont {Petry}},\
			and\ \bibinfo {author} {\bibfnamefont {A.}~\bibnamefont {Krimmel}},\
		}\bibfield  {title} {\bibinfo {title} {Phonon softening and martensitic
				transformation in $\alpha$-fe},\ }\href
		{https://doi.org/https://doi.org/10.1016/S0921-4526(96)01185-4} {\bibfield
			{journal} {\bibinfo  {journal} {Physica B}\ }\textbf {\bibinfo {volume}
				{234-236}},\ \bibinfo {pages} {897 } (\bibinfo {year} {1997})},\ \bibinfo
		{note} {proceedings of the First European Conference on Neutron
			Scattering}\BibitemShut {NoStop}%
		\bibitem [{\citenamefont {K\"ormann}\ \emph {et~al.}(2008)\citenamefont
			{K\"ormann}, \citenamefont {Dick}, \citenamefont {Grabowski}, \citenamefont
			{Hallstedt}, \citenamefont {Hickel},\ and\ \citenamefont
			{Neugebauer}}]{cn:Koermann08}%
		\BibitemOpen
		\bibfield  {author} {\bibinfo {author} {\bibfnamefont {F.}~\bibnamefont
				{K\"ormann}}, \bibinfo {author} {\bibfnamefont {A.}~\bibnamefont {Dick}},
			\bibinfo {author} {\bibfnamefont {B.}~\bibnamefont {Grabowski}}, \bibinfo
			{author} {\bibfnamefont {B.}~\bibnamefont {Hallstedt}}, \bibinfo {author}
			{\bibfnamefont {T.}~\bibnamefont {Hickel}},\ and\ \bibinfo {author}
			{\bibfnamefont {J.}~\bibnamefont {Neugebauer}},\ }\bibfield  {title}
		{\bibinfo {title} {Free energy of bcc iron: Integrated ab initio derivation
				of vibrational, electronic, and magnetic contributions},\ }\href
		{https://doi.org/10.1103/PhysRevB.78.033102} {\bibfield  {journal} {\bibinfo
				{journal} {Phys. Rev. B}\ }\textbf {\bibinfo {volume} {78}},\ \bibinfo
			{pages} {033102} (\bibinfo {year} {2008})}\BibitemShut {NoStop}%
		\bibitem [{\citenamefont {Leonov}\ \emph {et~al.}(2012)\citenamefont {Leonov},
			\citenamefont {Poteryaev}, \citenamefont {Anisimov},\ and\ \citenamefont
			{Vollhardt}}]{cn:Leonov12}%
		\BibitemOpen
		\bibfield  {author} {\bibinfo {author} {\bibfnamefont {I.}~\bibnamefont
				{Leonov}}, \bibinfo {author} {\bibfnamefont {A.~I.}\ \bibnamefont
				{Poteryaev}}, \bibinfo {author} {\bibfnamefont {V.~I.}\ \bibnamefont
				{Anisimov}},\ and\ \bibinfo {author} {\bibfnamefont {D.}~\bibnamefont
				{Vollhardt}},\ }\bibfield  {title} {\bibinfo {title} {Calculated phonon
				spectra of paramagnetic iron at the
				$\ensuremath{\alpha}$-$\ensuremath{\gamma}$ phase transition},\ }\href
		{https://doi.org/10.1103/PhysRevB.85.020401} {\bibfield  {journal} {\bibinfo
				{journal} {Phys. Rev. B}\ }\textbf {\bibinfo {volume} {85}},\ \bibinfo
			{pages} {020401} (\bibinfo {year} {2012})}\BibitemShut {NoStop}%
		\bibitem [{\citenamefont {Neuhaus}\ \emph {et~al.}(2014)\citenamefont
			{Neuhaus}, \citenamefont {Leitner}, \citenamefont {Nicolaus}, \citenamefont
			{Petry}, \citenamefont {Hennion},\ and\ \citenamefont
			{Hiess}}]{cn:Neuhaus14}%
		\BibitemOpen
		\bibfield  {author} {\bibinfo {author} {\bibfnamefont {J.}~\bibnamefont
				{Neuhaus}}, \bibinfo {author} {\bibfnamefont {M.}~\bibnamefont {Leitner}},
			\bibinfo {author} {\bibfnamefont {K.}~\bibnamefont {Nicolaus}}, \bibinfo
			{author} {\bibfnamefont {W.}~\bibnamefont {Petry}}, \bibinfo {author}
			{\bibfnamefont {B.}~\bibnamefont {Hennion}},\ and\ \bibinfo {author}
			{\bibfnamefont {A.}~\bibnamefont {Hiess}},\ }\bibfield  {title} {\bibinfo
			{title} {Role of vibrational entropy in the stabilization of the
				high-temperature phases of iron},\ }\href
		{https://doi.org/10.1103/PhysRevB.89.184302} {\bibfield  {journal} {\bibinfo
				{journal} {Phys. Rev. B}\ }\textbf {\bibinfo {volume} {89}},\ \bibinfo
			{pages} {184302} (\bibinfo {year} {2014})}\BibitemShut {NoStop}%
		\bibitem [{\citenamefont {{\'{S}}l{\k{e}}zak}\ \emph
			{et~al.}(2007)\citenamefont {{\'{S}}l{\k{e}}zak}, \citenamefont
			{{\L}a{\.{z}}ewski}, \citenamefont {Stankov}, \citenamefont {Parlinski},
			\citenamefont {Reitinger}, \citenamefont {Rennhofer}, \citenamefont
			{R\"{u}ffer}, \citenamefont {Sepiol}, \citenamefont {{\'{S}}l{\k{e}}zak},
			\citenamefont {Spiridis}, \citenamefont {Zaj{\k{a}}c}, \citenamefont
			{Chumakov},\ and\ \citenamefont {Korecki}}]{Slzak2007}%
		\BibitemOpen
		\bibfield  {author} {\bibinfo {author} {\bibfnamefont {T.}~\bibnamefont
				{{\'{S}}l{\k{e}}zak}}, \bibinfo {author} {\bibfnamefont {J.}~\bibnamefont
				{{\L}a{\.{z}}ewski}}, \bibinfo {author} {\bibfnamefont {S.}~\bibnamefont
				{Stankov}}, \bibinfo {author} {\bibfnamefont {K.}~\bibnamefont {Parlinski}},
			\bibinfo {author} {\bibfnamefont {R.}~\bibnamefont {Reitinger}}, \bibinfo
			{author} {\bibfnamefont {M.}~\bibnamefont {Rennhofer}}, \bibinfo {author}
			{\bibfnamefont {R.}~\bibnamefont {R\"{u}ffer}}, \bibinfo {author}
			{\bibfnamefont {B.}~\bibnamefont {Sepiol}}, \bibinfo {author} {\bibfnamefont
				{M.}~\bibnamefont {{\'{S}}l{\k{e}}zak}}, \bibinfo {author} {\bibfnamefont
				{N.}~\bibnamefont {Spiridis}}, \bibinfo {author} {\bibfnamefont
				{M.}~\bibnamefont {Zaj{\k{a}}c}}, \bibinfo {author} {\bibfnamefont {A.~I.}\
				\bibnamefont {Chumakov}},\ and\ \bibinfo {author} {\bibfnamefont
				{J.}~\bibnamefont {Korecki}},\ }\bibfield  {title} {\bibinfo {title} {Phonons
				at the fe(110) surface},\ }\bibfield  {journal} {\bibinfo  {journal} {Phys.
				Rev. Lett.}\ }\textbf {\bibinfo {volume} {99}},\ \href
		{https://doi.org/10.1103/physrevlett.99.066103}
		{10.1103/physrevlett.99.066103} (\bibinfo {year} {2007})\BibitemShut
		{NoStop}%
		\bibitem [{\citenamefont {{Roldan Cuenya}}\ \emph {et~al.}(2008)\citenamefont
			{{Roldan Cuenya}}, \citenamefont {Keune}, \citenamefont {Peters},
			\citenamefont {Schuster}, \citenamefont {Sahoo}, \citenamefont {von
				H{\"{o}}rsten}, \citenamefont {Sturhahn}, \citenamefont {Zhao}, \citenamefont
			{Toellner}, \citenamefont {Alp},\ and\ \citenamefont
			{Bader}}]{RoldanCuenya2008}%
		\BibitemOpen
		\bibfield  {author} {\bibinfo {author} {\bibfnamefont {B.}~\bibnamefont
				{{Roldan Cuenya}}}, \bibinfo {author} {\bibfnamefont {W.}~\bibnamefont
				{Keune}}, \bibinfo {author} {\bibfnamefont {R.}~\bibnamefont {Peters}},
			\bibinfo {author} {\bibfnamefont {E.}~\bibnamefont {Schuster}}, \bibinfo
			{author} {\bibfnamefont {B.}~\bibnamefont {Sahoo}}, \bibinfo {author}
			{\bibfnamefont {U.}~\bibnamefont {von H{\"{o}}rsten}}, \bibinfo {author}
			{\bibfnamefont {W.}~\bibnamefont {Sturhahn}}, \bibinfo {author}
			{\bibfnamefont {J.}~\bibnamefont {Zhao}}, \bibinfo {author} {\bibfnamefont
				{T.~S.}\ \bibnamefont {Toellner}}, \bibinfo {author} {\bibfnamefont {E.~E.}\
				\bibnamefont {Alp}},\ and\ \bibinfo {author} {\bibfnamefont {S.~D.}\
				\bibnamefont {Bader}},\ }\bibfield  {title} {\bibinfo {title} {{High-energy
					phonon confinement in nanoscale metallic multilayers}},\ }\href
		{https://doi.org/10.1103/PhysRevB.77.165410} {\bibfield  {journal} {\bibinfo
				{journal} {Phys. Rev. B}\ }\textbf {\bibinfo {volume} {77}},\ \bibinfo
			{pages} {165410} (\bibinfo {year} {2008})}\BibitemShut {NoStop}%
		\bibitem [{\citenamefont {Keune}\ \emph {et~al.}(2018)\citenamefont {Keune},
			\citenamefont {Hong}, \citenamefont {Hu}, \citenamefont {Zhao}, \citenamefont
			{Toellner}, \citenamefont {Alp}, \citenamefont {Sturhahn}, \citenamefont
			{Rahman},\ and\ \citenamefont {{Roldan Cuenya}}}]{Keune2018}%
		\BibitemOpen
		\bibfield  {author} {\bibinfo {author} {\bibfnamefont {W.}~\bibnamefont
				{Keune}}, \bibinfo {author} {\bibfnamefont {S.}~\bibnamefont {Hong}},
			\bibinfo {author} {\bibfnamefont {M.~Y.}\ \bibnamefont {Hu}}, \bibinfo
			{author} {\bibfnamefont {J.}~\bibnamefont {Zhao}}, \bibinfo {author}
			{\bibfnamefont {T.~S.}\ \bibnamefont {Toellner}}, \bibinfo {author}
			{\bibfnamefont {E.~E.}\ \bibnamefont {Alp}}, \bibinfo {author} {\bibfnamefont
				{W.}~\bibnamefont {Sturhahn}}, \bibinfo {author} {\bibfnamefont {T.~S.}\
				\bibnamefont {Rahman}},\ and\ \bibinfo {author} {\bibfnamefont
				{B.}~\bibnamefont {{Roldan Cuenya}}},\ }\bibfield  {title} {\bibinfo {title}
			{{Influence of interfaces on the phonon density of states of nanoscale
					metallic multilayers: Phonon confinement and localization}},\ }\href
		{https://doi.org/10.1103/PhysRevB.98.024308} {\bibfield  {journal} {\bibinfo
				{journal} {Phys. Rev. B}\ }\textbf {\bibinfo {volume} {98}},\ \bibinfo
			{pages} {024308} (\bibinfo {year} {2018})}\BibitemShut {NoStop}%
		\bibitem [{\citenamefont {Sternik}\ \emph {et~al.}(2006)\citenamefont
			{Sternik}, \citenamefont {Parlinski},\ and\ \citenamefont
			{Korecki}}]{Sternik2006}%
		\BibitemOpen
		\bibfield  {author} {\bibinfo {author} {\bibfnamefont {M.}~\bibnamefont
				{Sternik}}, \bibinfo {author} {\bibfnamefont {K.}~\bibnamefont {Parlinski}},\
			and\ \bibinfo {author} {\bibfnamefont {J.}~\bibnamefont {Korecki}},\
		}\bibfield  {title} {\bibinfo {title} {Fe m ∕ Au n multilayers from first
					principles},\ }\href {https://doi.org/10.1103/PhysRevB.74.195405} {\bibfield
			{journal} {\bibinfo  {journal} {Phys. Rev. B}\ }\textbf {\bibinfo {volume}
				{74}},\ \bibinfo {pages} {195405} (\bibinfo {year} {2006})}\BibitemShut
		{NoStop}%
		\bibitem [{\citenamefont {Rothenbach}\ \emph {et~al.}(2019)\citenamefont
			{Rothenbach}, \citenamefont {Gruner}, \citenamefont {Ollefs}, \citenamefont
			{Schmitz-Antoniak}, \citenamefont {Salamon}, \citenamefont {Zhou},
			\citenamefont {Li}, \citenamefont {Mo}, \citenamefont {Park}, \citenamefont
			{Shen}, \citenamefont {Weathersby}, \citenamefont {Yang}, \citenamefont
			{Wang}, \citenamefont {Pentcheva}, \citenamefont {Wende}, \citenamefont
			{Bovensiepen}, \citenamefont {Sokolowski-Tinten},\ and\ \citenamefont
			{Eschenlohr}}]{cn:Rothenbach19}%
		\BibitemOpen
		\bibfield  {author} {\bibinfo {author} {\bibfnamefont {N.}~\bibnamefont
				{Rothenbach}}, \bibinfo {author} {\bibfnamefont {M.~E.}\ \bibnamefont
				{Gruner}}, \bibinfo {author} {\bibfnamefont {K.}~\bibnamefont {Ollefs}},
			\bibinfo {author} {\bibfnamefont {C.}~\bibnamefont {Schmitz-Antoniak}},
			\bibinfo {author} {\bibfnamefont {S.}~\bibnamefont {Salamon}}, \bibinfo
			{author} {\bibfnamefont {P.}~\bibnamefont {Zhou}}, \bibinfo {author}
			{\bibfnamefont {R.}~\bibnamefont {Li}}, \bibinfo {author} {\bibfnamefont
				{M.}~\bibnamefont {Mo}}, \bibinfo {author} {\bibfnamefont {S.}~\bibnamefont
				{Park}}, \bibinfo {author} {\bibfnamefont {X.}~\bibnamefont {Shen}}, \bibinfo
			{author} {\bibfnamefont {S.}~\bibnamefont {Weathersby}}, \bibinfo {author}
			{\bibfnamefont {J.}~\bibnamefont {Yang}}, \bibinfo {author} {\bibfnamefont
				{X.~J.}\ \bibnamefont {Wang}}, \bibinfo {author} {\bibfnamefont
				{R.}~\bibnamefont {Pentcheva}}, \bibinfo {author} {\bibfnamefont
				{H.}~\bibnamefont {Wende}}, \bibinfo {author} {\bibfnamefont
				{U.}~\bibnamefont {Bovensiepen}}, \bibinfo {author} {\bibfnamefont
				{K.}~\bibnamefont {Sokolowski-Tinten}},\ and\ \bibinfo {author}
			{\bibfnamefont {A.}~\bibnamefont {Eschenlohr}},\ }\bibfield  {title}
		{\bibinfo {title} {Microscopic nonequilibrium energy transfer dynamics in a
				photoexcited metal/insulator heterostructure},\ }\bibfield  {journal}
		{\bibinfo  {journal} {Phys. Rev. B}\ }\textbf {\bibinfo {volume} {100}},\
		\href {https://doi.org/10.1103/physrevb.100.174301}
		{10.1103/physrevb.100.174301} (\bibinfo {year} {2019})\BibitemShut {NoStop}%
		\bibitem [{\citenamefont {Gruner}\ and\ \citenamefont
			{Pentcheva}(2019)}]{cn:GrunerRTTDDFT}%
		\BibitemOpen
		\bibfield  {author} {\bibinfo {author} {\bibfnamefont {M.~E.}\ \bibnamefont
				{Gruner}}\ and\ \bibinfo {author} {\bibfnamefont {R.}~\bibnamefont
				{Pentcheva}},\ }\bibfield  {title} {\bibinfo {title} {Dynamics of optical
				excitations in a fe/mgo(001) heterostructure from time-dependent density
				functional theory},\ }\href@noop {} {\bibfield  {journal} {\bibinfo
				{journal} {Phys. Rev. B}\ }\textbf {\bibinfo {volume} {99}} (\bibinfo {year}
			{2019})}\BibitemShut {NoStop}%
		\bibitem [{Sup()}]{Supplement}%
		\BibitemOpen
		\href@noop {} {\bibinfo {title} {See {S}upplemental {M}aterial at [{URL} will
				be inserted by publisher] for additional information concerning the sample
				preparation, structural characterisation and {NRIXS} data
				evaluation.}}\BibitemShut {Stop}%
		\bibitem [{\citenamefont {von Hörsten}(2019)}]{PiLink}%
		\BibitemOpen
		\bibfield  {author} {\bibinfo {author} {\bibfnamefont {U.}~\bibnamefont {von
					Hörsten}},\ }\href@noop {} {}\bibinfo {howpublished} {http://udue.de/Pi}
		(\bibinfo {year} {2019})\BibitemShut {NoStop}%
		\bibitem [{\citenamefont {Seto}\ \emph {et~al.}(1995)\citenamefont {Seto},
			\citenamefont {Yoda}, \citenamefont {Kikuta}, \citenamefont {Zhang},\ and\
			\citenamefont {Ando}}]{Seto1995}%
		\BibitemOpen
		\bibfield  {author} {\bibinfo {author} {\bibfnamefont {M.}~\bibnamefont
				{Seto}}, \bibinfo {author} {\bibfnamefont {Y.}~\bibnamefont {Yoda}}, \bibinfo
			{author} {\bibfnamefont {S.}~\bibnamefont {Kikuta}}, \bibinfo {author}
			{\bibfnamefont {X.~W.}\ \bibnamefont {Zhang}},\ and\ \bibinfo {author}
			{\bibfnamefont {M.}~\bibnamefont {Ando}},\ }\bibfield  {title} {\bibinfo
			{title} {Observation of nuclear resonant scattering accompanied by phonon
				excitation using synchrotron radiation},\ }\href
		{https://doi.org/10.1103/physrevlett.74.3828} {\bibfield  {journal} {\bibinfo
				{journal} {Phys. Rev. Lett.}\ }\textbf {\bibinfo {volume} {74}},\ \bibinfo
			{pages} {3828} (\bibinfo {year} {1995})}\BibitemShut {NoStop}%
		\bibitem [{\citenamefont {Sturhahn}\ \emph {et~al.}(1995)\citenamefont
			{Sturhahn}, \citenamefont {Toellner}, \citenamefont {Alp}, \citenamefont
			{Zhang}, \citenamefont {Ando}, \citenamefont {Yoda}, \citenamefont {Kikuta},
			\citenamefont {Seto}, \citenamefont {Kimball},\ and\ \citenamefont
			{Dabrowski}}]{Sturhahn1995}%
		\BibitemOpen
		\bibfield  {author} {\bibinfo {author} {\bibfnamefont {W.}~\bibnamefont
				{Sturhahn}}, \bibinfo {author} {\bibfnamefont {T.~S.}\ \bibnamefont
				{Toellner}}, \bibinfo {author} {\bibfnamefont {E.~E.}\ \bibnamefont {Alp}},
			\bibinfo {author} {\bibfnamefont {X.}~\bibnamefont {Zhang}}, \bibinfo
			{author} {\bibfnamefont {M.}~\bibnamefont {Ando}}, \bibinfo {author}
			{\bibfnamefont {Y.}~\bibnamefont {Yoda}}, \bibinfo {author} {\bibfnamefont
				{S.}~\bibnamefont {Kikuta}}, \bibinfo {author} {\bibfnamefont
				{M.}~\bibnamefont {Seto}}, \bibinfo {author} {\bibfnamefont {C.~W.}\
				\bibnamefont {Kimball}},\ and\ \bibinfo {author} {\bibfnamefont
				{B.}~\bibnamefont {Dabrowski}},\ }\bibfield  {title} {\bibinfo {title}
			{Phonon density of states measured by inelastic nuclear resonant
				scattering},\ }\href {https://doi.org/10.1103/physrevlett.74.3832} {\bibfield
			{journal} {\bibinfo  {journal} {Phys. Rev. Lett.}\ }\textbf {\bibinfo
				{volume} {74}},\ \bibinfo {pages} {3832} (\bibinfo {year}
			{1995})}\BibitemShut {NoStop}%
		\bibitem [{\citenamefont {Chumakov}\ \emph {et~al.}(1995)\citenamefont
			{Chumakov}, \citenamefont {R\"{u}ffer}, \citenamefont {Gr\"{u}nsteudel},
			\citenamefont {Gr\"{u}nsteudel}, \citenamefont {Gr\"{u}bel}, \citenamefont
			{Metge}, \citenamefont {Leupold},\ and\ \citenamefont
			{Goodwin}}]{Chumakov1995}%
		\BibitemOpen
		\bibfield  {author} {\bibinfo {author} {\bibfnamefont {A.~I.}\ \bibnamefont
				{Chumakov}}, \bibinfo {author} {\bibfnamefont {R.}~\bibnamefont
				{R\"{u}ffer}}, \bibinfo {author} {\bibfnamefont {H.}~\bibnamefont
				{Gr\"{u}nsteudel}}, \bibinfo {author} {\bibfnamefont {H.~F.}\ \bibnamefont
				{Gr\"{u}nsteudel}}, \bibinfo {author} {\bibfnamefont {G.}~\bibnamefont
				{Gr\"{u}bel}}, \bibinfo {author} {\bibfnamefont {J.}~\bibnamefont {Metge}},
			\bibinfo {author} {\bibfnamefont {O.}~\bibnamefont {Leupold}},\ and\ \bibinfo
			{author} {\bibfnamefont {H.~A.}\ \bibnamefont {Goodwin}},\ }\bibfield
		{title} {\bibinfo {title} {Energy dependence of nuclear recoil measured with
				incoherent nuclear scattering of synchrotron radiation},\ }\href
		{https://doi.org/10.1209/0295-5075/30/7/009} {\bibfield  {journal} {\bibinfo
				{journal} {Europhys. Lett.}\ }\textbf {\bibinfo {volume} {30}},\ \bibinfo
			{pages} {427} (\bibinfo {year} {1995})}\BibitemShut {NoStop}%
		\bibitem [{\citenamefont {R\"{o}hlsberger}(2005)}]{Rhlsberger2005}%
		\BibitemOpen
		\bibfield  {author} {\bibinfo {author} {\bibfnamefont {R.}~\bibnamefont
				{R\"{o}hlsberger}},\ }\href@noop {} {\emph {\bibinfo {title} {Nuclear
					Condensed Matter Physics with Synchrotron Radiation}}}\ (\bibinfo
		{publisher} {Springer Berlin Heidelberg},\ \bibinfo {year}
		{2005})\BibitemShut {NoStop}%
		\bibitem [{\citenamefont {Chen}\ and\ \citenamefont
			{Yang}(2007)}]{Chen2007LatticeDynamics}%
		\BibitemOpen
		\bibfield  {author} {\bibinfo {author} {\bibfnamefont {Y.-L.}\ \bibnamefont
				{Chen}}\ and\ \bibinfo {author} {\bibfnamefont {D.-P.}\ \bibnamefont
				{Yang}},\ }\href@noop {} {\emph {\bibinfo {title} {Mössbauer Effect in
					Lattice Dynamics: Experimental Techniques and Applications}}}\ (\bibinfo
		{publisher} {Wiley-VCH},\ \bibinfo {year} {2007})\BibitemShut {NoStop}%
		\bibitem [{\citenamefont {Sturhahn}(2004)}]{Sturhahn2004}%
		\BibitemOpen
		\bibfield  {author} {\bibinfo {author} {\bibfnamefont {W.}~\bibnamefont
				{Sturhahn}},\ }\bibfield  {title} {\bibinfo {title} {Nuclear resonant
				spectroscopy},\ }\href {https://doi.org/10.1088/0953-8984/16/5/009}
		{\bibfield  {journal} {\bibinfo  {journal} {J. Phys.: Condens. Matter}\
			}\textbf {\bibinfo {volume} {16}},\ \bibinfo {pages} {S497} (\bibinfo {year}
			{2004})}\BibitemShut {NoStop}%
		\bibitem [{PHO()}]{PHOENIX}%
		\BibitemOpen
		\href {http://www.NRIXS.com} {\bibinfo {title} {Download at
				www.nrixs.com.}}\BibitemShut {Stop}%
		\bibitem [{\citenamefont {Sturhahn}(2000)}]{Sturhahn2000}%
		\BibitemOpen
		\bibfield  {author} {\bibinfo {author} {\bibfnamefont {W.}~\bibnamefont
				{Sturhahn}},\ }\bibfield  {title} {\bibinfo {title} {{CONUSS and PHOENIX:
					Evaluation of nuclear resonant scattering data}},\ }\href
		{https://doi.org/10.1023/A:1012681503686} {\bibfield  {journal} {\bibinfo
				{journal} {Hyperfine Interact.}\ }\textbf {\bibinfo {volume} {125}},\
			\bibinfo {pages} {149} (\bibinfo {year} {2000})}\BibitemShut {NoStop}%
		\bibitem [{\citenamefont {Kresse}\ and\ \citenamefont
			{Furthm\"uller}(1996)}]{cn:VASP1}%
		\BibitemOpen
		\bibfield  {author} {\bibinfo {author} {\bibfnamefont {G.}~\bibnamefont
				{Kresse}}\ and\ \bibinfo {author} {\bibfnamefont {J.}~\bibnamefont
				{Furthm\"uller}},\ }\bibfield  {title} {\bibinfo {title} {Efficient iterative
				schemes for ab initio total-energy calculations using a plane-wave basis
				set},\ }\href@noop {} {\bibfield  {journal} {\bibinfo  {journal} {Phys. Rev.
					B}\ }\textbf {\bibinfo {volume} {54}},\ \bibinfo {pages} {11169} (\bibinfo
			{year} {1996})}\BibitemShut {NoStop}%
		\bibitem [{\citenamefont {Kresse}\ and\ \citenamefont
			{Joubert}(1999)}]{cn:VASP2}%
		\BibitemOpen
		\bibfield  {author} {\bibinfo {author} {\bibfnamefont {G.}~\bibnamefont
				{Kresse}}\ and\ \bibinfo {author} {\bibfnamefont {D.}~\bibnamefont
				{Joubert}},\ }\bibfield  {title} {\bibinfo {title} {From ultrasoft
				pseudopotentials to the projector augmented-wave method},\ }\href@noop {}
		{\bibfield  {journal} {\bibinfo  {journal} {Phys. Rev. B}\ }\textbf {\bibinfo
				{volume} {59}},\ \bibinfo {pages} {1758} (\bibinfo {year}
			{1999})}\BibitemShut {NoStop}%
		\bibitem [{\citenamefont {Perdew}\ \emph {et~al.}(1996)\citenamefont {Perdew},
			\citenamefont {Burke},\ and\ \citenamefont {Ernzerhof}}]{cn:Perdew96}%
		\BibitemOpen
		\bibfield  {author} {\bibinfo {author} {\bibfnamefont {J.~P.}\ \bibnamefont
				{Perdew}}, \bibinfo {author} {\bibfnamefont {K.}~\bibnamefont {Burke}},\ and\
			\bibinfo {author} {\bibfnamefont {M.}~\bibnamefont {Ernzerhof}},\ }\bibfield
		{title} {\bibinfo {title} {Generalized gradient approximation made simple},\
		}\href@noop {} {\bibfield  {journal} {\bibinfo  {journal} {Phys. Rev. Lett.}\
			}\textbf {\bibinfo {volume} {77}},\ \bibinfo {pages} {3865} (\bibinfo {year}
			{1996})}\BibitemShut {NoStop}%
		\bibitem [{\citenamefont {Bl\"ochl}\ \emph {et~al.}(1994)\citenamefont
			{Bl\"ochl}, \citenamefont {Jepsen},\ and\ \citenamefont
			{Andersen}}]{cn:Bloechl94}%
		\BibitemOpen
		\bibfield  {author} {\bibinfo {author} {\bibfnamefont {P.~E.}\ \bibnamefont
				{Bl\"ochl}}, \bibinfo {author} {\bibfnamefont {O.}~\bibnamefont {Jepsen}},\
			and\ \bibinfo {author} {\bibfnamefont {O.~K.}\ \bibnamefont {Andersen}},\
		}\bibfield  {title} {\bibinfo {title} {Improved tetrahedron method for
				brillouin-zone integrations},\ }\href
		{https://doi.org/10.1103/PhysRevB.49.16223} {\bibfield  {journal} {\bibinfo
				{journal} {Phys. Rev. B}\ }\textbf {\bibinfo {volume} {49}},\ \bibinfo
			{pages} {16223} (\bibinfo {year} {1994})}\BibitemShut {NoStop}%
		\bibitem [{\citenamefont {Alf\`e}(2009)}]{cn:Alfe09PHON}%
		\BibitemOpen
		\bibfield  {author} {\bibinfo {author} {\bibfnamefont {D.}~\bibnamefont
				{Alf\`e}},\ }\bibfield  {title} {\bibinfo {title} {Phon: A program to
				calculate phonons using the small displacement method},\ }\href@noop {}
		{\bibfield  {journal} {\bibinfo  {journal} {Comp. Phys. Commun.}\ }\textbf
			{\bibinfo {volume} {180}},\ \bibinfo {pages} {2622} (\bibinfo {year}
			{2009})}\BibitemShut {NoStop}%
		\bibitem [{\citenamefont {Dewhurst}\ \emph {et~al.}(2019)\citenamefont
			{Dewhurst}, \citenamefont {Sharma}, \citenamefont {Nordstr\"om},
			\citenamefont {Cricchio}, \citenamefont {Gran\"as}, \citenamefont {Gross},
			\citenamefont {Ambrosch-Draxl}, \citenamefont {Persson}, \citenamefont
			{Bultmark}, \citenamefont {Brouder}, \citenamefont {Armiento}, \citenamefont
			{Chizmeshya}, \citenamefont {Anderson}, \citenamefont {Nekrasov},
			\citenamefont {Wagner}, \citenamefont {Kalarasse}, \citenamefont {Spitaler},
			\citenamefont {Pittalis}, \citenamefont {Lathiotakis}, \citenamefont
			{Burnus}, \citenamefont {Sagmeister}, \citenamefont {Meisenbichler},
			\citenamefont {Leb\`egue}, \citenamefont {Zhang}, \citenamefont {K\"ormann},
			\citenamefont {Baranov}, \citenamefont {Kozhevnikov}, \citenamefont
			{Suehara}, \citenamefont {Essenberger}, \citenamefont {Sanna}, \citenamefont
			{McQueen}, \citenamefont {Baldsiefen}, \citenamefont {Blaber}, \citenamefont
			{Filanovich}, \citenamefont {Bj\"orkman}, \citenamefont {Stankovski},
			\citenamefont {Goraus}, \citenamefont {Meinert}, \citenamefont {Rohr},
			\citenamefont {Nazarov}, \citenamefont {Krieger}, \citenamefont {Floyd},
			\citenamefont {Davydov}, \citenamefont {Eich}, \citenamefont {Castro},
			\citenamefont {Kitahara}, \citenamefont {Glasbrenner}, \citenamefont
			{Bussmann}, \citenamefont {Mazin}, \citenamefont {Verstraete}, \citenamefont
			{Ernsting}, \citenamefont {Dugdale}, \citenamefont {Elliott}, \citenamefont
			{Dulak}, \citenamefont {Livas}, \citenamefont {Cottenier}, \citenamefont
			{Shinohara}, \citenamefont {Fechner}, \citenamefont {Kvashnin}, \citenamefont
			{M\"uller}, \citenamefont {Gerasimov}, \citenamefont {Le}, \citenamefont
			{Bartolom\'e},\ and\ \citenamefont {Wirnata}}]{cn:Elk628}%
		\BibitemOpen
		\bibfield  {author} {\bibinfo {author} {\bibfnamefont {K.}~\bibnamefont
				{Dewhurst}}, \bibinfo {author} {\bibfnamefont {S.}~\bibnamefont {Sharma}},
			\bibinfo {author} {\bibfnamefont {L.}~\bibnamefont {Nordstr\"om}}, \bibinfo
			{author} {\bibfnamefont {F.}~\bibnamefont {Cricchio}}, \bibinfo {author}
			{\bibfnamefont {O.}~\bibnamefont {Gran\"as}}, \bibinfo {author}
			{\bibfnamefont {E.~K.~U.}\ \bibnamefont {Gross}}, \bibinfo {author}
			{\bibfnamefont {C.}~\bibnamefont {Ambrosch-Draxl}}, \bibinfo {author}
			{\bibfnamefont {C.}~\bibnamefont {Persson}}, \bibinfo {author} {\bibfnamefont
				{F.}~\bibnamefont {Bultmark}}, \bibinfo {author} {\bibfnamefont
				{C.}~\bibnamefont {Brouder}}, \bibinfo {author} {\bibfnamefont
				{R.}~\bibnamefont {Armiento}}, \bibinfo {author} {\bibfnamefont
				{A.}~\bibnamefont {Chizmeshya}}, \bibinfo {author} {\bibfnamefont
				{P.}~\bibnamefont {Anderson}}, \bibinfo {author} {\bibfnamefont
				{I.}~\bibnamefont {Nekrasov}}, \bibinfo {author} {\bibfnamefont
				{F.}~\bibnamefont {Wagner}}, \bibinfo {author} {\bibfnamefont
				{F.}~\bibnamefont {Kalarasse}}, \bibinfo {author} {\bibfnamefont
				{J.}~\bibnamefont {Spitaler}}, \bibinfo {author} {\bibfnamefont
				{S.}~\bibnamefont {Pittalis}}, \bibinfo {author} {\bibfnamefont
				{N.}~\bibnamefont {Lathiotakis}}, \bibinfo {author} {\bibfnamefont
				{T.}~\bibnamefont {Burnus}}, \bibinfo {author} {\bibfnamefont
				{S.}~\bibnamefont {Sagmeister}}, \bibinfo {author} {\bibfnamefont
				{C.}~\bibnamefont {Meisenbichler}}, \bibinfo {author} {\bibfnamefont
				{S.}~\bibnamefont {Leb\`egue}}, \bibinfo {author} {\bibfnamefont
				{Y.}~\bibnamefont {Zhang}}, \bibinfo {author} {\bibfnamefont
				{F.}~\bibnamefont {K\"ormann}}, \bibinfo {author} {\bibfnamefont
				{A.}~\bibnamefont {Baranov}}, \bibinfo {author} {\bibfnamefont
				{A.}~\bibnamefont {Kozhevnikov}}, \bibinfo {author} {\bibfnamefont
				{S.}~\bibnamefont {Suehara}}, \bibinfo {author} {\bibfnamefont
				{F.}~\bibnamefont {Essenberger}}, \bibinfo {author} {\bibfnamefont
				{A.}~\bibnamefont {Sanna}}, \bibinfo {author} {\bibfnamefont
				{T.}~\bibnamefont {McQueen}}, \bibinfo {author} {\bibfnamefont
				{T.}~\bibnamefont {Baldsiefen}}, \bibinfo {author} {\bibfnamefont
				{M.}~\bibnamefont {Blaber}}, \bibinfo {author} {\bibfnamefont
				{A.}~\bibnamefont {Filanovich}}, \bibinfo {author} {\bibfnamefont
				{T.}~\bibnamefont {Bj\"orkman}}, \bibinfo {author} {\bibfnamefont
				{M.}~\bibnamefont {Stankovski}}, \bibinfo {author} {\bibfnamefont
				{J.}~\bibnamefont {Goraus}}, \bibinfo {author} {\bibfnamefont
				{M.}~\bibnamefont {Meinert}}, \bibinfo {author} {\bibfnamefont
				{D.}~\bibnamefont {Rohr}}, \bibinfo {author} {\bibfnamefont {V.}~\bibnamefont
				{Nazarov}}, \bibinfo {author} {\bibfnamefont {K.}~\bibnamefont {Krieger}},
			\bibinfo {author} {\bibfnamefont {P.}~\bibnamefont {Floyd}}, \bibinfo
			{author} {\bibfnamefont {A.}~\bibnamefont {Davydov}}, \bibinfo {author}
			{\bibfnamefont {F.}~\bibnamefont {Eich}}, \bibinfo {author} {\bibfnamefont
				{A.~R.}\ \bibnamefont {Castro}}, \bibinfo {author} {\bibfnamefont
				{K.}~\bibnamefont {Kitahara}}, \bibinfo {author} {\bibfnamefont
				{J.}~\bibnamefont {Glasbrenner}}, \bibinfo {author} {\bibfnamefont
				{K.}~\bibnamefont {Bussmann}}, \bibinfo {author} {\bibfnamefont
				{I.}~\bibnamefont {Mazin}}, \bibinfo {author} {\bibfnamefont
				{M.}~\bibnamefont {Verstraete}}, \bibinfo {author} {\bibfnamefont
				{D.}~\bibnamefont {Ernsting}}, \bibinfo {author} {\bibfnamefont
				{S.}~\bibnamefont {Dugdale}}, \bibinfo {author} {\bibfnamefont
				{P.}~\bibnamefont {Elliott}}, \bibinfo {author} {\bibfnamefont
				{M.}~\bibnamefont {Dulak}}, \bibinfo {author} {\bibfnamefont {J.~A.~F.}\
				\bibnamefont {Livas}}, \bibinfo {author} {\bibfnamefont {S.}~\bibnamefont
				{Cottenier}}, \bibinfo {author} {\bibfnamefont {Y.}~\bibnamefont
				{Shinohara}}, \bibinfo {author} {\bibfnamefont {M.}~\bibnamefont {Fechner}},
			\bibinfo {author} {\bibfnamefont {Y.}~\bibnamefont {Kvashnin}}, \bibinfo
			{author} {\bibfnamefont {T.}~\bibnamefont {M\"uller}}, \bibinfo {author}
			{\bibfnamefont {A.}~\bibnamefont {Gerasimov}}, \bibinfo {author}
			{\bibfnamefont {M.~D.}\ \bibnamefont {Le}}, \bibinfo {author} {\bibfnamefont
				{J.~L.}\ \bibnamefont {Bartolom\'e}},\ and\ \bibinfo {author} {\bibfnamefont
				{R.}~\bibnamefont {Wirnata}},\ }\href@noop {} {\bibinfo {title} {The elk
				code, version 6.2.8}},\ \bibinfo {howpublished} {http://elk.sourceforge.net/}
		(\bibinfo {year} {2019})\BibitemShut {NoStop}%
		\bibitem [{\citenamefont {Gütlich}\ \emph {et~al.}(2011)\citenamefont
			{Gütlich}, \citenamefont {Bill},\ and\ \citenamefont
			{Trautwein}}]{G_tlich_2011}%
		\BibitemOpen
		\bibfield  {author} {\bibinfo {author} {\bibfnamefont {P.}~\bibnamefont
				{Gütlich}}, \bibinfo {author} {\bibfnamefont {E.}~\bibnamefont {Bill}},\
			and\ \bibinfo {author} {\bibfnamefont {A.~X.}\ \bibnamefont {Trautwein}},\
		}\href@noop {} {\emph {\bibinfo {title} {Mössbauer Spectroscopy and
					Transition Metal Chemistry}}}\ (\bibinfo  {publisher} {Springer Berlin
			Heidelberg},\ \bibinfo {year} {2011})\BibitemShut {NoStop}%
		\bibitem [{\citenamefont {Vincze}\ \emph {et~al.}(1994)\citenamefont {Vincze},
			\citenamefont {Kaptás}, \citenamefont {Kemény}, \citenamefont {Kiss},\ and\
			\citenamefont {Balogh}}]{Vincze_Kaptas_Kemeny_Kiss_Balogh_1994}%
		\BibitemOpen
		\bibfield  {author} {\bibinfo {author} {\bibfnamefont {I.}~\bibnamefont
				{Vincze}}, \bibinfo {author} {\bibfnamefont {D.}~\bibnamefont {Kaptás}},
			\bibinfo {author} {\bibfnamefont {T.}~\bibnamefont {Kemény}}, \bibinfo
			{author} {\bibfnamefont {L.~F.}\ \bibnamefont {Kiss}},\ and\ \bibinfo
			{author} {\bibfnamefont {J.}~\bibnamefont {Balogh}},\ }\bibfield  {title}
		{\bibinfo {title} {Field induced magnetic moments in amorphous fe-zr
				spin-glass-like alloys},\ }\href {https://doi.org/10.1103/physrevlett.73.496}
		{\bibfield  {journal} {\bibinfo  {journal} {Phys. Rev. Lett.}\ }\textbf
			{\bibinfo {volume} {73}},\ \bibinfo {pages} {496–499} (\bibinfo {year}
			{1994})}\BibitemShut {NoStop}%
		\bibitem [{\citenamefont {Greenwood}\ and\ \citenamefont
			{Gibb}(1971)}]{Greenwood_1971}%
		\BibitemOpen
		\bibfield  {author} {\bibinfo {author} {\bibfnamefont {N.~N.}\ \bibnamefont
				{Greenwood}}\ and\ \bibinfo {author} {\bibfnamefont {T.~C.}\ \bibnamefont
				{Gibb}},\ }\href@noop {} {\emph {\bibinfo {title} {Mössbauer
					Spectroscopy}}}\ (\bibinfo  {publisher} {Springer Netherlands},\ \bibinfo
		{year} {1971})\BibitemShut {NoStop}%
		\bibitem [{\citenamefont {McCammon}\ and\ \citenamefont
			{Price}(1985)}]{McCammon1985}%
		\BibitemOpen
		\bibfield  {author} {\bibinfo {author} {\bibfnamefont {C.~A.}\ \bibnamefont
				{McCammon}}\ and\ \bibinfo {author} {\bibfnamefont {D.~C.}\ \bibnamefont
				{Price}},\ }\bibfield  {title} {\bibinfo {title} {M\"ossbauer spectra of fe x
				o (x$>$0.95)},\ }\href {https://doi.org/10.1007/bf00307402} {\bibfield
			{journal} {\bibinfo  {journal} {Phys. Chem. Miner.}\ }\textbf {\bibinfo
				{volume} {11}},\ \bibinfo {pages} {250} (\bibinfo {year} {1985})}\BibitemShut
		{NoStop}%
		\bibitem [{\citenamefont {Danan}\ \emph {et~al.}(1968)\citenamefont {Danan},
			\citenamefont {Herr},\ and\ \citenamefont {Meyer}}]{Danan_1968}%
		\BibitemOpen
		\bibfield  {author} {\bibinfo {author} {\bibfnamefont {H.}~\bibnamefont
				{Danan}}, \bibinfo {author} {\bibfnamefont {A.}~\bibnamefont {Herr}},\ and\
			\bibinfo {author} {\bibfnamefont {A.~J.~P.}\ \bibnamefont {Meyer}},\
		}\bibfield  {title} {\bibinfo {title} {New determinations of the saturation
				magnetization of nickel and iron},\ }\href
		{https://doi.org/10.1063/1.2163571} {\bibfield  {journal} {\bibinfo
				{journal} {Journal of Applied Physics}\ }\textbf {\bibinfo {volume} {39}},\
			\bibinfo {pages} {669} (\bibinfo {year} {1968})}\BibitemShut {NoStop}%
		\bibitem [{\citenamefont {Kozio{\l}-Rachwa{\l}}\ \emph
			{et~al.}(2013)\citenamefont {Kozio{\l}-Rachwa{\l}}, \citenamefont
			{Skowro{\'{n}}ski}, \citenamefont {{\'{S}}l{\c{e}}zak}, \citenamefont
			{Wilgocka-{\'{S}}l{\c{e}}zak}, \citenamefont {Przewo{\'{z}}nik},
			\citenamefont {Stobiecki}, \citenamefont {Qin}, \citenamefont {van Dijken},\
			and\ \citenamefont {Korecki}}]{Kozio-Rachwa2013}%
		\BibitemOpen
		\bibfield  {author} {\bibinfo {author} {\bibfnamefont {A.}~\bibnamefont
				{Kozio{\l}-Rachwa{\l}}}, \bibinfo {author} {\bibfnamefont {W.}~\bibnamefont
				{Skowro{\'{n}}ski}}, \bibinfo {author} {\bibfnamefont {T.}~\bibnamefont
				{{\'{S}}l{\c{e}}zak}}, \bibinfo {author} {\bibfnamefont {D.}~\bibnamefont
				{Wilgocka-{\'{S}}l{\c{e}}zak}}, \bibinfo {author} {\bibfnamefont
				{J.}~\bibnamefont {Przewo{\'{z}}nik}}, \bibinfo {author} {\bibfnamefont
				{T.}~\bibnamefont {Stobiecki}}, \bibinfo {author} {\bibfnamefont {Q.~H.}\
				\bibnamefont {Qin}}, \bibinfo {author} {\bibfnamefont {S.}~\bibnamefont {van
					Dijken}},\ and\ \bibinfo {author} {\bibfnamefont {J.}~\bibnamefont
				{Korecki}},\ }\bibfield  {title} {\bibinfo {title} {{Room-temperature
					perpendicular magnetic anisotropy of MgO/Fe/MgO ultrathin films}},\ }\href
		{https://doi.org/10.1063/1.4843675} {\bibfield  {journal} {\bibinfo
				{journal} {J. Appl. Phys.}\ }\textbf {\bibinfo {volume} {114}},\ \bibinfo
			{pages} {224307} (\bibinfo {year} {2013})}\BibitemShut {NoStop}%
		\bibitem [{\citenamefont {Duff}(1974)}]{cn:Duff66}%
		\BibitemOpen
		\bibfield  {author} {\bibinfo {author} {\bibfnamefont {K.~J.}\ \bibnamefont
				{Duff}},\ }\bibfield  {title} {\bibinfo {title} {Calibration of the isomer
				shift for $^{57}\mathrm{Fe}$},\ }\href
		{https://doi.org/10.1103/PhysRevB.9.66} {\bibfield  {journal} {\bibinfo
				{journal} {Phys. Rev. B}\ }\textbf {\bibinfo {volume} {9}},\ \bibinfo {pages}
			{66} (\bibinfo {year} {1974})}\BibitemShut {NoStop}%
		\bibitem [{\citenamefont {Blaha}\ \emph {et~al.}(1992)\citenamefont {Blaha},
			\citenamefont {Schwarz},\ and\ \citenamefont {Ray}}]{cn:Blaha92}%
		\BibitemOpen
		\bibfield  {author} {\bibinfo {author} {\bibfnamefont {P.}~\bibnamefont
				{Blaha}}, \bibinfo {author} {\bibfnamefont {K.}~\bibnamefont {Schwarz}},\
			and\ \bibinfo {author} {\bibfnamefont {A.}~\bibnamefont {Ray}},\ }\bibfield
		{title} {\bibinfo {title} {Isomer shifts and electric field gradients in
				y(fe1-xalx)2},\ }\href
		{https://doi.org/https://doi.org/10.1016/0304-8853(92)90983-U} {\bibfield
			{journal} {\bibinfo  {journal} {J. Magn. Magn. Mater.}\ }\textbf {\bibinfo
				{volume} {104-107}},\ \bibinfo {pages} {683} (\bibinfo {year}
			{1992})}\BibitemShut {NoStop}%
		\bibitem [{\citenamefont {Neese}(2002)}]{Neese2002}%
		\BibitemOpen
		\bibfield  {author} {\bibinfo {author} {\bibfnamefont {F.}~\bibnamefont
				{Neese}},\ }\bibfield  {title} {\bibinfo {title} {{Prediction and
					interpretation of the 57Fe isomer shift in M{\"{o}}ssbauer spectra by density
					functional theory}},\ }\href {https://doi.org/10.1016/S0020-1693(02)01031-9}
		{\bibfield  {journal} {\bibinfo  {journal} {Inorg. Chim. Acta}\ }\textbf
			{\bibinfo {volume} {337}},\ \bibinfo {pages} {181} (\bibinfo {year}
			{2002})}\BibitemShut {NoStop}%
		\bibitem [{\citenamefont {Wdowik}\ and\ \citenamefont
			{Ruebenbauer}(2007)}]{cn:Wdowik07}%
		\BibitemOpen
		\bibfield  {author} {\bibinfo {author} {\bibfnamefont {U.~D.}\ \bibnamefont
				{Wdowik}}\ and\ \bibinfo {author} {\bibfnamefont {K.}~\bibnamefont
				{Ruebenbauer}},\ }\bibfield  {title} {\bibinfo {title} {Calibration of the
				isomer shift for the $14.4\text{\ensuremath{-}}\mathrm{keV}$ transition in
				$^{57}\mathrm{Fe}$ using the full-potential linearized augmented plane-wave
				method},\ }\href {https://doi.org/10.1103/PhysRevB.76.155118} {\bibfield
			{journal} {\bibinfo  {journal} {Phys. Rev. B}\ }\textbf {\bibinfo {volume}
				{76}},\ \bibinfo {pages} {155118} (\bibinfo {year} {2007})}\BibitemShut
		{NoStop}%
		\bibitem [{\citenamefont {Spiel}\ \emph {et~al.}(2009)\citenamefont {Spiel},
			\citenamefont {Blaha},\ and\ \citenamefont {Schwarz}}]{cn:Spiel09}%
		\BibitemOpen
		\bibfield  {author} {\bibinfo {author} {\bibfnamefont {C.}~\bibnamefont
				{Spiel}}, \bibinfo {author} {\bibfnamefont {P.}~\bibnamefont {Blaha}},\ and\
			\bibinfo {author} {\bibfnamefont {K.}~\bibnamefont {Schwarz}},\ }\bibfield
		{title} {\bibinfo {title} {Density functional calculations on the
				charge-ordered and valence-mixed modification of
				${\text{ybafe}}_{2}{\text{o}}_{5}$},\ }\href
		{https://doi.org/10.1103/PhysRevB.79.115123} {\bibfield  {journal} {\bibinfo
				{journal} {Phys. Rev. B}\ }\textbf {\bibinfo {volume} {79}},\ \bibinfo
			{pages} {115123} (\bibinfo {year} {2009})}\BibitemShut {NoStop}%
		\bibitem [{\citenamefont {Kurian}\ and\ \citenamefont
			{Filatov}(2010)}]{Kurian2010Isomer}%
		\BibitemOpen
		\bibfield  {author} {\bibinfo {author} {\bibfnamefont {R.}~\bibnamefont
				{Kurian}}\ and\ \bibinfo {author} {\bibfnamefont {M.}~\bibnamefont
				{Filatov}},\ }\bibfield  {title} {\bibinfo {title} {Calibration of 57fe
				isomer shift from ab initio calculations: can theory and experiment reach an
				agreement?},\ }\href {https://doi.org/10.1039/B918655G} {\bibfield  {journal}
			{\bibinfo  {journal} {Phys. Chem. Chem. Phys.}\ }\textbf {\bibinfo {volume}
				{12}},\ \bibinfo {pages} {2758} (\bibinfo {year} {2010})}\BibitemShut
		{NoStop}%
		\bibitem [{\citenamefont {Dufek}\ \emph {et~al.}(1995)\citenamefont {Dufek},
			\citenamefont {Blaha},\ and\ \citenamefont {Schwarz}}]{cn:Dufek95}%
		\BibitemOpen
		\bibfield  {author} {\bibinfo {author} {\bibfnamefont {P.}~\bibnamefont
				{Dufek}}, \bibinfo {author} {\bibfnamefont {P.}~\bibnamefont {Blaha}},\ and\
			\bibinfo {author} {\bibfnamefont {K.}~\bibnamefont {Schwarz}},\ }\bibfield
		{title} {\bibinfo {title} {Determination of the nuclear quadrupole moment of
				${}^{57}$fe},\ }\href {https://doi.org/10.1103/PhysRevLett.75.3545}
		{\bibfield  {journal} {\bibinfo  {journal} {Phys. Rev. Lett.}\ }\textbf
			{\bibinfo {volume} {75}},\ \bibinfo {pages} {3545} (\bibinfo {year}
			{1995})}\BibitemShut {NoStop}%
		\bibitem [{\citenamefont {Petrilli}\ \emph {et~al.}(1998)\citenamefont
			{Petrilli}, \citenamefont {Bl\"ochl}, \citenamefont {Blaha},\ and\
			\citenamefont {Schwarz}}]{cn:Petrilli98}%
		\BibitemOpen
		\bibfield  {author} {\bibinfo {author} {\bibfnamefont {H.~M.}\ \bibnamefont
				{Petrilli}}, \bibinfo {author} {\bibfnamefont {P.~E.}\ \bibnamefont
				{Bl\"ochl}}, \bibinfo {author} {\bibfnamefont {P.}~\bibnamefont {Blaha}},\
			and\ \bibinfo {author} {\bibfnamefont {K.}~\bibnamefont {Schwarz}},\
		}\bibfield  {title} {\bibinfo {title} {Electric-field-gradient calculations
				using the projector augmented wave method},\ }\href
		{https://doi.org/10.1103/PhysRevB.57.14690} {\bibfield  {journal} {\bibinfo
				{journal} {Phys. Rev. B}\ }\textbf {\bibinfo {volume} {57}},\ \bibinfo
			{pages} {14690} (\bibinfo {year} {1998})}\BibitemShut {NoStop}%
		\bibitem [{\citenamefont {Mart\'{\i}nez-Pinedo}\ \emph
			{et~al.}(2001)\citenamefont {Mart\'{\i}nez-Pinedo}, \citenamefont
			{Schwerdtfeger}, \citenamefont {Caurier}, \citenamefont {Langanke},
			\citenamefont {Nazarewicz},\ and\ \citenamefont {S\"ohnel}}]{cn:Martinez01}%
		\BibitemOpen
		\bibfield  {author} {\bibinfo {author} {\bibfnamefont {G.}~\bibnamefont
				{Mart\'{\i}nez-Pinedo}}, \bibinfo {author} {\bibfnamefont {P.}~\bibnamefont
				{Schwerdtfeger}}, \bibinfo {author} {\bibfnamefont {E.}~\bibnamefont
				{Caurier}}, \bibinfo {author} {\bibfnamefont {K.}~\bibnamefont {Langanke}},
			\bibinfo {author} {\bibfnamefont {W.}~\bibnamefont {Nazarewicz}},\ and\
			\bibinfo {author} {\bibfnamefont {T.}~\bibnamefont {S\"ohnel}},\ }\bibfield
		{title} {\bibinfo {title} {Nuclear quadrupole moment of $^{57}fe$ from
				microscopic nuclear and atomic calculations},\ }\href
		{https://doi.org/10.1103/PhysRevLett.87.062701} {\bibfield  {journal}
			{\bibinfo  {journal} {Phys. Rev. Lett.}\ }\textbf {\bibinfo {volume} {87}},\
			\bibinfo {pages} {062701} (\bibinfo {year} {2001})}\BibitemShut {NoStop}%
		\bibitem [{\citenamefont {Ebert}\ and\ \citenamefont
			{Battocletti}(2005)}]{cn:Ebert05}%
		\BibitemOpen
		\bibfield  {author} {\bibinfo {author} {\bibfnamefont {H.}~\bibnamefont
				{Ebert}}\ and\ \bibinfo {author} {\bibfnamefont {M.}~\bibnamefont
				{Battocletti}},\ }\bibfield  {title} {\bibinfo {title} {Spin-orbit induced
				electric field gradients in magnetic solids},\ }in\ \href@noop {} {\emph
			{\bibinfo {booktitle} {HFI/NQI 2004}}},\ \bibinfo {editor} {edited by\
			\bibinfo {editor} {\bibfnamefont {K.}~\bibnamefont {Maier}}\ and\ \bibinfo
			{editor} {\bibfnamefont {R.}~\bibnamefont {Vianden}}}\ (\bibinfo  {publisher}
		{Springer},\ \bibinfo {address} {Berlin, Heidelberg},\ \bibinfo {year}
		{2005})\ p.~\bibinfo {pages} {25}\BibitemShut {NoStop}%
		\bibitem [{\citenamefont {Kaufmann}\ and\ \citenamefont
			{Vianden}(1979)}]{Kaufmann1979EFG}%
		\BibitemOpen
		\bibfield  {author} {\bibinfo {author} {\bibfnamefont {E.~N.}\ \bibnamefont
				{Kaufmann}}\ and\ \bibinfo {author} {\bibfnamefont {R.~J.}\ \bibnamefont
				{Vianden}},\ }\bibfield  {title} {\bibinfo {title} {The electric field
				gradient in noncubic metals},\ }\href
		{https://doi.org/10.1103/RevModPhys.51.161} {\bibfield  {journal} {\bibinfo
				{journal} {Rev. Mod. Phys.}\ }\textbf {\bibinfo {volume} {51}},\ \bibinfo
			{pages} {161} (\bibinfo {year} {1979})}\BibitemShut {NoStop}%
		\bibitem [{\citenamefont {Chumakov}\ and\ \citenamefont
			{Sturhahn}(1999)}]{Chumakov1999}%
		\BibitemOpen
		\bibfield  {author} {\bibinfo {author} {\bibfnamefont {A.}~\bibnamefont
				{Chumakov}}\ and\ \bibinfo {author} {\bibfnamefont {W.}~\bibnamefont
				{Sturhahn}},\ }\bibfield  {title} {\bibinfo {title} {{Experimental aspects of
					inelastic nuclear resonance scattering}},\ }\href
		{https://doi.org/10.1023/A:1017052730094} {\bibfield  {journal} {\bibinfo
				{journal} {Hyperfine Interact.}\ }\textbf {\bibinfo {volume} {123/124}},\
			\bibinfo {pages} {781} (\bibinfo {year} {1999})}\BibitemShut {NoStop}%
		\bibitem [{\citenamefont {Singwi}\ and\ \citenamefont
			{Sj{\"{o}}lander}(1960)}]{Singwi1960}%
		\BibitemOpen
		\bibfield  {author} {\bibinfo {author} {\bibfnamefont {K.~S.}\ \bibnamefont
				{Singwi}}\ and\ \bibinfo {author} {\bibfnamefont {A.}~\bibnamefont
				{Sj{\"{o}}lander}},\ }\bibfield  {title} {\bibinfo {title} {{Resonance
					Absorption of Nuclear Gamma Rays and the Dynamics of Atomic Motions}},\
		}\href {https://doi.org/10.1103/PhysRev.120.1093} {\bibfield  {journal}
			{\bibinfo  {journal} {Phys. Rev.}\ }\textbf {\bibinfo {volume} {120}},\
			\bibinfo {pages} {1093} (\bibinfo {year} {1960})}\BibitemShut {NoStop}%
		\bibitem [{\citenamefont {Kohn}\ \emph {et~al.}(1998)\citenamefont {Kohn},
			\citenamefont {Chumakov},\ and\ \citenamefont {R{\"{u}}ffer}}]{Kohn1998}%
		\BibitemOpen
		\bibfield  {author} {\bibinfo {author} {\bibfnamefont {V.~G.}\ \bibnamefont
				{Kohn}}, \bibinfo {author} {\bibfnamefont {A.~I.}\ \bibnamefont {Chumakov}},\
			and\ \bibinfo {author} {\bibfnamefont {R.}~\bibnamefont {R{\"{u}}ffer}},\
		}\bibfield  {title} {\bibinfo {title} {{Nuclear resonant inelastic absorption
					of synchrotron radiation in an anisotropic single crystal}},\ }\href
		{https://doi.org/10.1103/PhysRevB.58.8437} {\bibfield  {journal} {\bibinfo
				{journal} {Phys. Rev. B}\ }\textbf {\bibinfo {volume} {58}},\ \bibinfo
			{pages} {8437} (\bibinfo {year} {1998})}\BibitemShut {NoStop}%
		\bibitem [{\citenamefont {Grimvall}(1999)}]{Grimvall1999}%
		\BibitemOpen
		\bibfield  {author} {\bibinfo {author} {\bibfnamefont {G.}~\bibnamefont
				{Grimvall}},\ }\href@noop {} {\emph {\bibinfo {title} {{Thermophysical
						properties of materials}}}}\ (\bibinfo  {publisher} {Elsevier},\ \bibinfo
		{year} {1999})\BibitemShut {NoStop}%
		\bibitem [{\citenamefont {Fultz}(2010)}]{Fultz_2010}%
		\BibitemOpen
		\bibfield  {author} {\bibinfo {author} {\bibfnamefont {B.}~\bibnamefont
				{Fultz}},\ }\bibfield  {title} {\bibinfo {title} {Vibrational thermodynamics
				of materials},\ }\href {https://doi.org/10.1016/j.pmatsci.2009.05.002}
		{\bibfield  {journal} {\bibinfo  {journal} {Prog. Mater Sci.}\ }\textbf
			{\bibinfo {volume} {55}},\ \bibinfo {pages} {247} (\bibinfo {year}
			{2010})}\BibitemShut {NoStop}%
		\bibitem [{\citenamefont {Stankov}\ \emph {et~al.}(2010)\citenamefont
			{Stankov}, \citenamefont {Miglierini}, \citenamefont {Chumakov},
			\citenamefont {Sergueev}, \citenamefont {Yue}, \citenamefont {Sepiol},
			\citenamefont {Svec}, \citenamefont {Hu},\ and\ \citenamefont
			{R{\"{u}}ffer}}]{Stankov2010}%
		\BibitemOpen
		\bibfield  {author} {\bibinfo {author} {\bibfnamefont {S.}~\bibnamefont
				{Stankov}}, \bibinfo {author} {\bibfnamefont {M.}~\bibnamefont {Miglierini}},
			\bibinfo {author} {\bibfnamefont {A.~I.}\ \bibnamefont {Chumakov}}, \bibinfo
			{author} {\bibfnamefont {I.}~\bibnamefont {Sergueev}}, \bibinfo {author}
			{\bibfnamefont {Y.~Z.}\ \bibnamefont {Yue}}, \bibinfo {author} {\bibfnamefont
				{B.}~\bibnamefont {Sepiol}}, \bibinfo {author} {\bibfnamefont
				{P.}~\bibnamefont {Svec}}, \bibinfo {author} {\bibfnamefont {L.}~\bibnamefont
				{Hu}},\ and\ \bibinfo {author} {\bibfnamefont {R.}~\bibnamefont
				{R{\"{u}}ffer}},\ }\bibfield  {title} {\bibinfo {title} {{Vibrational
					thermodynamics of Fe 90 Zr 7 B 3 nanocrystalline alloy from nuclear inelastic
					scattering}},\ }\href {https://doi.org/10.1103/PhysRevB.82.144301} {\bibfield
			{journal} {\bibinfo  {journal} {Phys. Rev. B}\ }\textbf {\bibinfo {volume}
				{82}},\ \bibinfo {pages} {144301} (\bibinfo {year} {2010})}\BibitemShut
		{NoStop}%
		\bibitem [{\citenamefont {Hu}\ \emph {et~al.}(2013)\citenamefont {Hu},
			\citenamefont {Toellner}, \citenamefont {Dauphas}, \citenamefont {Alp},\ and\
			\citenamefont {Zhao}}]{Hu2013}%
		\BibitemOpen
		\bibfield  {author} {\bibinfo {author} {\bibfnamefont {M.~Y.}\ \bibnamefont
				{Hu}}, \bibinfo {author} {\bibfnamefont {T.~S.}\ \bibnamefont {Toellner}},
			\bibinfo {author} {\bibfnamefont {N.}~\bibnamefont {Dauphas}}, \bibinfo
			{author} {\bibfnamefont {E.~E.}\ \bibnamefont {Alp}},\ and\ \bibinfo {author}
			{\bibfnamefont {J.}~\bibnamefont {Zhao}},\ }\bibfield  {title} {\bibinfo
			{title} {{Moments in nuclear resonant inelastic x-ray scattering and their
					applications}},\ }\href {https://doi.org/10.1103/PhysRevB.87.064301}
		{\bibfield  {journal} {\bibinfo  {journal} {Phys. Rev. B}\ }\textbf {\bibinfo
				{volume} {87}},\ \bibinfo {pages} {064301} (\bibinfo {year}
			{2013})}\BibitemShut {NoStop}%
		\bibitem [{\citenamefont {Gruner}\ \emph {et~al.}(2015)\citenamefont {Gruner},
			\citenamefont {Keune}, \citenamefont {Cuenya}, \citenamefont {Weis},
			\citenamefont {Landers}, \citenamefont {Makarov}, \citenamefont {Klar},
			\citenamefont {Hu}, \citenamefont {Alp}, \citenamefont {Zhao}, \citenamefont
			{Krautz}, \citenamefont {Gutfleisch},\ and\ \citenamefont
			{Wende}}]{cn:Gruner15PRL}%
		\BibitemOpen
		\bibfield  {author} {\bibinfo {author} {\bibfnamefont {M.~E.}\ \bibnamefont
				{Gruner}}, \bibinfo {author} {\bibfnamefont {W.}~\bibnamefont {Keune}},
			\bibinfo {author} {\bibfnamefont {B.~R.}\ \bibnamefont {Cuenya}}, \bibinfo
			{author} {\bibfnamefont {C.}~\bibnamefont {Weis}}, \bibinfo {author}
			{\bibfnamefont {J.}~\bibnamefont {Landers}}, \bibinfo {author} {\bibfnamefont
				{S.~I.}\ \bibnamefont {Makarov}}, \bibinfo {author} {\bibfnamefont
				{D.}~\bibnamefont {Klar}}, \bibinfo {author} {\bibfnamefont {M.~Y.}\
				\bibnamefont {Hu}}, \bibinfo {author} {\bibfnamefont {E.~E.}\ \bibnamefont
				{Alp}}, \bibinfo {author} {\bibfnamefont {J.}~\bibnamefont {Zhao}}, \bibinfo
			{author} {\bibfnamefont {M.}~\bibnamefont {Krautz}}, \bibinfo {author}
			{\bibfnamefont {O.}~\bibnamefont {Gutfleisch}},\ and\ \bibinfo {author}
			{\bibfnamefont {H.}~\bibnamefont {Wende}},\ }\bibfield  {title} {\bibinfo
			{title} {Element-resolved thermodynamics of magnetocaloric {\rm
					lafe$_{13-x}$si$_x$}},\ }\href@noop {} {\bibfield  {journal} {\bibinfo
				{journal} {Phys. Rev. Lett.}\ }\textbf {\bibinfo {volume} {114}},\ \bibinfo
			{pages} {057202} (\bibinfo {year} {2015})}\BibitemShut {NoStop}%
		\bibitem [{\citenamefont {Landers}\ \emph {et~al.}(2018)\citenamefont
			{Landers}, \citenamefont {Salamon}, \citenamefont {Keune}, \citenamefont
			{Gruner}, \citenamefont {Krautz}, \citenamefont {Zhao}, \citenamefont {Hu},
			\citenamefont {Toellner}, \citenamefont {Alp}, \citenamefont {Gutfleisch},\
			and\ \citenamefont {Wende}}]{cn:Landers18}%
		\BibitemOpen
		\bibfield  {author} {\bibinfo {author} {\bibfnamefont {J.}~\bibnamefont
				{Landers}}, \bibinfo {author} {\bibfnamefont {S.}~\bibnamefont {Salamon}},
			\bibinfo {author} {\bibfnamefont {W.}~\bibnamefont {Keune}}, \bibinfo
			{author} {\bibfnamefont {M.~E.}\ \bibnamefont {Gruner}}, \bibinfo {author}
			{\bibfnamefont {M.}~\bibnamefont {Krautz}}, \bibinfo {author} {\bibfnamefont
				{J.}~\bibnamefont {Zhao}}, \bibinfo {author} {\bibfnamefont {M.~Y.}\
				\bibnamefont {Hu}}, \bibinfo {author} {\bibfnamefont {T.}~\bibnamefont
				{Toellner}}, \bibinfo {author} {\bibfnamefont {E.~E.}\ \bibnamefont {Alp}},
			\bibinfo {author} {\bibfnamefont {O.}~\bibnamefont {Gutfleisch}},\ and\
			\bibinfo {author} {\bibfnamefont {H.}~\bibnamefont {Wende}},\ }\bibfield
		{title} {\bibinfo {title} {Determining the vibrational entropy change in
				giant magnetocaloric {LaFe$_{11.6}$Si$_{1.4}$} by nuclear resonant inelastic
				x-ray scattering},\ }\href@noop {} {\bibfield  {journal} {\bibinfo  {journal}
				{Phys. Rev. B}\ }\textbf {\bibinfo {volume} {98}},\ \bibinfo {pages} {024417}
			(\bibinfo {year} {2018})}\BibitemShut {NoStop}%
		\bibitem [{\citenamefont {Achterhold}\ \emph {et~al.}(2002)\citenamefont
			{Achterhold}, \citenamefont {Keppler}, \citenamefont {Ostermann},
			\citenamefont {van B{\"{u}}rck}, \citenamefont {Sturhahn}, \citenamefont
			{Alp},\ and\ \citenamefont {Parak}}]{Achterhold2002}%
		\BibitemOpen
		\bibfield  {author} {\bibinfo {author} {\bibfnamefont {K.}~\bibnamefont
				{Achterhold}}, \bibinfo {author} {\bibfnamefont {C.}~\bibnamefont {Keppler}},
			\bibinfo {author} {\bibfnamefont {A.}~\bibnamefont {Ostermann}}, \bibinfo
			{author} {\bibfnamefont {U.}~\bibnamefont {van B{\"{u}}rck}}, \bibinfo
			{author} {\bibfnamefont {W.}~\bibnamefont {Sturhahn}}, \bibinfo {author}
			{\bibfnamefont {E.~E.}\ \bibnamefont {Alp}},\ and\ \bibinfo {author}
			{\bibfnamefont {F.~G.}\ \bibnamefont {Parak}},\ }\bibfield  {title} {\bibinfo
			{title} {{Vibrational dynamics of myoglobin determined by the phonon-assisted
					M{\"{o}}ssbauer effect}},\ }\href
		{https://doi.org/10.1103/PhysRevE.65.051916} {\bibfield  {journal} {\bibinfo
				{journal} {Physical Review E}\ }\textbf {\bibinfo {volume} {65}},\ \bibinfo
			{pages} {051916} (\bibinfo {year} {2002})}\BibitemShut {NoStop}%
		\bibitem [{\citenamefont {Hu}\ \emph {et~al.}(2003)\citenamefont {Hu},
			\citenamefont {Sturhahn}, \citenamefont {Toellner}, \citenamefont {Mannheim},
			\citenamefont {{E. Brown}}, \citenamefont {Zhao},\ and\ \citenamefont
			{Alp}}]{Hu2003}%
		\BibitemOpen
		\bibfield  {author} {\bibinfo {author} {\bibfnamefont {M.~Y.}\ \bibnamefont
				{Hu}}, \bibinfo {author} {\bibfnamefont {W.}~\bibnamefont {Sturhahn}},
			\bibinfo {author} {\bibfnamefont {T.~S.}\ \bibnamefont {Toellner}}, \bibinfo
			{author} {\bibfnamefont {P.~D.}\ \bibnamefont {Mannheim}}, \bibinfo {author}
			{\bibfnamefont {D.}~\bibnamefont {{E. Brown}}}, \bibinfo {author}
			{\bibfnamefont {J.}~\bibnamefont {Zhao}},\ and\ \bibinfo {author}
			{\bibfnamefont {E.~E.}\ \bibnamefont {Alp}},\ }\bibfield  {title} {\bibinfo
			{title} {{Measuring velocity of sound with nuclear resonant inelastic x-ray
					scattering}},\ }\href {https://doi.org/10.1103/PhysRevB.67.094304} {\bibfield
			{journal} {\bibinfo  {journal} {Phys. Rev. B}\ }\textbf {\bibinfo {volume}
				{67}},\ \bibinfo {pages} {094304} (\bibinfo {year} {2003})}\BibitemShut
		{NoStop}%
		\bibitem [{\citenamefont {Morrison}\ \emph {et~al.}(2019)\citenamefont
			{Morrison}, \citenamefont {Jackson}, \citenamefont {Sturhahn}, \citenamefont
			{Zhao},\ and\ \citenamefont {Toellner}}]{Morrison_2019}%
		\BibitemOpen
		\bibfield  {author} {\bibinfo {author} {\bibfnamefont {R.~A.}\ \bibnamefont
				{Morrison}}, \bibinfo {author} {\bibfnamefont {J.~M.}\ \bibnamefont
				{Jackson}}, \bibinfo {author} {\bibfnamefont {W.}~\bibnamefont {Sturhahn}},
			\bibinfo {author} {\bibfnamefont {J.}~\bibnamefont {Zhao}},\ and\ \bibinfo
			{author} {\bibfnamefont {T.~S.}\ \bibnamefont {Toellner}},\ }\bibfield
		{title} {\bibinfo {title} {High pressure thermoelasticity and sound
				velocities of fe-ni-si alloys},\ }\href
		{https://doi.org/10.1016/j.pepi.2019.05.011} {\bibfield  {journal} {\bibinfo
				{journal} {Physics of the Earth and Planetary Interiors}\ }\textbf {\bibinfo
				{volume} {294}},\ \bibinfo {pages} {106268} (\bibinfo {year}
			{2019})}\BibitemShut {NoStop}%
		\bibitem [{\citenamefont {Toellner}\ \emph {et~al.}(1997)\citenamefont
			{Toellner}, \citenamefont {Hu}, \citenamefont {Sturhahn}, \citenamefont
			{Quast},\ and\ \citenamefont {Alp}}]{Toellner1997}%
		\BibitemOpen
		\bibfield  {author} {\bibinfo {author} {\bibfnamefont {T.~S.}\ \bibnamefont
				{Toellner}}, \bibinfo {author} {\bibfnamefont {M.~Y.}\ \bibnamefont {Hu}},
			\bibinfo {author} {\bibfnamefont {W.}~\bibnamefont {Sturhahn}}, \bibinfo
			{author} {\bibfnamefont {K.}~\bibnamefont {Quast}},\ and\ \bibinfo {author}
			{\bibfnamefont {E.~E.}\ \bibnamefont {Alp}},\ }\bibfield  {title} {\bibinfo
			{title} {Inelastic nuclear resonant scattering with sub-{meV} energy
				resolution},\ }\href {https://doi.org/10.1063/1.120448} {\bibfield  {journal}
			{\bibinfo  {journal} {Appl. Phys. Lett.}\ }\textbf {\bibinfo {volume} {71}},\
			\bibinfo {pages} {2112} (\bibinfo {year} {1997})}\BibitemShut {NoStop}%
	\end{thebibliography}
	%
	
\end{document}